\pdfoutput=1
\documentclass[usenatbib]{mnras}
\usepackage{epsfig}
\usepackage{color}
\usepackage{graphicx}
\usepackage{tabularx}
\usepackage{mathtools}

\usepackage{xcolor}

\newcommand{\Ms}{\mathrm{M}_\odot}

\newcommand{\eagle}{{\sc Eagle }}
\newcommand{\aquarius}{{\sc Aquarius }}
\newcommand{\vialactea}{{\sc Via Lactea II }}
\newcommand{\apostle}{{\sc Apostle }}
\newcommand{\subfind}{{\sc Subfind }}

\newcommand\eqnref[1]{Eqn.~(\ref{#1})} 
\newcommand\eqnrefp[1]{(Eqn.~\ref{#1})} 
\newcommand{\ts}{\textsuperscript} 

\title[Shaken and Stirred: The Milky Way’s Dark Substructures]{Shaken and Stirred: The Milky Way's Dark Substructures}

\author[Sawala et al.]  {\parbox{\textwidth}{Till
    Sawala$^1$\thanks{E-mail: \texttt{till.sawala@helsinki.fi}}, Pauli
    Pihajoki$^1$, Peter~H.~Johansson$^1$, 
    \mbox{Carlos~S.~Frenk$^2$}, Julio~F.~Navarro$^{3,4}$, Kyle~A.~Oman$^3$ and Simon~D.~M.~White$^5$}\vspace{0.4cm}\\
  \parbox{\textwidth}{
    $^{1}$Department of Physics, University of Helsinki, Gustaf
    H\"allstr\"omin katu 2a,
    FI-00014 Helsinki, Finland\\
    $^{2}$Institute for Computational Cosmology, Department of Physics,   
    University of Durham, South Road, Durham DH1 3LE, UK \\
    $^{3}$Department of Physics and Astronomy, University of Victoria,
    3800 Finnerty Road, Victoria, British Columbia V8P 5C2, Canada\\
    $^{4}$Senior CIfAR Fellow\\
    $^{5}$Max-Planck Institute for Astrophysics,
    Karl-Schwarzschild-Straße 1, 85741 Garching, Germany }}
\begin{document}

\date{Accepted 2016 ***. Received 2016 ***; in original
  form 2016}

\pagerange{\pageref{firstpage}--\pageref{lastpage}} \pubyear{2016}

\maketitle

\label{firstpage}

\begin{abstract}
  The predicted abundance and properties of the low-mass substructures
  embedded inside larger dark matter haloes differ sharply among
  alternative dark matter models. Too small to host galaxies
  themselves, these subhaloes may still be detected via gravitational
  lensing, or via perturbations of the Milky Way's globular cluster
  streams and its stellar disk. Here we use the \apostle cosmological
  simulations to predict the abundance and the spatial and velocity
  distributions of subhaloes in the range $10^{6.5}-10^{8.5}\Ms$
  inside haloes of mass $\sim10^{12} \Ms$ in $\Lambda$CDM. Although
  these subhaloes are themselves devoid of baryons, we find that
  baryonic effects are important. Compared to corresponding dark
  matter only simulations, the loss of baryons from subhaloes and
  stronger tidal disruption due to the presence of baryons near the
  centre of the main halo, reduce the number of subhaloes by $\sim
  1/4$ to $1/2$, independently of subhalo mass, but increasingly
  towards the host halo centre. We also find that subhaloes have
  non-Maxwellian orbital velocity distributions, with centrally rising
  velocity anisotropy and positive velocity bias which reduces the
  number of low-velocity subhaloes, particularly near the halo
  centre. We parameterise the predicted population of subhaloes in
  terms of mass, galactocentric distance, and velocities. We discuss
  implications of our results for the prospects of detecting dark
  matter substructures and for possible inferences about the nature of
  dark matter.
\end{abstract}


\section{Introduction} \label{sec:Introduction} While the $\Lambda$
Cold Dark Matter (hereafter $\Lambda$CDM) model explains many large
scale observations, from the anisotropy of the microwave background
radiation \citep[e.g.][]{Wright-1992} to the distribution of galaxies
in the cosmic web \citep{Davis-1985}, inferences about the particle
nature of dark matter or its possible (self)-interactions require
observations require observations on far smaller scales. Warm Dark
Matter (WDM) particles, such as sterile neutrinos with masses of a few
keV, have free-streaming scales of less than 100 kpc, and differ from
CDM in terms of the halo mass functions at mass scales on the order of
$10^9 \Ms$ and below \citep[e.g.][]{Avila-Reese-2001, Bose-2016},
while weak self-interactions would produce shallow cores of the order
of several kpc in the centre of dark matter haloes
\citep[e.g.][]{Spergel-2000}. In principle, there is no shortage of
observations that probe these small scales. They include the
structures seen in the Lyman-$\alpha$ forest
\citep[e.g.][]{Croft-2002, Viel-2013}, the abundance of dwarf galaxies
in deep HI surveys \citep{Tikhonov-2009, Papastergis-2011}, and the
abundance \citep[e.g.][]{Klypin-1999, Boylan-Kolchin-2011,
  Lovell-2012, Kennedy-2014} as well as internal kinematics that probe
the density profiles \citep[e.g.][]{Walker-2011, Strigari-2014} of
Local Group dwarf galaxies.

While these studies have progressively narrowed the parameter space of
viable dark matter candidates, inferences about the non-baryonic
nature of dark matter from observations of the Universe's baryonic
components are inherently limited by uncertainties in our
understanding of complex astrophysical processes, such as radiative
hydrodynamics, gas cooling, star formation, metal-enrichment, stellar
winds, supernova and AGN feedback, and cosmic reionisation. For simple
number counts, the effects of baryons in suppressing the formation of
dwarf galaxies in CDM can be degenerate with the effects of warm dark
matter \citep[e.g.][]{Sawala-2013}. As of 2016, a plethora of studies
have also offered baryonic solutions to the various problems for
$\Lambda$CDM that had previously been identified in Dark Matter Only
(hereafter DMO) simulations \citep[e.g.][]{Okamoto-2008,
  Governato-2010, Zolotov-2012, Brooks-2013, Arraki-2014, Chan-2015,
  Sawala-2015, Dutton-2016}.

In addition, in the $\Lambda$CDM cosmological model, the majority of
low-mass substructures which would most easily discriminate between
different dark-matter models are predicted to be completely dark
\citep{Bullock-2000, Benson-2002, Okamoto-2008, Sawala-2016a,
  Ocvirk-2015}, and hence unobservable through starlight. Fortunately,
alternative methods exist that can reveal small structures and
substructures purely through their gravitational effect and detect
even pure dark matter haloes, thereby potentially breaking the
degeneracy with baryonic physics:

\begin{itemize}
\item Gravitational lensing directly probes the projected mass
  distribution in and around galaxies and can reveal their luminous
  and non-luminous components. Weak gravitational lensing has
  confirmed the existence of massive dark haloes surrounding galaxies
  down to Milky-Way scales, or masses of $\sim 10^{12}\Ms$
  \citep[e.g.][]{Mandelbaum-2006}. While these provide strong evidence
  for the existence of non-baryonic dark matter, they cannot
  distinguish between different currently viable models of cold, warm
  or self-interacting dark matter that deviate on mass scales below
  $\sim10^{9}\Ms$. However, much lower masses, down to $\sim 10^6\Ms$,
  may be probed through strong gravitational lensing, either via
  flux-ratio anomalies \citep[e.g.][]{Mao-1998, Xu-2009, Xu-2015}, or
  detectable perturbations of observed Einstein rings by
  substructures in the lens itself or along the line of sight
  \citep{Mao-1998, Metcalf-2001, Dalal-2002, Vegetti-2012,
    Vegetti-2014}. On these scales, different dark matter models may
  be clearly distinguished, provided that the expected abundances and
  distributions of substructures for different models can be reliably
  predicted.

\item Gaps in stellar streams originating from the tidal disruption of
  either globular clusters or dwarf galaxies can also provide evidence
  for substructures. In particular, globular cluster streams in the
  Milky Way, such as Palomar-5 (hereafter Pal-5, discovered by
  \citealt{Odenkirchen-2001}) and GD-1 (discovered by
  \citealt{Grillmair-2006}) can be stretched out over many kiloparsecs
  along their orbit while conserving their phase-space
  volume. Compared to dwarf galaxies, globular clusters have much
  lower internal velocity dispersions resulting in much narrower
  streams, making them very sensitive tracers both of the
  Galactic potential, and of perturbations by low-mass
  substructures \citep[e.g.][]{Ibata-2002, Carlberg-2013}. Based on
  the \vialactea DMO simulations, \cite{Yoon-2011} have calculated the
  interaction frequency of the Pal-5 stream with dark substructures
  during its assumed lifetime of 550 Myrs; they predicted $\sim 20$
  direct encounters with subhaloes of $10^6-10^7\Ms$, and $\sim 5$
  with subhaloes above $10^7\Ms$. \cite{Erkal-2015a, Erkal-2015b} have
  computed the properties of predicted gaps in streams such as Pal-5
  and GD-1 in $\Lambda$CDM. They show that the improved photometry,
  greater depth, and more precise radial velocity and proper motion
  measurements of upcoming surveys such as Gaia \citep{Perryman-2001,
    Gaia-2012}, DES \citep{DES-2005} and LSST \citep{LSST-2009} should
  allow a characterisation of perturbers in terms of mass,
  concentration, impact time, and 3D velocity, for subhaloes above
  $10^7\Ms$, albeit with an irreducible degeneracy between mass and
  velocity. Recently, \cite{Bovy-2016} have used the density data of
  Pal-5 to infer the number of subhaloes in the mass range $M =
  10^{6.5} - 10^{9} \Ms$ inside the central 20 kpc of the Milky Way to
  be $10^{+11}_{-6}$. However, they also noted the uncertainty due to
  unaccounted baryonic effects, and due to the required assumptions in
  the subhalo velocity distribution.

\item The cold thin stellar disk of the Milky Way is another sensitive
  probe of the interactions with orbiting low-mass
  substructures. Satellite substructures passing through the Milky Way
  disk are expected to cause small but detectable changes in both the
  radial and vertical velocity distribution of stars in the disk,
  resulting in a thickening of the thin disk \citep[e.g.][]{Toth-1992,
    Quinn-1993, Navarro-1994, Walker-1996, Sellwood-1998, Benson-2004,
    Kazantzidis-2008}. The thinness and long-term stability of the
  Milky Way stellar disk could thus potentially put strong limits on
  the number of allowed massive dark substructures in the vicinity of
  the disk, and recent work by \cite{Feldmann-2015} suggest that the
  expected increase in the vertical velocity dispersion of disk stars
  due to the impact of dark substructures should be detectable with
  Gaia. However, the vertical heating and thickening of the disk by
  dark substructures are severely reduced in simulations that include
  dissipational gas physics. The inclusion of gas reduces disk heating
  mainly through two mechanisms: the absorption of kinetic impact
  energy by the gas and/or the formation of a new thin stellar disk
  that can recontract heated stars towards the disk plane
  \citep[e.g.][]{ Stewart-2009, Hopkins-2009, Moster-2010}.
\end{itemize}

While the above phenomena have a gravitational origin, they still fall
short of providing a complete census of dark matter
substructures. Instead, inferences about dark matter models based on
the number of detected perturbations must be made statistically, and
in each case, require an accurate prediction of the abundance,
properties and distribution of dark matter substructures inside the
central $\sim 10-20$~kpc of galaxy or group-sized dark matter haloes.

Previous work has relied on very high resolution DMO simulations, such
as \vialactea \citep{Diemand-2007} and \aquarius
\citep{Springel-2008}. These have shown that tidal stripping reduces
the mass fraction of dark matter contained in self-bound
substructures towards the halo centre
\citep[e.g.][]{Springel-2008}. It has also been argued that the
presence of a stellar disk and adiabatic contraction of the halo can
lead to enhanced tidal disruption of substructures. Based on DMO
simulations with an additional massive disk-like potential,
\cite{Donghia-2010} quantified the disruption of substructures through
tidal stripping due to the smooth halo, tidal stirring near
pericentre, and ``disk shocking'' by the passage of a substructure
through the dense stellar disk. For their parameters, this led to a
depletion of substructures by up to a factor of 3 for a subhaloes of
mass $10^{7}\Ms$. Similarly, \cite{Yurin-2015} imposed a less massive
disk inside a DMO simulation, and found a reduction in subhalo
abundance by a factor of 2 in the centre.

In addition to the enhanced tidal disruption studied by these authors,
the loss of baryons reduces the masses and abundances of low-mass
subhaloes relative to DMO simulations \citep{Sawala-2013,
  Schaller-2015, Sawala-2015}. While earlier work has focussed on the
haloes of star-forming dwarf galaxies, here we use high resolution
simulations which capture the full baryonic effects to explore the
extent to which baryonic physics can change the abundance of even
completely dark substructures deeply inside the MW halo, and discuss
possible implications for the detection of substructures through
lensing, stream gaps, and disk heating.

This paper is organised as follows: in Section~\ref{sec:methods} we
briefly describe the simulations used in this work, the selection of
haloes and substructures, and the reconstruction of orbits. In
Section~\ref{sec:results} we discuss how baryons affect the abundance
and distribution of substructures inside dark matter haloes, as a
function of satellite mass, galactocentric radius, and time. In
Section~\ref{sec:dynamics} we examine the subhalo energy, angular
momenta, orbital velocity profiles and orbital anisotropy, and, in
Section~\ref{sec:velocities} we describe the non-Maxwellian subhalo
velocity distributions. We discuss the implications of our results for
different observables in Section~\ref{sec:observables}, and conclude
with a summary in Section~\ref{sec:conclusion}. Additional details
about the orbital interpolation and a comparison of the measured
velocity distributions to standard Maxwellian fits are given in the
Appendix.

\section{Methods} \label{sec:methods} We test the impact of baryons on
substructures in Milky-Way sized $\Lambda$CDM haloes by comparing
cosmological simulations of Local Group analogues with and without
baryons but otherwise identical initial conditions. In this section,
we describe our simulations (Section~\ref{sec:simulations}), the
identification of substructures (Section~\ref{sec:substructures}), and
the reconstruction of their orbits (Section~\ref{sec:orbits}).

\subsection{The \apostle simulations} \label{sec:simulations} Our
results are based on {\it A Project Of Simulating The Local
  Environment} ({\sc Apostle}, \citealt{Sawala-2016b}), a suite of
cosmological hydrodynamic zoom-in simulations of Local Group regions
using the code developed for the {\it Evolution and Assembly of
  GaLaxies and their Environments} ({\sc Eagle}, \citealt{Schaye-2014,
  Crain-2015}) project. The simulations are performed in a WMAP-7
cosmology \citep{Komatsu-2011}, with density parameters at $z=0$ for
matter, baryons and dark energy of $\Omega_\mathrm{M}=0.272$,
$\Omega_\mathrm{b}=0.0455$ and $\Omega_\mathrm{\lambda}=0.728$,
respectively, a Hubble parameter of H$_0=70.4$~km/s Mpc$^{-1}$, a
power spectrum of (linear) amplitude on the scale of 8$h^{-1}$Mpc of
$\sigma_8=0.81$ and a power-law spectral index $n_s=0.967$.  Regions
were selected from a 100$^3$Mpc$^3$ simulation (identified as {\sc
  Dove} in \citealt{Jenkins-2013}), to resemble the observed dynamical
constraints in terms of distance and relative velocity between the MW
and M31, and the isolation of the Local Group
\citep{Fattahi-2015}. Zoom initial conditions were constructed using
2\ts{nd} order Lagrangian perturbation theory \citep{Jenkins-2010}, at
three different resolution levels, with gas (dark matter) particle
masses of $\sim1.0 (5.0)\times10^4\Ms$ (labelled L1), $\sim1.2 (5.9)
\times 10^5\Ms$ (labelled L2), and $\sim1.5 (7.5) \times 10^6\Ms$
(labelled L3), respectively. The gravitational softening lengths are
initially fixed in comoving coordinates, and limited in physical
coordinates to 134 pc, 307 pc and 711 pc. Except to check for
convergence in Figure~\ref{fig:mf}, we only use the L1 simulations in
this work. Each volume has also been resimulated as a DMO simulation,
with identical initial conditions, and dark matter particle masses
larger by a factor of $\left(\Omega_{b}+\Omega_{DM}\right) /
\Omega_{DM}$.

The \eagle code is based on {\sc P-Gadget-3}, an improved version of
the publicly available {\sc Gadget-2} code
\citep{Springel-2005}. Gravitational accelerations are computed using
the Tree-PM scheme of {\sc P-Gadget-3}, while hydrodynamic forces are
computed with the smoothed particle hydrodynamics (SPH) scheme {\sc
  Anarchy} described in Dalla-Vecchia et al. (in prep.)  and
\cite{Schaller-2015b}, which uses the pressure-entropy formalism
introduced by \cite{Hopkins-2013}. The \eagle subgrid physics model
has been calibrated to reproduce the $z=0.1$ stellar mass function and
galaxy sizes in the stellar mass range $10^8-10^{11}\Ms$ in a
cosmological volume of $100^3$ Mpc$^3$.  It includes radiative
metallicity-dependent cooling following \cite{Wiersma-2009}, star
formation with a pressure-dependent efficiency and a
metallicity-dependent density threshold \citep{Schaye-2008}, stellar
evolution and stellar mass loss, and thermal feedback that captures
the collective effects of stellar winds, radiation pressure and
supernova explosions, using the stochastic, thermal prescription of
\cite{DallaVecchia-2012}. Reionisation of hydrogen is assumed to be
instantaneous at $z=11.5$, while He~{\sc II} reionisation follows a
Gaussian centred at $z=3.5$ with $\sigma(z) = 0.5$, to reproduce the
observed thermal history \citep{Schaye-2000, Wiersma-2009b}. The
\eagle model also includes black hole growth fuelled by gas accretion
and mergers and feedback from active galactic nuclei
\citep[AGN,][]{Booth-2009, Johansson-2009, Rosas-Guevara-2013}. In
this work, we use the ``Reference'' choice of subgrid parameters
\citep{Crain-2015} at all resolutions. Further details of the \eagle
and \apostle simulations and comparison of results to observations can
be found in the references above.

\begin{table}   \label{tab:haloes}
\setlength\tabcolsep{10pt} 
  \caption{Haloes used in this study}
\begin{tabularx}{\columnwidth}{c|ccc}
  \hline
  \hline
  & DMO &  \multicolumn{2}{>{\setlength{\hsize}{2\hsize}\addtolength{\hsize}{2\tabcolsep}}c}{
    Hydrodynamic} \\
  & M$_{200} [\Ms]$ &  M$_{200} [\Ms]$ & M$_{*} [\Ms]$\\
  \hline
  AP-1-1 & $1.65 \times 10^{12}$ & $1.57 \times 10^{12}$ & $2.75\times 10^{10}$ \\
  AP-1-2 & $1.10 \times 10^{12}$ & $1.01 \times 10^{12}$ &$1.20\times 10^{10}$ \\
  AP-4-1 & $1.34 \times 10^{12}$ & $1.16 \times 10^{12}$  & $1.23\times 10^{10}$ \\
  AP-4-2 & $1.39 \times 10^{12}$ & $1.13 \times 10^{12}$ & $1.88\times 10^{10}$ \\
  \hline \hline
\end{tabularx}
\vspace{.3cm}\\
Structural parameters of the four \apostle haloes used in this
study at $z=0$ and resolution L1, in the DMO and hydrodynamic
simulations. All values are in physical units. $M_{200}$ is
computed for the total halo, including substructures, while stellar
masses are those of the central subhalo only, excluding satellites.
\end{table}

\begin{figure*}
  \begin{center}
    \vspace{-.4cm}
    \includegraphics*[trim = 10mm 32mm 15mm 16mm, clip, width = .495\textwidth]{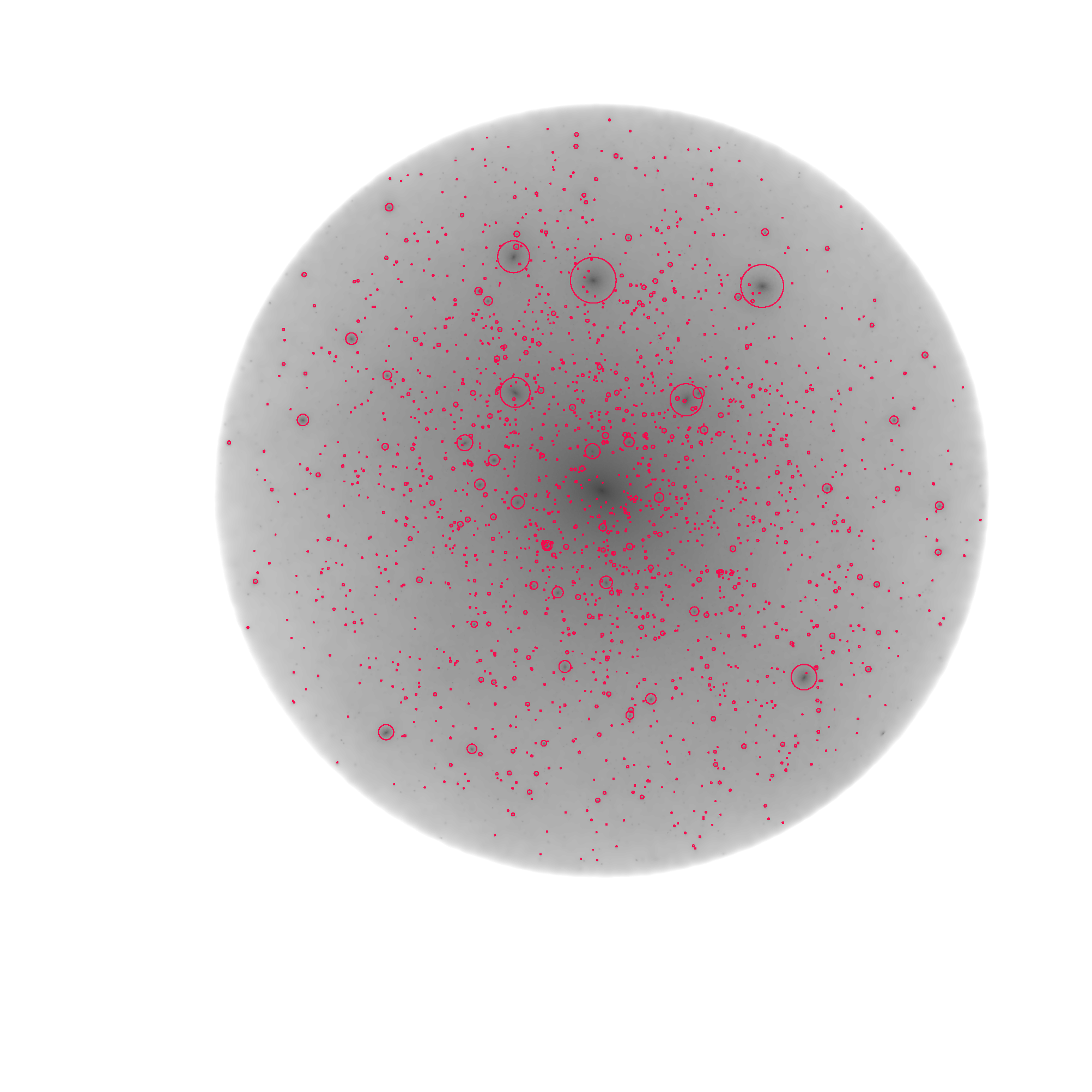}
    \includegraphics*[trim = 10mm 32mm 15mm 16mm, clip, width = .495\textwidth]{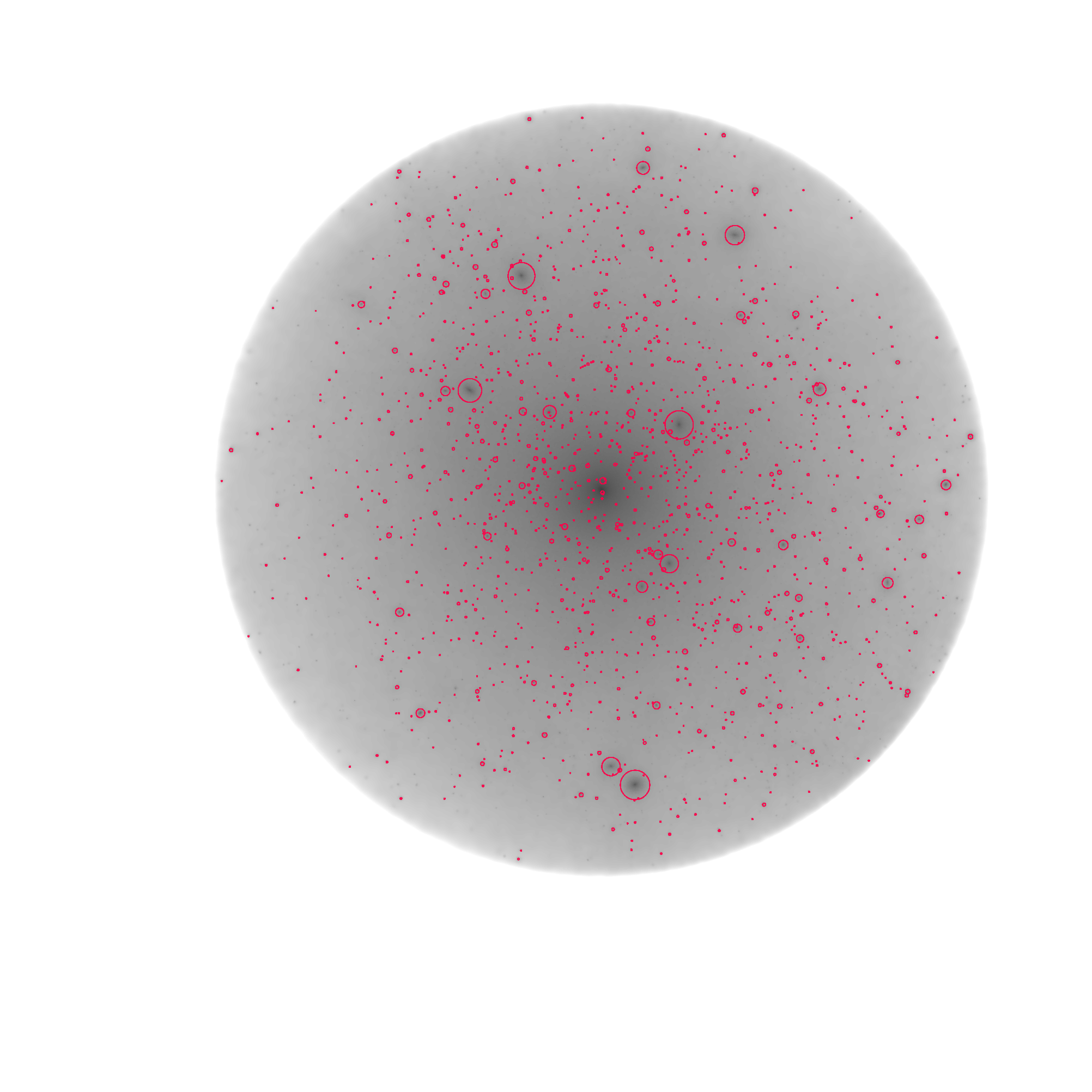}
  \end{center}
\vspace{-.2cm}
  \caption{Projected dark matter density at $z=0$ in the MW-mass halo
    {\sc AP-1-1} at resolution L1, in matched DMO (left) and
    hydrodynamic (right) simulations inside r$_{200}$. Red circles
    indicate the positions of subhaloes with masses above
    $10^{6.5}\Ms$ inside the respective regions. The hydrodynamic
    simulation contains fewer subhaloes, and the dark matter in the
    central region is visibly rounder.} \label{fig:distribution}
\end{figure*}

\subsection{Halo and subhalo selection} \label{sec:substructures}
Structures (haloes) are identified using a Friends-of-Friends
algorithm \citep{Davis-1985}, and substructures (subhaloes) are
identified using the \subfind algorithm (\citealt{Springel-2001}, with
the extension of \citealt{Dolag-2009}) for 18 snapshots up to a
lookback time of 5 Gyr ($z\sim0.5$). We identify haloes and subhaloes
at each snapshot, and find their progenitors at earlier times using a
subhalo merger tree (as described in the appendix of
\citealt{Jiang-2014}).

We denote the radius inside which the mean density is $200$ times the
critical density at the time as $r_{200}$, and the enclosed mass as
$M_{200}$. For substructures, we quote the total mass bound to a
subhalo: in the hydrodynamic simulation, this includes dark matter,
stellar and gas particles, although in the mass range
$10^{6.5}-10^{8.5}\Ms$ we study here, subhaloes are almost entirely
devoid of baryons.

The number of identified subhaloes and the assigned masses depend on
the substructure identification algorithm (see \citealt{Onions-2012}
for a comparison). For subhaloes of $10^4$ particles,
\cite{Springel-2008} find that the mass assigned by the \subfind
algorithm closely follows the mass enclosed within the tidal radius,
while \cite{Onions-2012} find that substructures can be reliably
identified with at least 20 particles and their basic properties
recovered with at least 100 particles. As discussed in
Section~\ref{sec:total-abundance}, we find that the subhalo mass
function converges with resolution in both the hydrodynamic and DMO
simulations. It should be noted that even if the subhalo mass function
is numerically converged, by construction, the {\sc subfind} mass
depends on the local overdensity. Part of the central decline in
subhalo number density within a given mass interval is therefore
attributable not directly to stripping, but to the increasing
background density. However, to first order, as long as the background
densities are similar, this should not affect the relative difference
in subhalo number density between the DMO and hydrodynamic
simulations.

In this work we limit our analysis to subhaloes with mass above
$10^{6.5}\Ms$, corresponding to at least 50 particles in the L1 DMO
simulation. With a gravitational softening length limited to $<134$~pc
at resolution L1, the main haloes are unaffected by softening in the
regions of interest here. The dark matter mass profiles of the main
haloes and their relation to the disk are discussed further in
\cite{Schaller-2016}.

\subsection{Orbits} \label{sec:orbits} All three observational probes
introduced in Section~\ref{sec:Introduction} are sensitive to
substructures within the central $\sim 10-20$~kpc, equivalent to $\sim
0.05-0.1 \times r_{200}$ of the host halo at $z=0$. Throughout this
work, we use the minimum of the host halo potential to define the
origin of our reference frame, and the minimum of each satellite's
potential to define its position.

Because most subhaloes found near the halo centre at any time have
orbits with large apocentres and cross the central regions at high
speed (see Section~\ref{sec:energies}), any single snapshot only
captures a small fraction of all the subhaloes that come near the halo
centre. To obtain a complete measurement of the expected subhalo
distribution, we therefore interpolate all orbits using cubic splines,
and integrate all quantities over time to determine their expected
probability density during a given finite time interval.

\begin{figure*}
  \begin{center}
    \includegraphics*[trim = 0mm 00mm 0mm 0mm, clip, height =  0.49\textwidth]{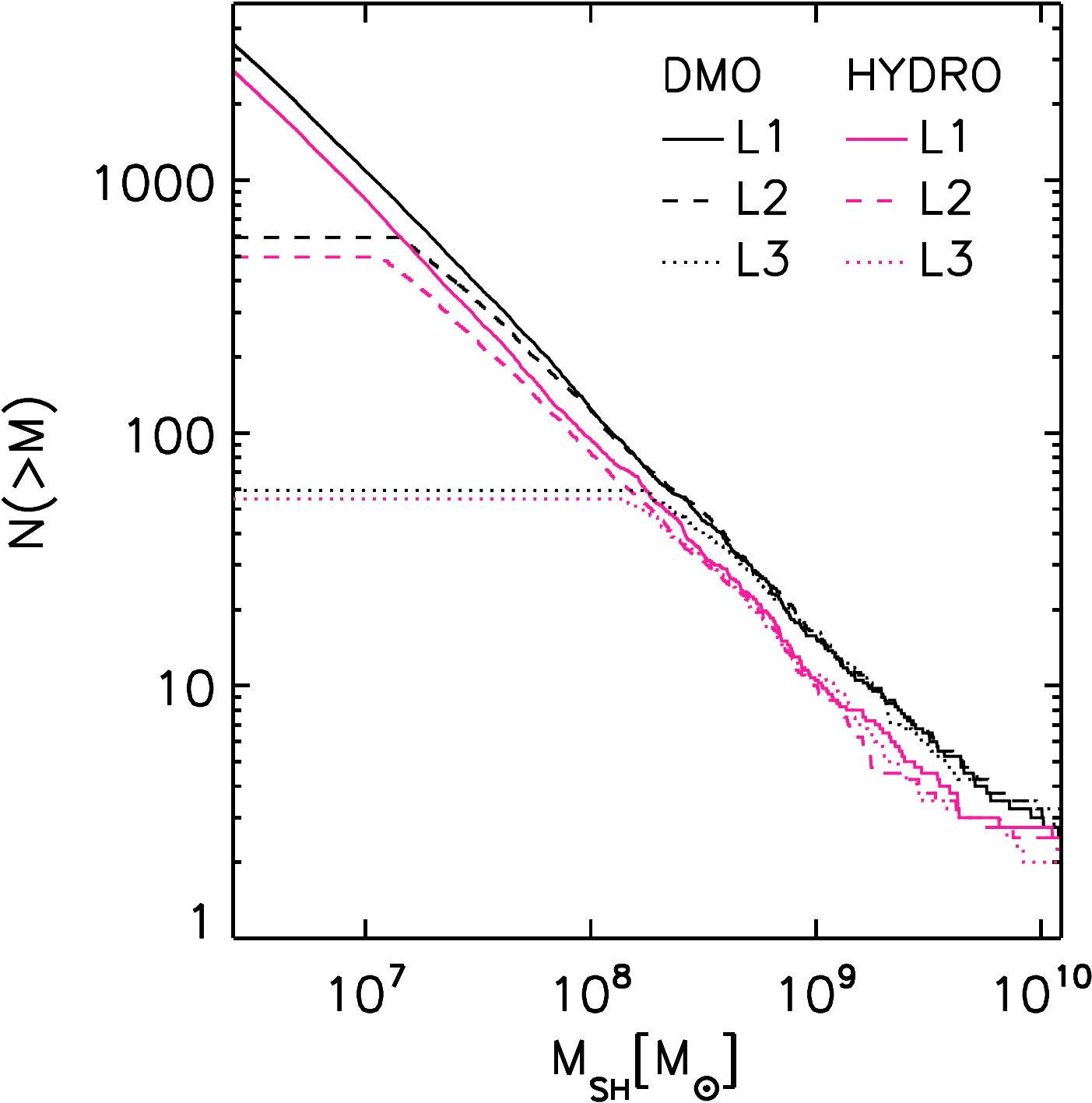} \hspace{3mm}
    \includegraphics*[trim = 0mm 0mm 0mm 0mm, clip, height =  0.49\textwidth]{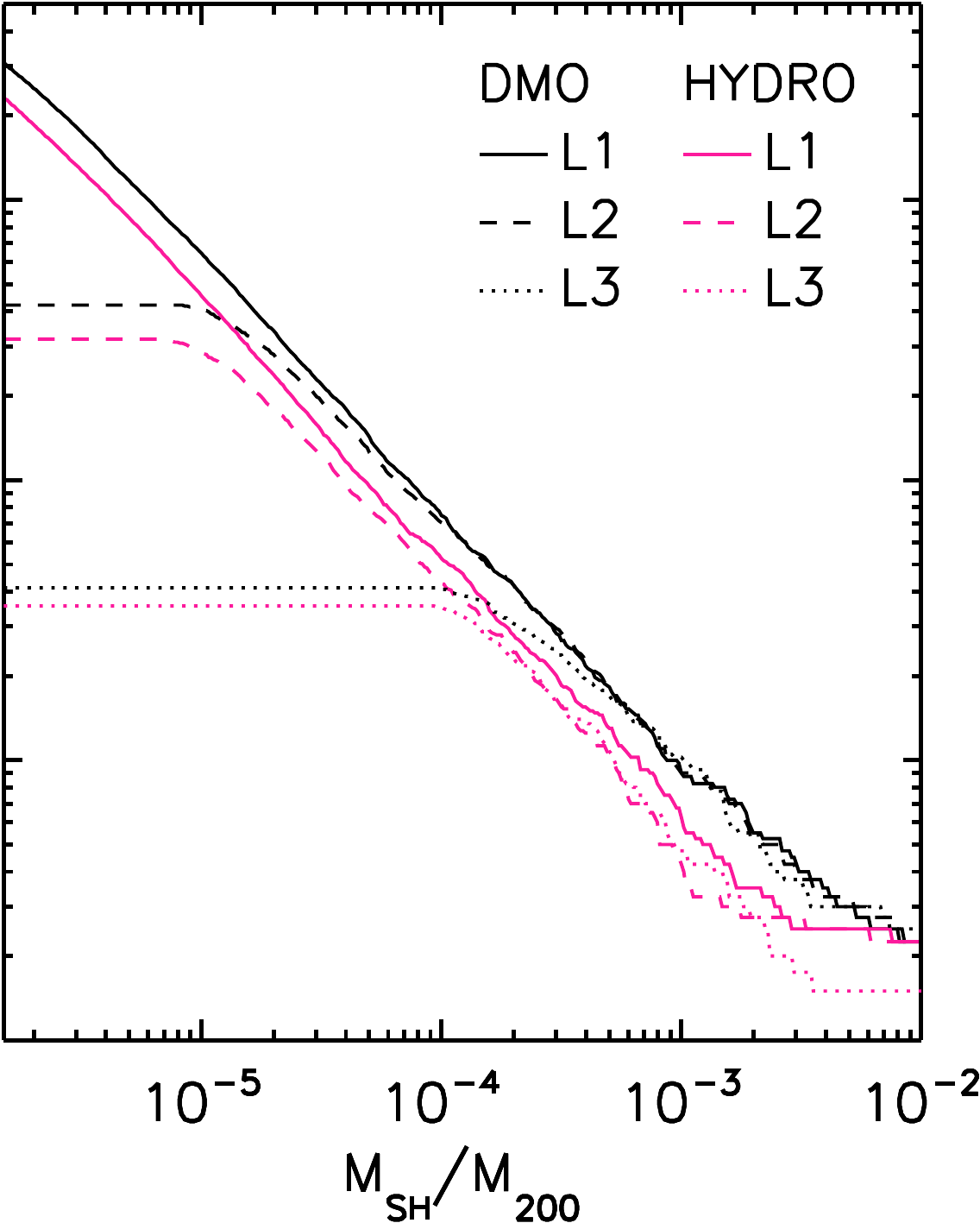} \\ 
  \end{center}
  \caption{Cumulative abundance of substructures in Milky-Way mass
    haloes at the present time. Each panel presents results averaged
    over the four haloes listed in Table~\ref{tab:haloes} simulated
    as DMO (black lines) or hydrodynamically (red lines), at three
    different resolutions, from L3 (dotted, lowest), through L2 (dashed,
    intermediate) to L1 (solid, highest).  The left panel shows
    subhaloes within 300~kpc of each host, while the right panel
    includes subhaloes within $r_{200}$, with the mass expressed
    relative to the hosts' $M_{200}$. Convergence of the DMO and
    hydrodynamic simulations is similar and the relative difference between
    the hydrodynamic and DMO simulations is similar at different
    resolution levels.
    \label{fig:mf}}
\end{figure*}

Subhalo velocities are commonly measured using a mass-weighted average
of the particle velocities, and thus defined relative to the
centre-of-mass frame. However, because the host halo potential can be
offset from the centre of mass by $\sim10$~kpc, subhalo velocities
measured in this way cannot be used directly for our purpose. Instead,
we establish velocities consistent with our centre-of-potential
reference frame from the interpolated positions. Details are described
in Appendix~\ref{appendix:orbits}.

Where we average our results over the haloes listed in
Table~\ref{tab:haloes}, we first compute the properties of subhaloes
relative to the individual host halo's properties such as $r_{200}$,
potential, where appropriate, and then combine the results of all
orbits from all haloes to compute the arithmetic mean.

\section{Subhalo Abundance}\label{sec:results}
Figure~\ref{fig:distribution} illustrates the spatial distribution and
the effect of baryons on the number of substructures by comparing the
present-day projected mass distribution and the location of
substructures with masses above $10^{6.5}\Ms$ in one of our Milky-Way
mass haloes in DMO and hydrodynamic simulations (identified as halo
{\sc AP-1-1} in Table~\ref{tab:haloes}). In the DMO simulation, shown
on left, the halo has a total mass of $M_{200}=1.65\times10^{12}\Ms$
and a corresponding $r_{200} = 236$~kpc, reducing slightly to
$M_{200}=1.57\times10^{12}\Ms$ and $r_{200}=232$~kpc in the
hydrodynamic simulation shown on the right. For this particular halo,
and at this particular snapshot, a reduction in substructures is
barely noticeable by eye, and robust quantitative statements require a
more detailed analysis.

\subsection{Total subhalo abundance}\label{sec:total-abundance}
Figure~\ref{fig:mf} shows the cumulative abundance of substructures as
a function of subhalo mass, averaging over four MW mass haloes in both
DMO and hydrodynamic simulations, at our three resolution levels from
L3 (lowest), through L2 (intermediate) to L1 (highest). In the left
panel, all subhaloes are included out to a distance of 300~kpc. It can
be seen that, for subhaloes of mass $ < 10^{9.5}\Ms$, there is a
near-constant decrease in abundance by $\sim 1/3$ in the hydrodynamic
relative to the DMO simulation. In the right panel, subhalo masses are
expressed relative to the $M_{200}$ of the host, and subhaloes are
selected inside the hosts' $r_{200}$. Although the decrease in
abundance in the hydrodynamic simulation is slightly enhanced by the
reduction of $r_{200}$, the principal difference in abundance between
the DMO and hydrodynamic simulation persists. Clearly, baryons affect
the masses of subhaloes below $10^{9.5}\Ms$ more than those of their
$10^{12}\Ms$ hosts, destroying the scale-free nature of pure dark
matter simulations. On the other hand, below $\sim10^{9.5}\Ms$, the
offset in the abundance is nearly constant, as the baryon loss of
subhaloes in this mass range is nearly constant.

\begin{figure*}
  \begin{center}
   \includegraphics*[trim = 20mm 160mm 25mm 15mm, clip, width=0.493\textwidth]{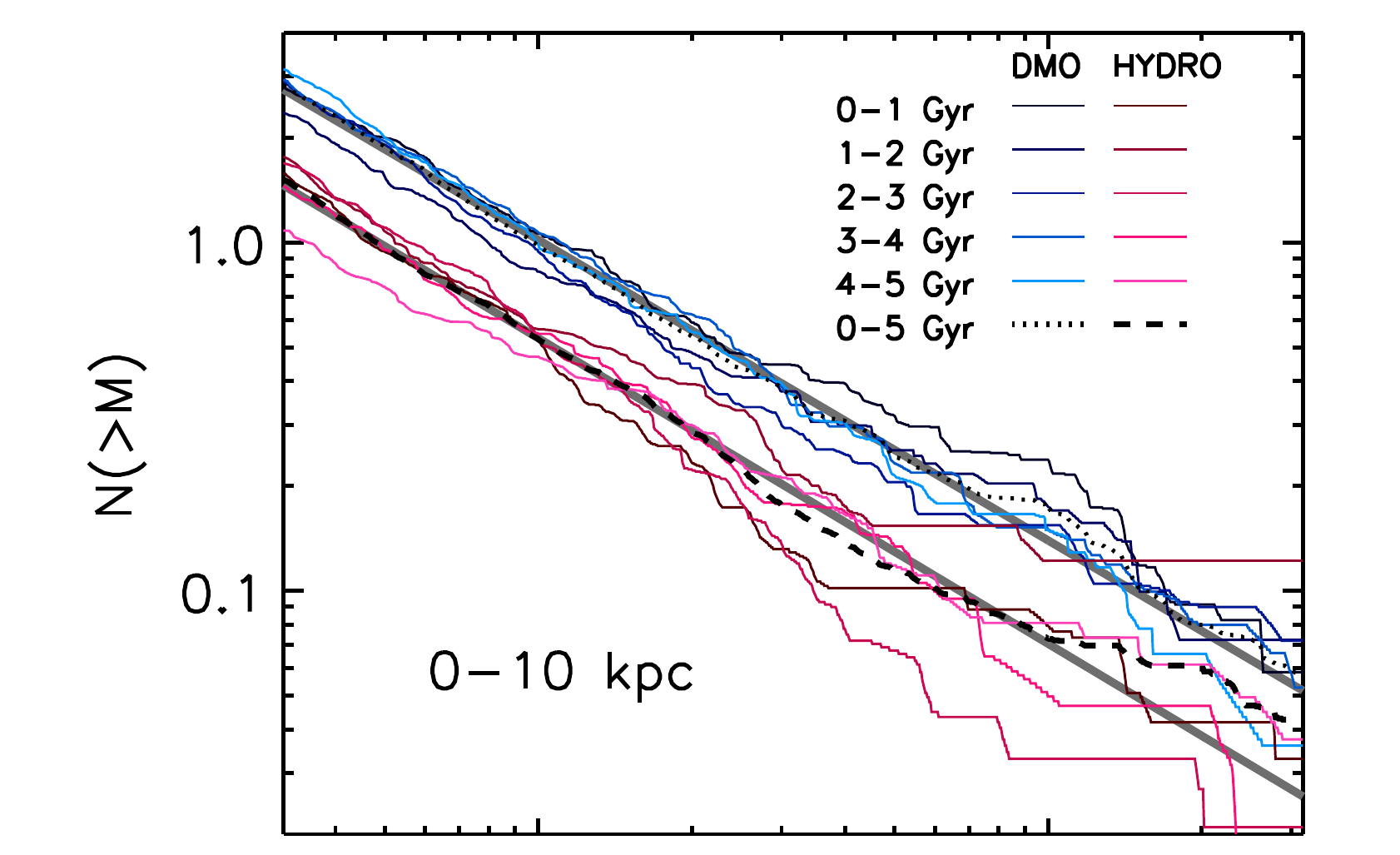} 
   \includegraphics*[trim = 20mm 160mm 25mm 15mm, clip, width=0.493\textwidth]{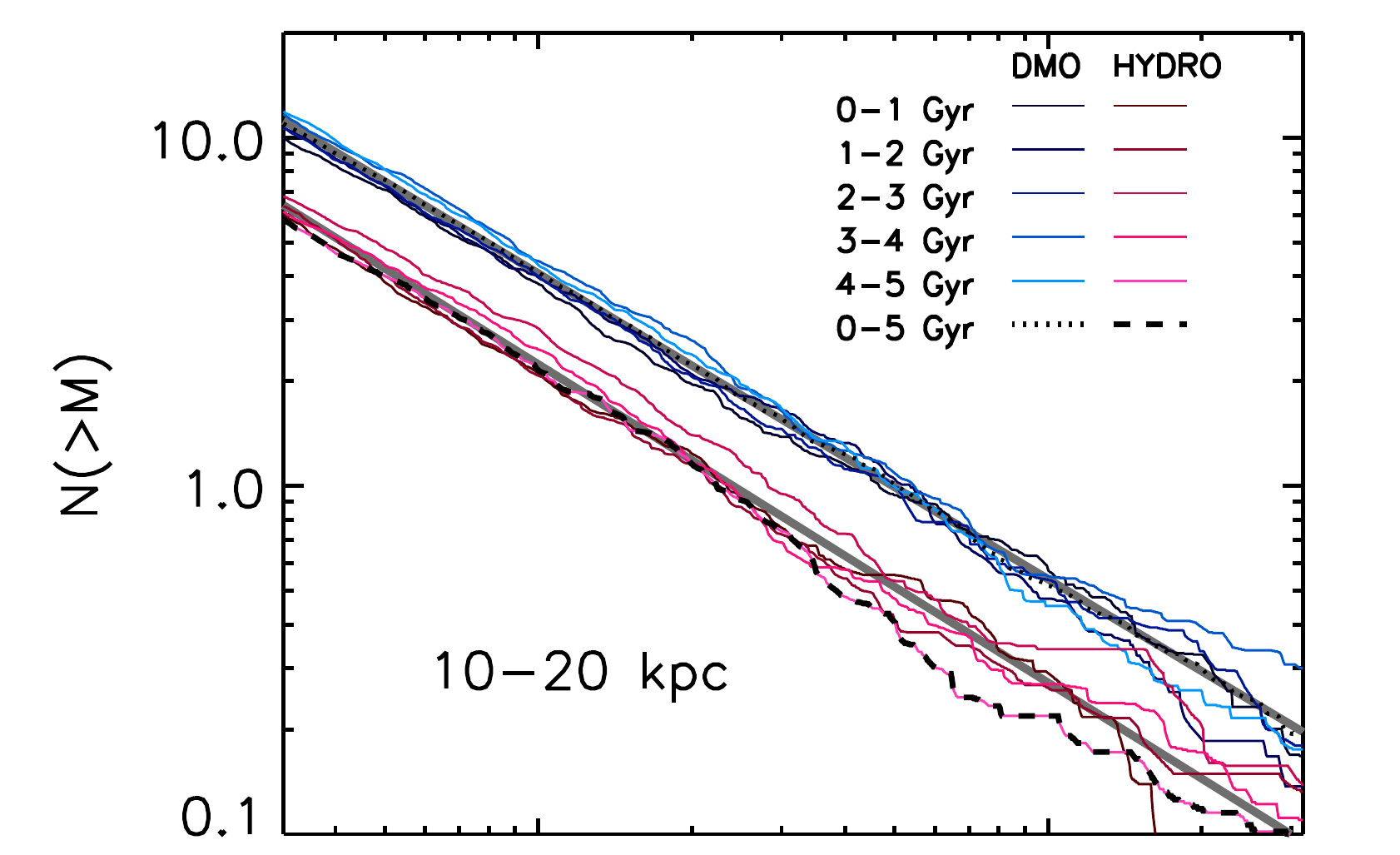} \\
   \includegraphics*[trim = 20mm 140mm 25mm 60mm, clip, width=0.493\textwidth]{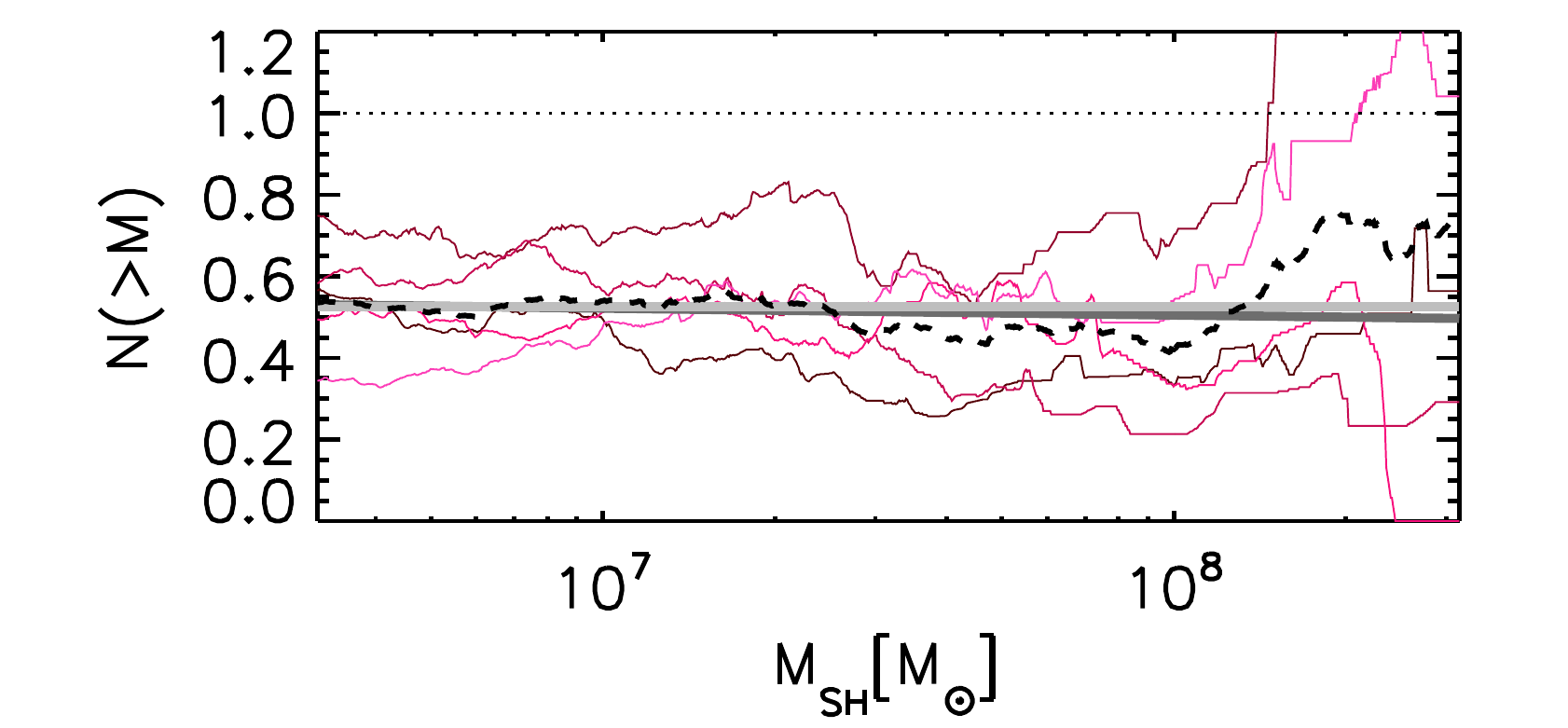} 
   \includegraphics*[trim = 20mm 140mm 25mm 60mm, clip, width=0.493\textwidth]{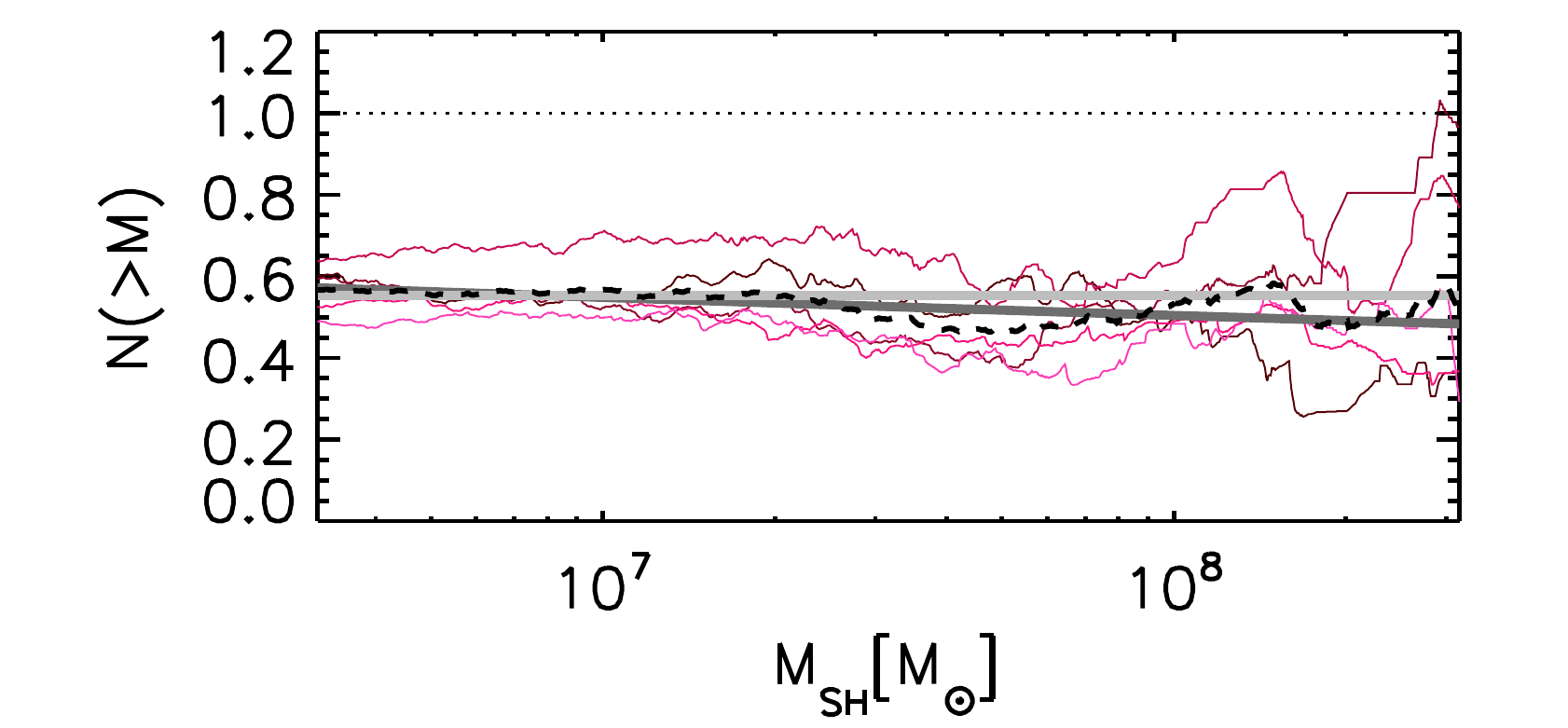} \\
   \includegraphics*[trim = 20mm 160mm 25mm 15mm, clip, width=0.493\textwidth]{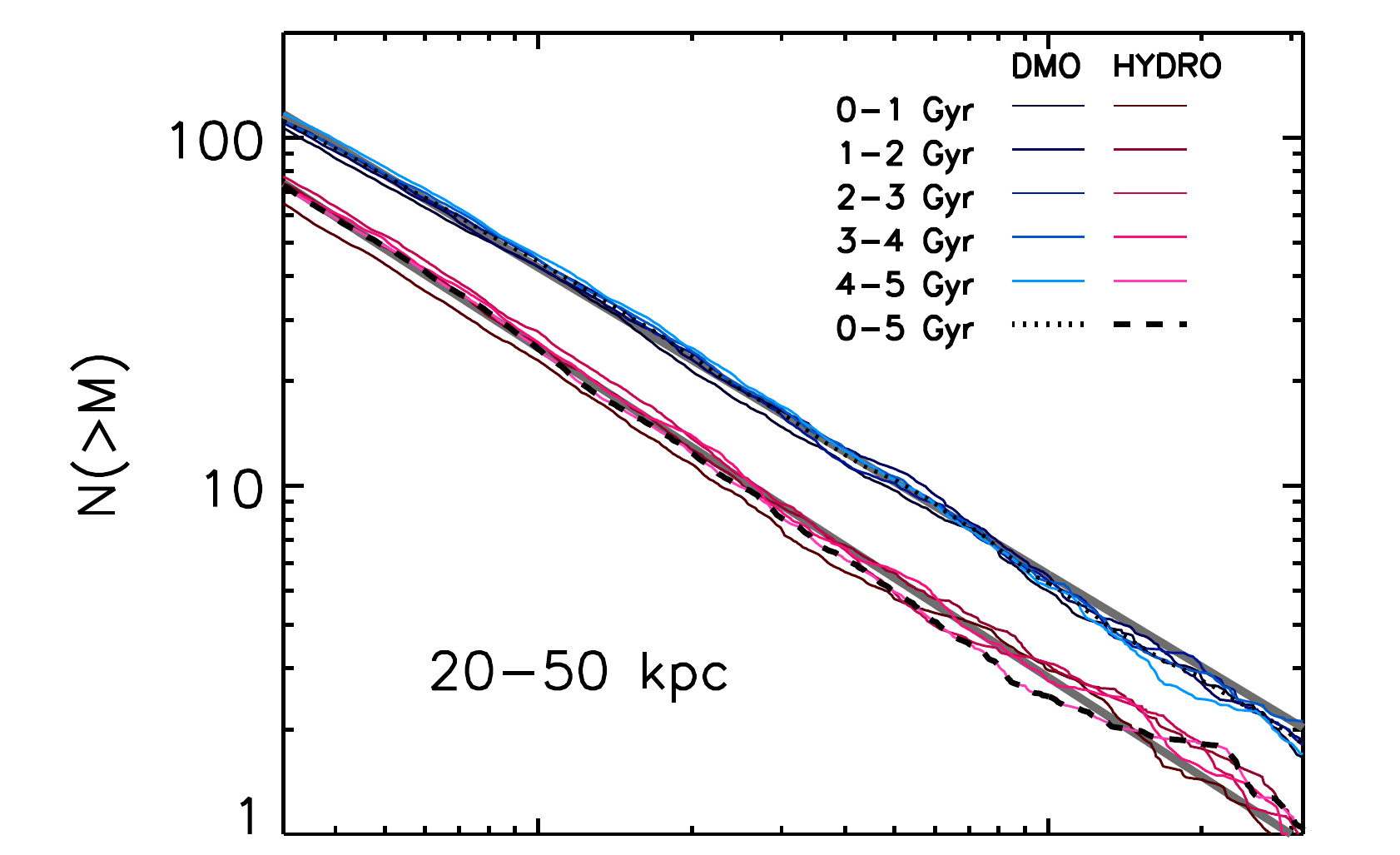} 
   \includegraphics*[trim = 20mm 160mm 25mm 15mm, clip, width=0.493\textwidth]{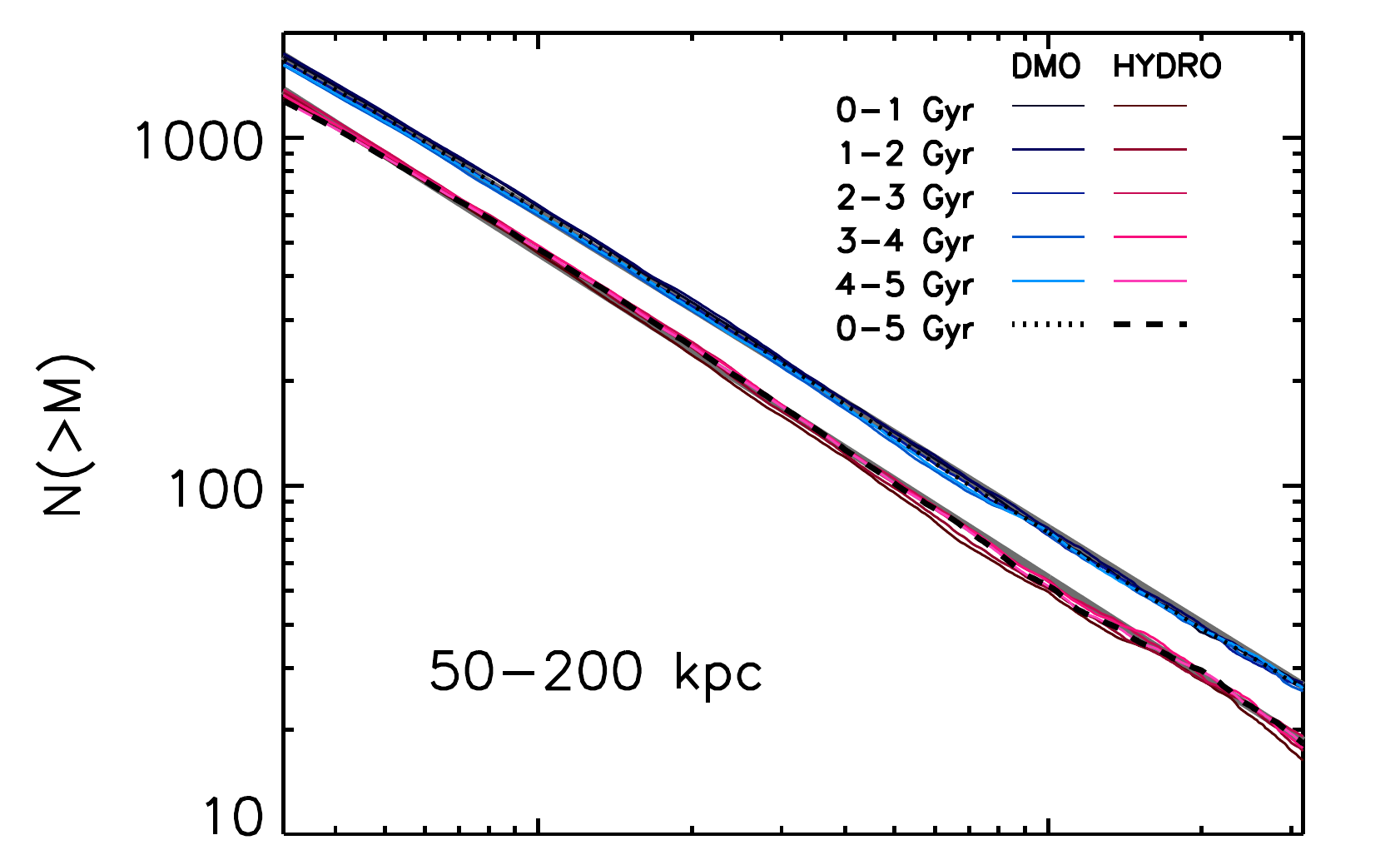} \\
   \includegraphics*[trim = 20mm 140mm 25mm 60mm, clip, width=0.493\textwidth]{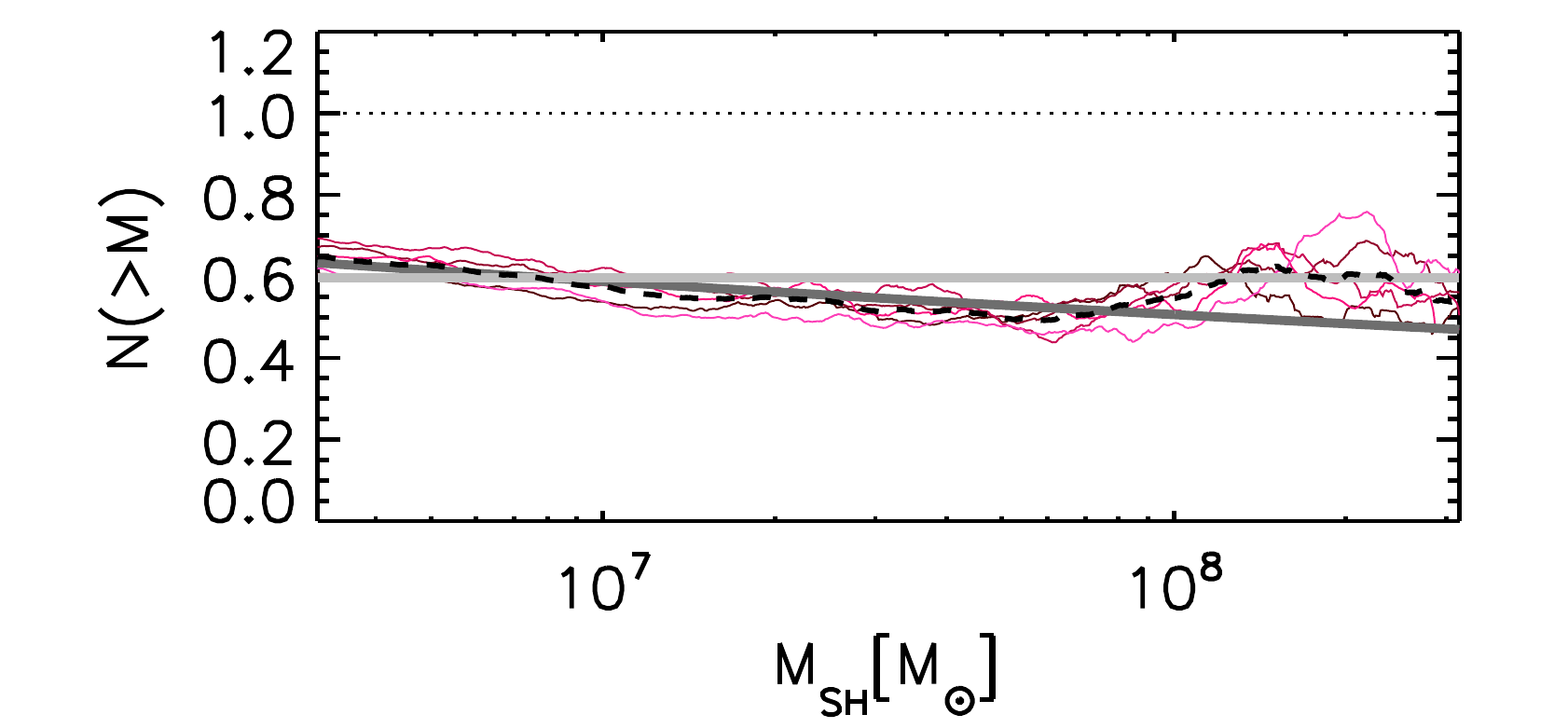} 
   \includegraphics*[trim = 20mm 140mm 25mm 60mm, clip, width=0.493\textwidth]{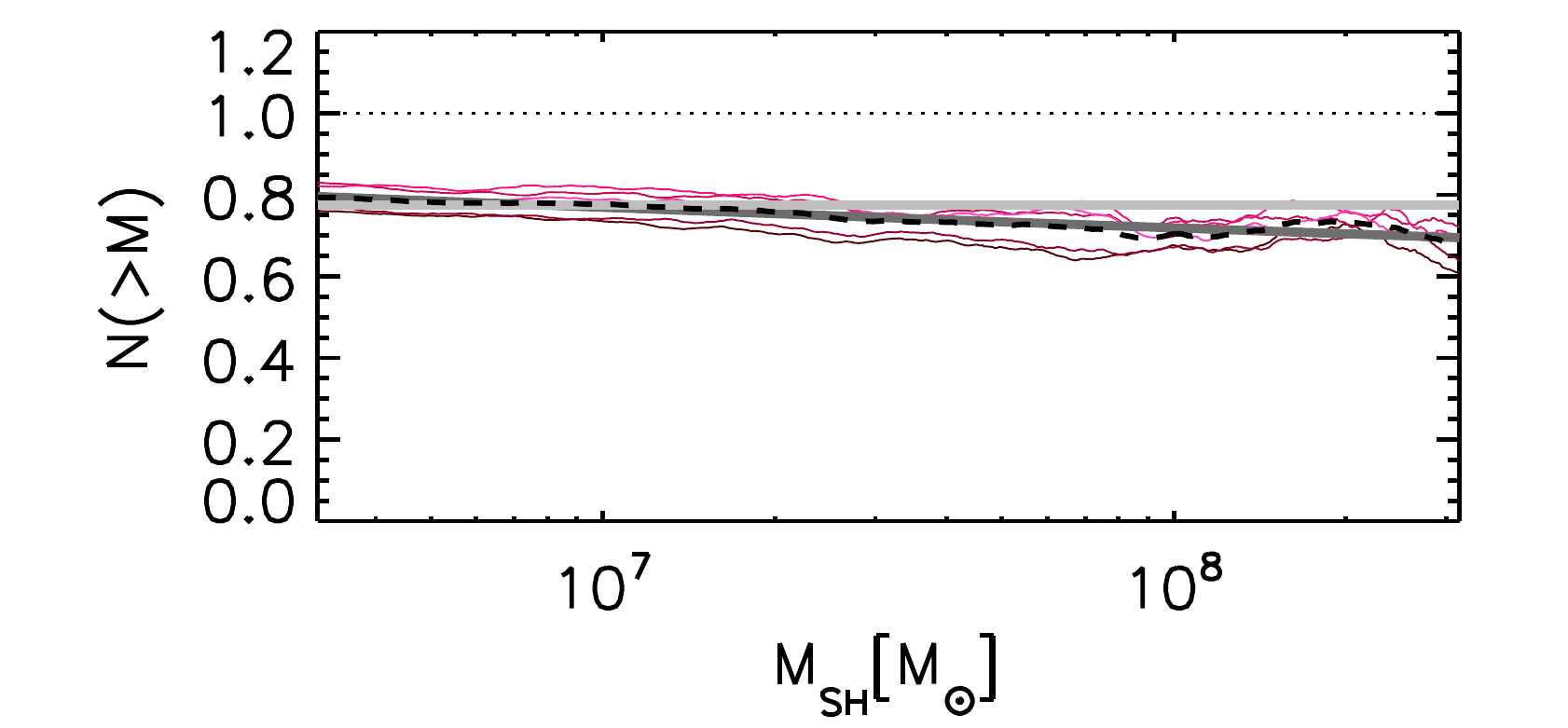} 
  \end{center}
  \caption{Large panels: cumulative substructure mass functions in
    spherical shells, in the DMO and hydrodynamic simulations. Blue
    and red solid lines indicate results from the DMO and hydrodynamic
    simulations over successive 1 Gyr time intervals, respectively,
    while dotted and dashed lines show the results averaged over the
    entire 5 Gyr period. Dark grey lines are power-law fits to the
    mass functions over the mass interval shown. Small panels: ratio
    between the cumulative substructure mass functions in the DMO and
    hydrodynamic simulations. Solid dark grey lines show the ratios
    between the power-law fits to the DMO and hydrodynamic mass
    functions, solid light grey lines are constant values. Differences
    between the hydrodynamic and DMO simulation are present at all
    radii, but increase towards the centre. For substructures in the
    range $10^{6.5}-10^{8.5}\Ms$, there is little evidence of a mass
    or time dependence.\label{fig:radial-mf}}
\end{figure*}

\subsection{Baryon effects on subhalo  abundance}\label{sec:distribution}
In Figure~\ref{fig:radial-mf} we show the cumulative mass functions of
substructures in four spherical shells, increasing in radius, from
$0-10$ to $10-20$, $20-50$, and $50-200$~kpc. The results are averaged
over all four haloes at resolution L1, and time-averaged in lookback
time over either 5 intervals of 1 Gyr each, or over a 5 Gyr period.

\begin{table}
\setlength\tabcolsep{7pt} 

\caption{Subhalo abundance parameters}
\label{tab:fits}

\begin{tabularx}{\columnwidth}{l|cccc}
  \hline \hline
  &0-10 kpc \hspace{-20em}& 10-20 kpc \hspace{-20em}& 20-50 kpc \hspace{-02em}& 50-200 kpc \\
  \hline
  &\multicolumn{4}{c}{power law slope $n^{[1]}$} \\
  \hline
  DMO & -1.86 & -1.88 & -1.88 & -1.90 \\
  Hydro & -1.88 & -1.91 & -1.94 & -1.93 \\
  \hline
  &\multicolumn{4}{c}{$N_{Hydr}(r)/N_{DMO}$(r)$^{[2]}$} \\
  \hline
  & 0.52 & 0.55 & 0.60 & 0.77\\
  \hline \hline
\end{tabularx}
\vspace{.3cm}\\
$^{[1]}$Power-law slopes for the subhalo mass functions in the DMO
and hydrodynamic simulation in the mass range
$10^{6.5}-10^{8.5}\Ms$. $^{[2]}$Suppression of the number of substructures in
the hydrodynamic
relative to the DMO simulation, assuming a constant factor,
independent of mass.
\end{table}

Comparing the results from the hydrodynamic and DMO simulations, it
can be seen that, in all shells, the abundance of substructures is
reduced in the hydrodynamic simulation. The difference increases with
decreasing radius, indicating stronger tidal stripping near the centre
in the hydrodynamic simulation. 

We fit the subhalo mass functions in all four shells by power laws,
$dn/dm \propto m^n$, and overplot the fits as dark grey lines in the
large panels of Figure~\ref{fig:radial-mf}. In both the DMO and
hydrodynamic simulations, the results are similar to those reported in
the \aquarius simulations by \cite{Springel-2008}, who found values
between $-1.93$ and $-1.87$ for the slope, with the steepest values
found for the lowest mass range. We find slightly shallower profiles
in the innermost bins, but no significant differences in slope between
the DMO and hydrodynamic simulations, indicating that the additional
disruption of substructures due to baryonic effects in the mass range
$10^{6.5}-10^{8.5}\Ms$ is not strongly mass-dependent.

In the bottom panels of Figure~\ref{fig:radial-mf}, we show the ratios
between the subhalo abundances in the hydrodynamic and DMO simulations
in the different radial shells. We overplot, in dark grey, the ratio
between the two respective power-law fits and, in light grey, a fit to
a constant value over the entire mass range shown. We find that, in
the subhalo mass range $10^{6.5}-10^{8.5}\Ms$, a factor constant in
mass that varies only with radius gives an almost equally good fit to
the suppression of substructures: by $23\%$ for $r=50-200$~kpc, $40\%$
for $r=20-50$~kpc, $45\%$ for $r=10-20$~kpc, and $ 48\%$ for
$r<10$~kpc. We list the best-fitting power-law slopes, and the
constant reduction factors in Table~\ref{tab:fits}.

As discussed in \cite{Sawala-2013} and \cite{Schaller-2015}, the
mass-loss of isolated subhaloes due to the complete loss of baryons
relative to a DMO simulation is nearly constant below $\sim 10^9 \Ms$,
and the reduction in abundance by $\sim 23\%$ in the outermost shell
is consistent with the results expected for isolated subhaloes. Note
that this does not mean that these subhaloes do not experience tidal
stripping, but merely that, at these large radii, there is little
difference in tidal stripping between the DMO and hydrodynamic
simulations.

\subsection{Substructure and mass profiles}\label{sec:substructure-bias}
\begin{figure}
\vspace{2mm}
  \begin{center}
    \includegraphics*[trim = 0mm 16.5mm 0mm 0mm, clip, width = 0.49\textwidth]{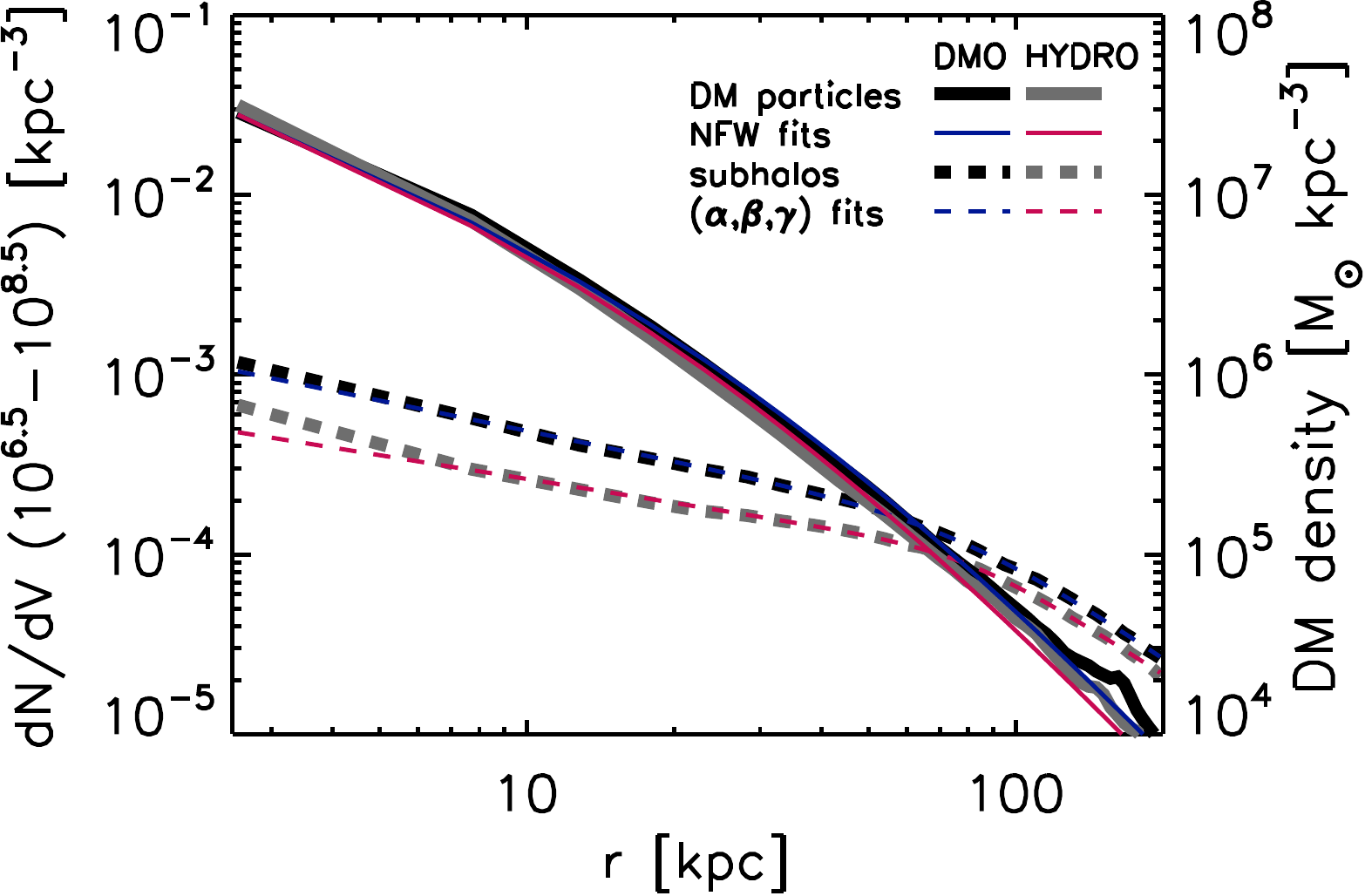} 
    \includegraphics*[trim = -5mm 0mm -23mm 0mm, clip, width = 0.49\textwidth]{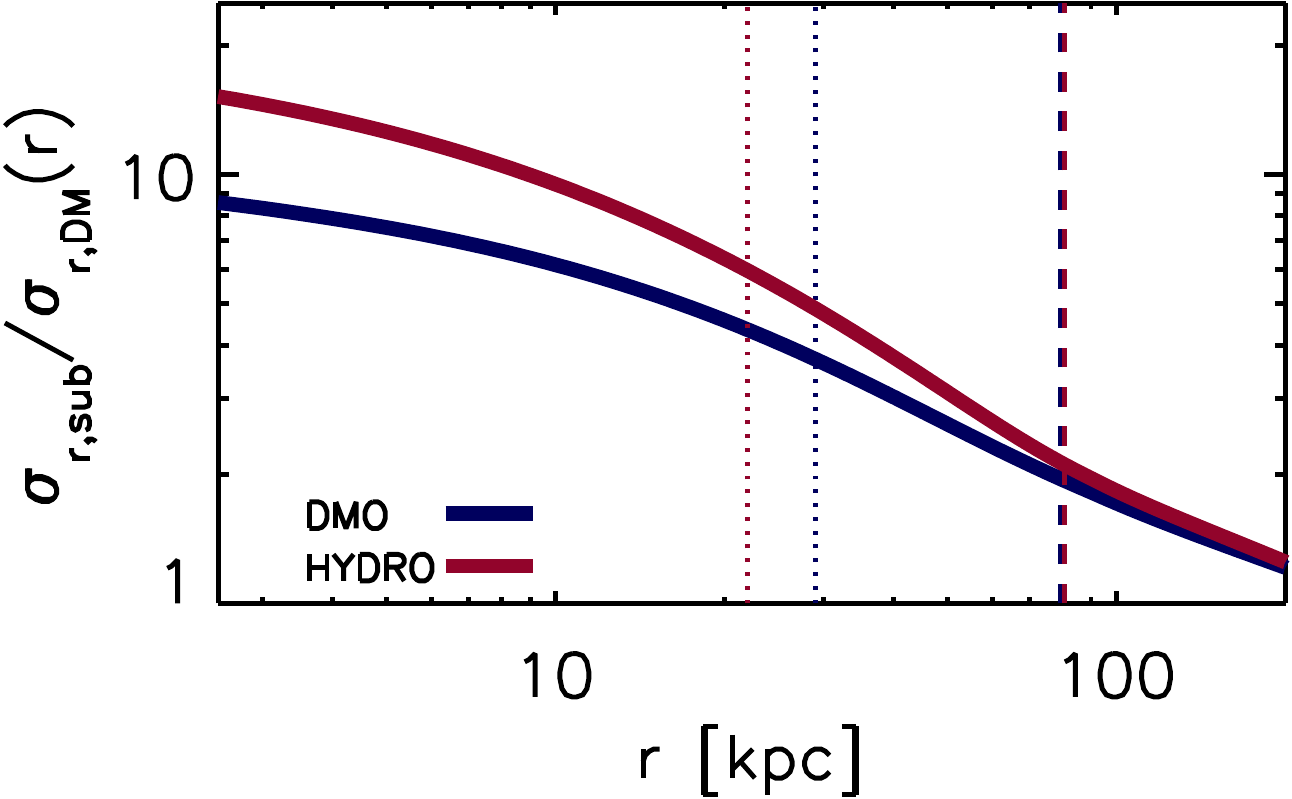} 
  \end{center}
  \caption{Top: number density profiles of substructures in the mass range
    $10^{6.5}-10^{8.5}\Ms$ (dashed lines, left axis) and dark matter
    mass density profiles (solid lines, right axis). Black and grey
    lines show results of the DMO and hydrodynamic simulations,
    respectively, blue and red lines show fits to the two sets of
    simulation data; dashed for $(\alpha, \beta, \gamma)$ fits to
    the subhalo number densities, solid for NFW fits for the DM mass
    densities, at $z=0$. Bottom: expected radial velocity dispersion of
    subhaloes relative to velocity dispersion of DM particles at $z=0$
    from \protect{\eqnref{eqn:velocity-bias}} given the above radial density
    profiles, normalised to the respective values at 300 kpc. Dotted and dashed lines
    indicate the scale radii of the NFW fit to the particle densities, and of
    the $(\alpha, \beta, \gamma)$ profiles for the subhalo number densities, respectively.
    \label{fig:radial}}
\end{figure}

In the top panel of Figure~\ref{fig:radial}, we compare the mass
density profiles of dark matter at $z=0$ to the number density
profiles of subhaloes in the mass range $10^{6.5}-10^{8.5}\Ms$ in our
DMO and hydrodynamic simulations, each averaged over four haloes.

We find that the averaged mass density profiles, represented by solid
lines, are well described by NFW-profiles \citep{Navarro-1996} of the
form
\begin{equation} \label{eqn:nfw}
\rho(r) = \rho_s \left(\frac{r}{r_s}\right)^{-1} \left(1+
    \frac{r}{r_s}\right)^{-2}
\end{equation}
with values for the scale radii, $r_s$, of $29$~kpc and $22$~kpc, and
densities at the scale radii, $\rho_s$, of $3.08 \times 10^6 \Ms$
kpc$^{-3}$ and $4.58 \times^6 \Ms$ kpc$^{-3}$ for the DMO and
hydrodynamic simulations, respectively. As expected, since the total
dark matter mass is lower in the hydrodynamic simulations, the average
DM density in the haloes is slightly below that of the DMO
counterparts. However, due to adiabatic contraction, the average
central DM density in the hydrodynamic simulations rises above that of
the DMO simulations.

Compared to the DM mass density profiles, the subhalo number density
profiles, represented by dashed lines in Figure~\ref{fig:radial}, are
much shallower towards the centre. We fit these by more general,
double power law models (sometimes called $\alpha, \beta, \gamma$-
models, e.g.~\citealt{Zhao-1996}) of the form
\begin{equation}  \label{eqn:alphabetagamma}
  \rho(r) = \rho_s 2^{(\beta-\gamma)/\alpha} \left(\frac{r}{r_s} \right)^{-\gamma} \left(1+
    \left(\frac{r}{r_s}\right)^{\alpha} \right)^{(\gamma-\beta)/\alpha} 
\end{equation}
Here, $\alpha$ determines the transition between an inner power law
with asymptotic slope $-\gamma$ and an outer power law with asymptotic
slope $-\beta$, centred on the scale radius $r_s$, where the density
is $\rho_s$. The 2-parameter NFW model (\eqnref{eqn:nfw}) is a special
case of this 5-parameter model for $(\alpha, \beta, \gamma) = (1, 3,
1)$. 

For the substructure number density profile in the mass range
$10^{6.5}-10^{8.5}\Ms$, averaged over 4 haloes in each simulation, we
obtain best fits of
$$ (r_s, \rho_s, \alpha, \beta, \gamma) = (79.5~\rm{kpc}, 1.06 \times10^{-3} kpc^{-3},
3.06, 0.99, 0.56) $$ and
$$ (r_s, \rho_s, \alpha, \beta, \gamma) = (80.7~\rm{kpc}, 2.02 \times10^{-3} kpc^{-3},
4.82, 0.71, 0.44) $$ for the DMO and hydrodynamic simulations,
respectively. Note that because the subhalo mass function does not
significantly change with radius, the subhalo number density and
subhalo mass density have the same radial dependence.

\begin{figure*}
  \begin{center}
    \includegraphics*[trim = 0mm 0mm 0mm 0mm, clip, height = 0.295\textwidth]{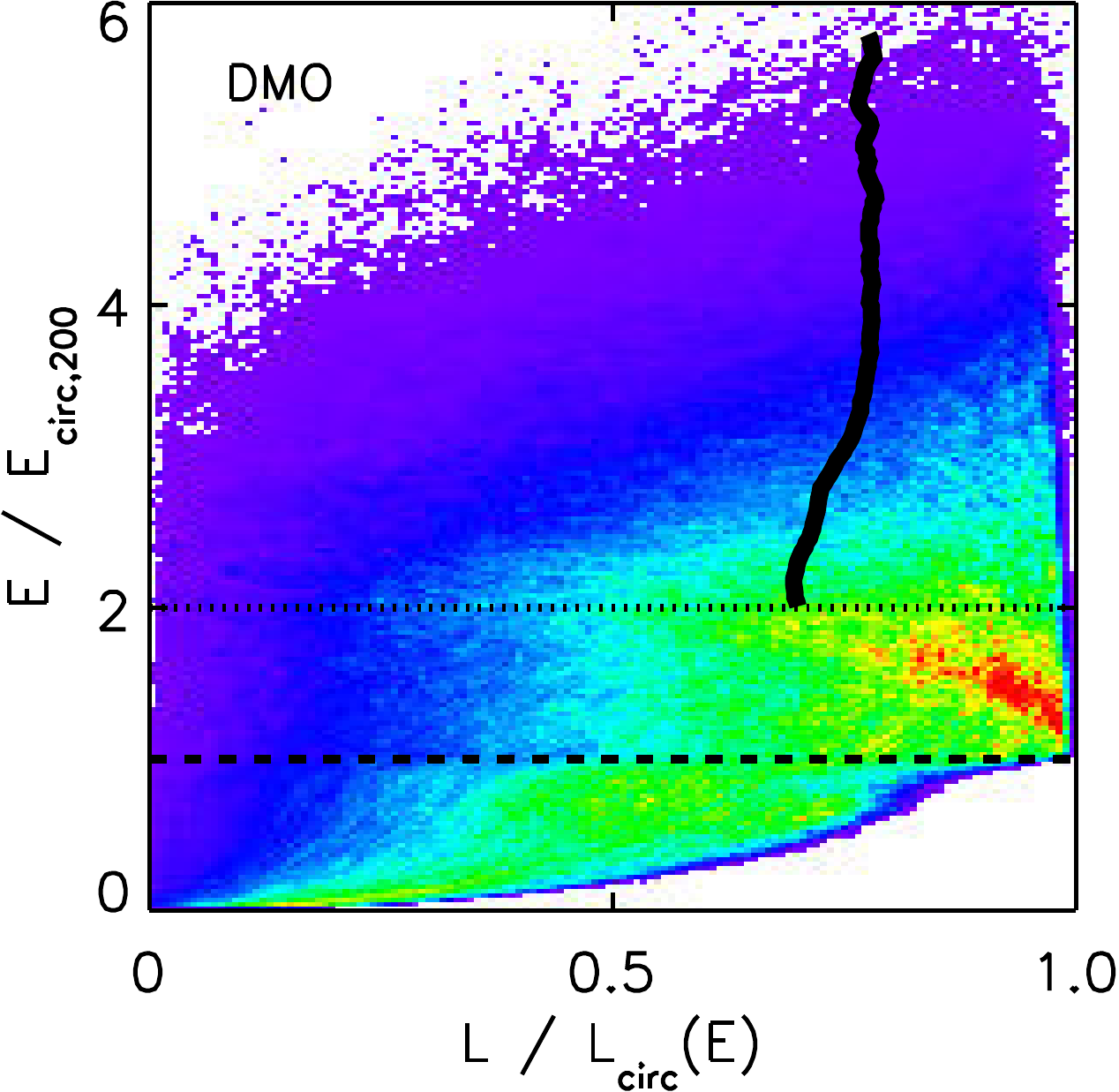}\hspace{1mm}
    \includegraphics*[trim = 0mm 0mm 0mm 0mm, clip, height = 0.295\textwidth]{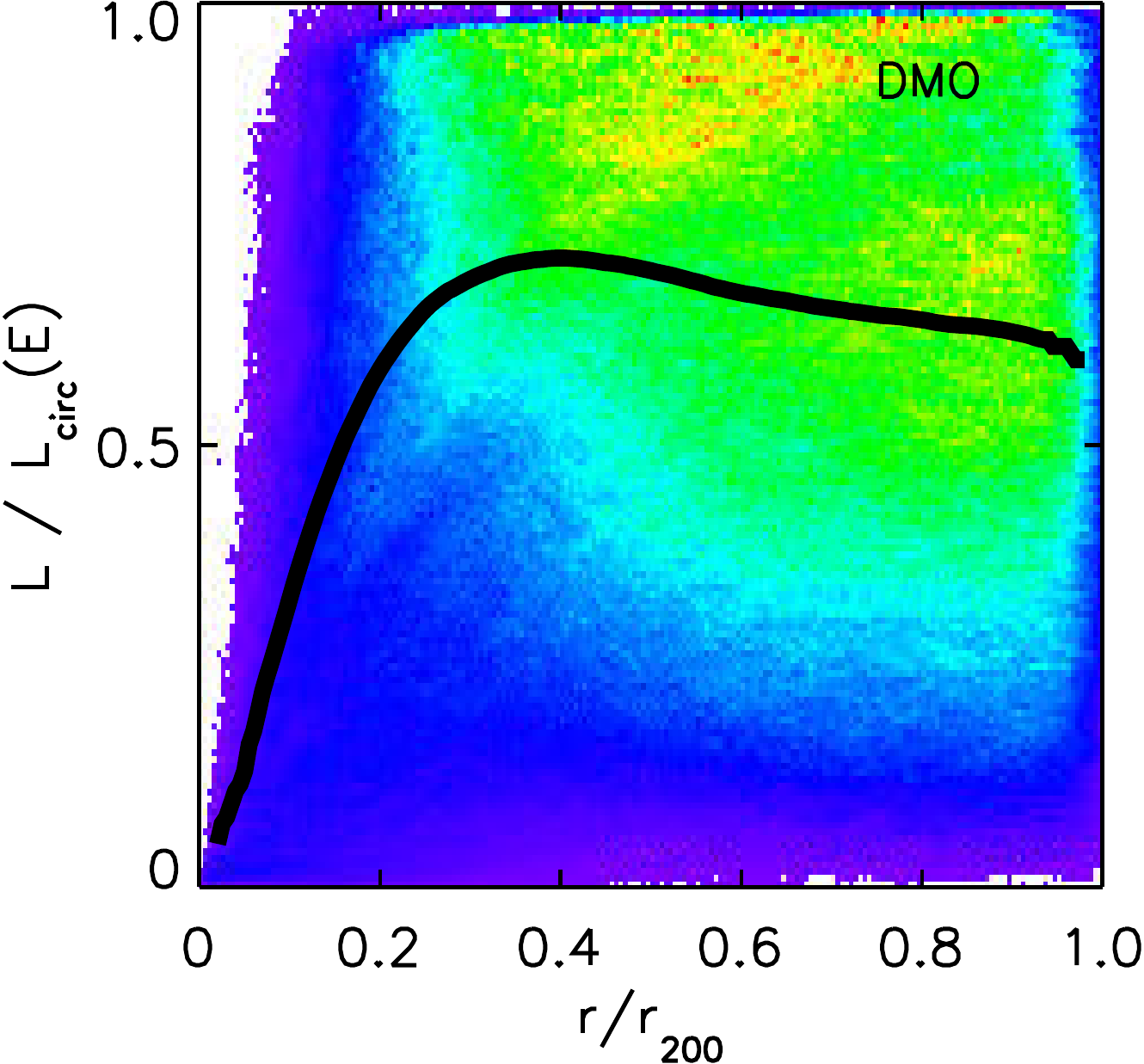}\hspace{1mm}
    \includegraphics*[trim = 0mm 0mm 0mm 0mm, clip, height = 0.295\textwidth]{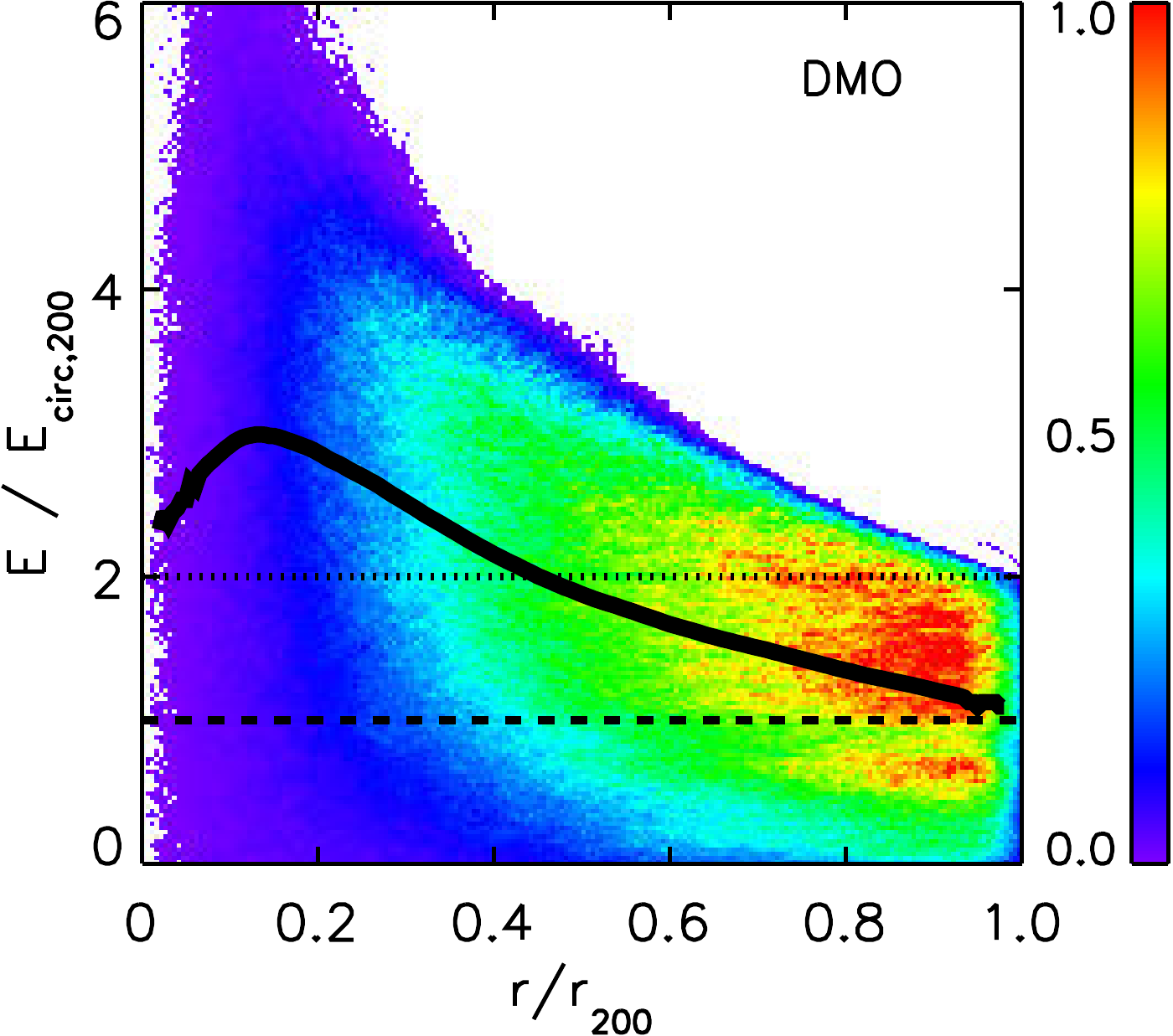} \\
\vspace{2mm}
    \includegraphics*[trim = 0mm 0mm 0mm 0mm, clip, height = 0.295\textwidth]{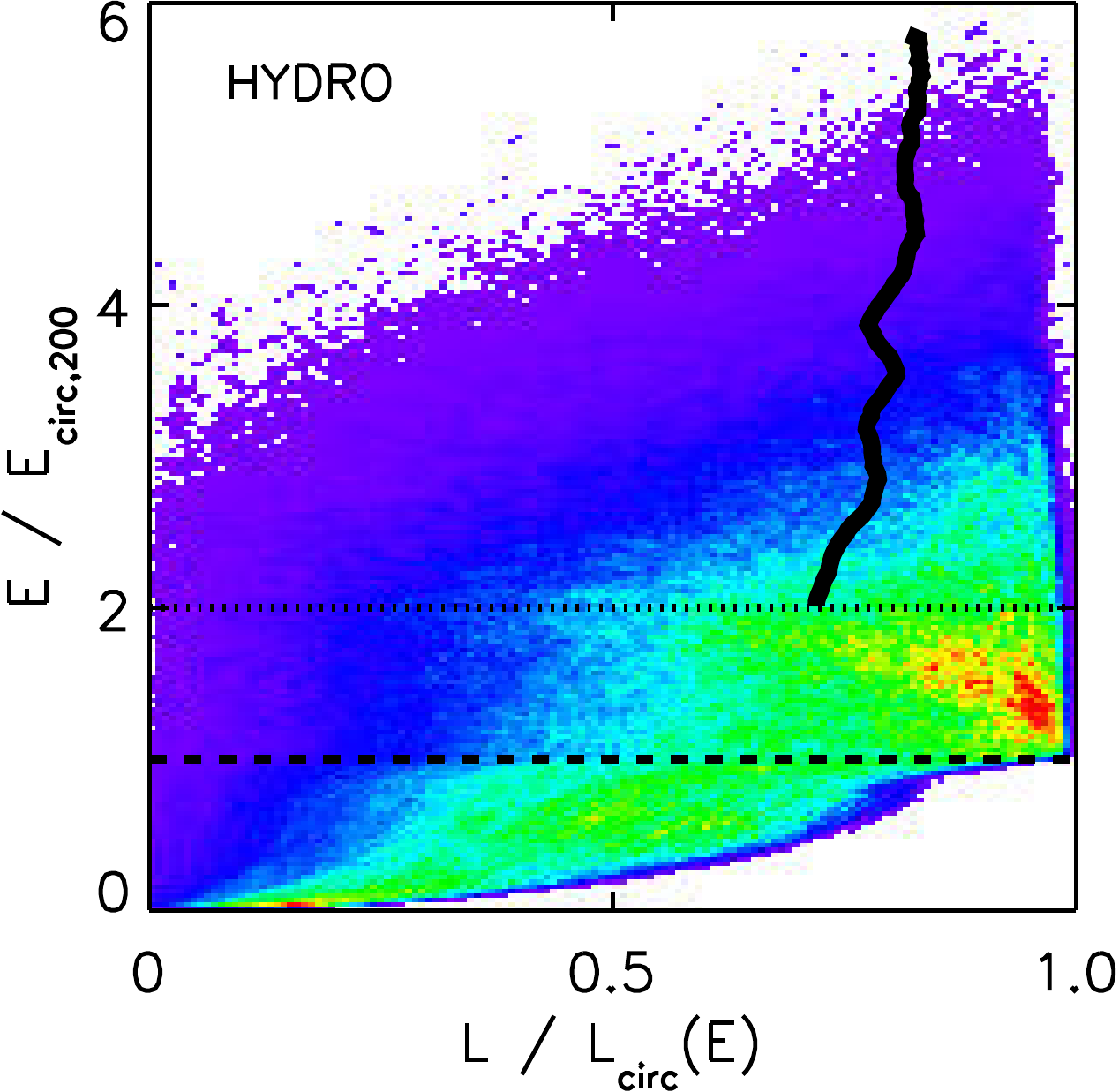}\hspace{1mm}
    \includegraphics*[trim = 0mm 0mm 0mm 0mm, clip, height = 0.295\textwidth]{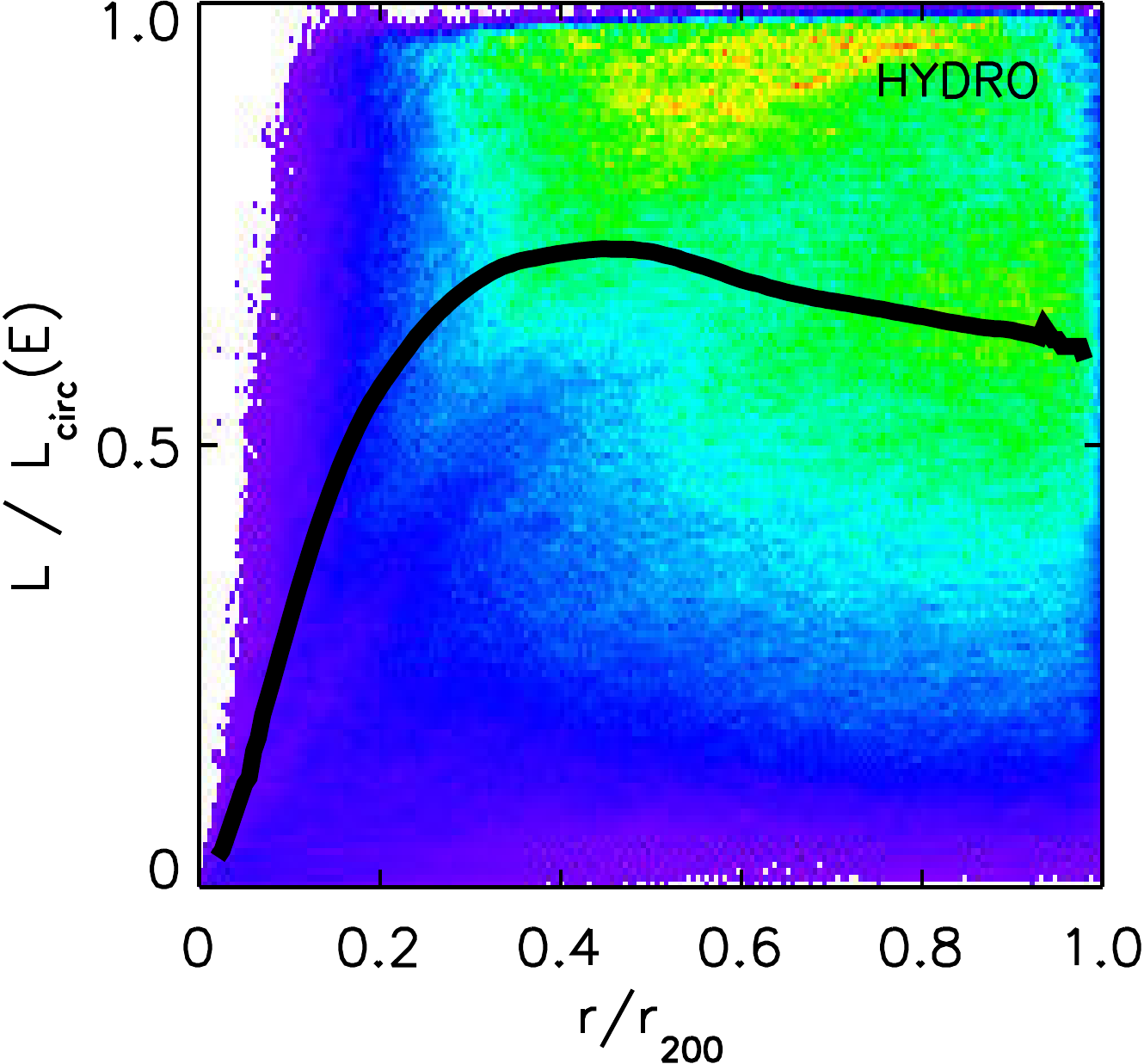}\hspace{1mm}
    \includegraphics*[trim = 0mm 0mm 0mm 0mm, clip, height = 0.295\textwidth]{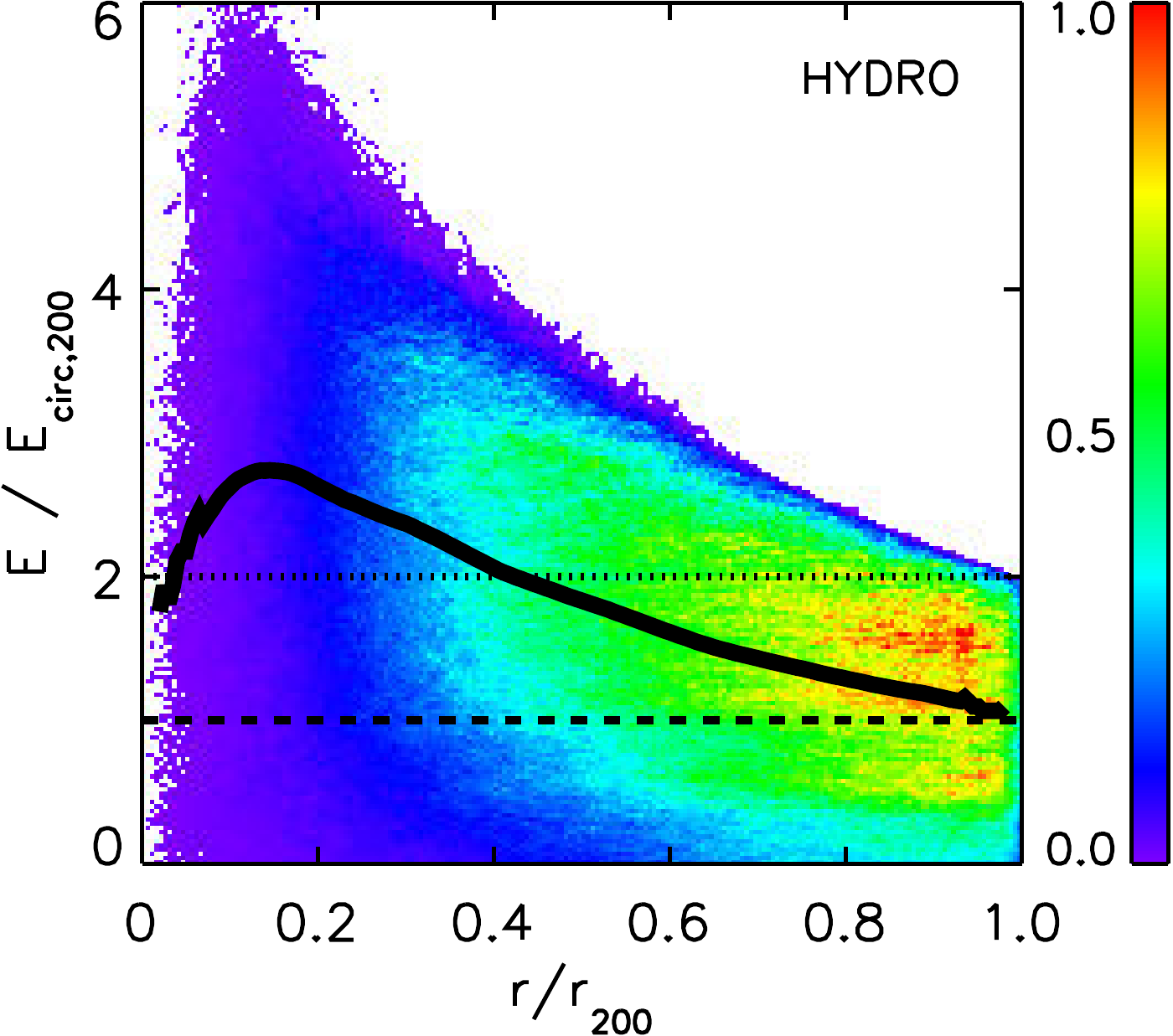} \\
  \end{center}
  \caption{Left: subhalo orbital angular momentum, normalised by the
    angular momentum of a circular orbit of the same energy, versus
    subhalo energy, normalised by the energy of a circular orbit at
    $r_{200}$. Middle: subhalo distance from the centre normalised by
    $r_{200}$, versus normalised angular momentum.  Right: normalised
    subhalo distance versus normalised energy. Note that because the
    total energy of a circular orbit at $r_{200}$ is negative, bound
    haloes appear with positive normalised energies. The top row shows
    results for the DMO simulations, the bottom row shows results for
    the hydrodynamic simulations. On the left and right panels the
    dashed and dotted lines at $E/E_{circ,200} = 1$ and $2$ indicate
    the energy for a halo on a circular orbit at $r_{200}$, and the
    potential energy for a halo at $r_{200}$,
    respectively. Overplotted onto the left panel is the median of
    $L/L_{circ}(E)$ as a function of $E/E_{circ,200}$, overplotted on
    the middle and right panels are the median of $L/L_{circ,200}$ and
    $E/E_{circ,200}$, both as a function of $r/r_{200}$. See footnote
    on page~\pageref{2d-explanation} for details of the 2D
    histograms.
    \label{fig:2d-energy} }

\end{figure*}

The most important differences between the subhalo number density
profiles and the mass density profiles are the inner slopes,
$-\gamma$, and the associated scale radii, $r_s$. In both the DMO and
hydrodynamic simulations, the substructure number density profiles
transition to much shallower profiles at much greater scale radii than
the DM mass density profiles. The difference in inner slope and scale
radius between the DMO and hydrodynamic simulations is less
significant, but as seen in Section~\ref{sec:distribution}, the
subhalo number density at a given radius is lower in the hydrodynamic
simulations.  Consequently, we find that the ``substructure bias'',
the relative underdensity of subhaloes compared to DM particles
towards the centre, already identified by \cite{Ghigna-2000} based on
DMO simulations, is even stronger in the hydrodynamic simulations,
where the central DM density is higher and the central subhalo density
lower compared to in the DMO counterparts.  The outer slope, $\beta$,
is quite poorly constrained, and the differences are not significant
for the central subhalo deficit.

\section{Subhalo Velocities} \label{sec:dynamics} The disruption of
substructures, and the impact of baryons, are also reflected in the
subhalo velocities. In Section~\ref{sec:velocity-bias}, we compute the
expected velocity bias of subhaloes relative to DM particles. In
Section~\ref{sec:energies}, we discuss the distributions of energies
and angular momenta, and in Section~\ref{sec:anisotropy}, we present
the subhalo anisotropy profiles.

\subsection{Subhalo velocity bias} \label{sec:velocity-bias} For a
spherical halo of size $R$ containing populations of particles in
equilibrium, assuming isotropy, the radial velocity dispersion,
$\sigma_r(r)$, of each population is related to its density,
$\rho(r)$, via
\begin{equation}
  \rho(r) \sigma_r^2(r) -
  \rho(R) \sigma_r^2(R) = \int_r^R \rho(r) \frac{GM(r)}{r^2} dr 
\end{equation}
where $M(r)$ is the enclosed mass. For $r~\ll~R$, $\rho(R)~\ll~\rho(r)$,
and the second term on the LHS can be ignored. Using the results for
the substructure density profiles for both DM particles and subhaloes
in Section~\ref{sec:substructure-bias}, as suggested by
\cite{Diemand-2004}, we can thus calculate the expected velocity bias
of the subhaloes relative to the DM particles.
\begin{equation}\label{eqn:velocity-bias}
  \frac{\sigma_{r,sub}(r)}{\sigma_{r,DM}(r)} = \left(
    \frac{\rho_{DM}(r)}{\rho_{sub}(r)}  \frac{\int_r^R \rho_{sub}(r) \frac{M(r)}{r^2} dr}{\int_r^R \rho_{DM}(r)\frac{M(r)}{r^2} dr}   \right)^{1/2}
\end{equation}
With the parametrisation for $\rho_{DM}(r)$ and $\rho_{sub}(r)$ given
by \eqnref{eqn:nfw} and \eqnref{eqn:alphabetagamma}, respectively, and
assuming that the velocity bias vanishes beyond $=300$ kpc, where the
subhalo number density and DM mass density are small, we can compute
the expected velocity bias of subhaloes relative to DM particles.

The expected velocity biases for the DMO and hydrodynamic simulations
are shown in the bottom panel of Figure~\ref{fig:radial}. It can be
seen that for both simulations, the velocity bias rises towards the
centre, most steeply between the (larger) scale radius of the
$(\alpha, \beta, \gamma)$ subhaloes number density profiles and the
(smaller) scale radius of the (NFW) DM density profiles, where the
difference between the two slopes is maximal. Because of the stronger
substructure bias, the expected velocity bias is likewise stronger in
the hydrodynamic simulation.

\subsection{Orbital energy and angular momentum}\label{sec:energies}
Assuming spherical symmetry about the centre of potential and
truncation at $r_{200}$, we compute the halo potential $\Phi(r)$ from
the density $\rho(r)$ of all particles at each snapshot:
$$
\Phi(r) = - 4 \pi G \left( \frac{1}{r} \int_0^r \rho(r') r'^2 dr' + \int_r^{r_{200}}
\rho(r') r' dr' \right)
$$

In Figure~\ref{fig:2d-energy}, we show the three 2D density
distributions\footnote{In Figures~\ref{fig:2d-energy}
  and~\ref{fig:2d-velocities} we use the interpolated orbits of all
  subhaloes in the mass-range $10^{6.5}-10^{8.5}$ and within the
  specified radii and time intervals to construct time-averaged
  2D-histograms. The histograms are normalised by the maximum
  occupation value for each pair of otherwise identical DMO and
  hydrodynamic panels, and coloured using the linear colour scales,
  indicated by the colour bars to the right of both
  figures. \label{2d-explanation}} of specific orbital energies,
specific orbital angular momenta, and radii, of subhaloes in the mass
range $10^{6.5}-10^{8.5}\Ms$ inside $r_{200}$ from orbits interpolated
over 5 Gyr in lookback time. We normalise the energies, $E$, by the
total energy of a circular orbit at $r_{200}$, $E_{circ,200}$, the
angular momenta, $L$, by the angular momentum of a circular orbit of
the same energy, $L_{circ}(E)$, and the radius $r$ by the virial
radius, $r_{200}$. Note that since our potential definition has the
zero-point at infinity (neglecting all mass beyond $r_{200}$) the
total energy of a circular orbit at $r_{200}$ is negative. As a
result, subhalo orbits which are more bound, corresponding to more
negative total energies, have higher values of $E/E_{circ,200}$.

The left column of Figure~\ref{fig:2d-energy} shows the E-L
probability density.  Because the energy of a circular orbit increases
monotonically with radius, subhaloes close to $L/L_{circ}(E)=1$ are
ordered by radius: those located at $r_{200}$ are located at
$(L/L_{circ}(E), E/E_{circ,200}) = (1,1)$. Circular orbits with
smaller radii have more negative energies, and line up above this
point.

In the middle and right columns of Figure~\ref{fig:2d-energy}, we show
the distributions of $L/L_{circ}(E)$ and $E/E_{circ,200}$,
respectively, both versus $r/r_{200}$. In the L-R plane, we see that
the average circularity is relatively constant at radii beyond $\sim
0.2 r_{200}$ (corresponding to $\sim 50$ kpc) and declines sharply for
smaller radii, indicating a transition towards more radial orbits near
the centre. In the E-R plane, we see that the average specific orbital
energy becomes more negative towards the centre, but peaks at $\sim
0.1 r_{200}$ (corresponding to $\sim 25$ kpc), where the increase in
the average kinetic energy of subhaloes compensates for the
continuously more negative potential energy. Because the average
circularity also declines towards the centre, the increase in kinetic
energy indicates an increase in radial velocities of subhaloes at
small radii. This effect is slightly stronger in the hydrodynamic
simulations.

Since the minimum total energy of a subhalo is given by the potential
energy at its radius, the value of $E/E_{circ,200}$ is limited from
above, explaining the ``forbidden'' region for high values of
$E/E_{circ,200}$ in the E-R plane, seen in the right column of
Figure~\ref{fig:2d-energy}.

For guidance, on the E-L and E-R planes in Figure~\ref{fig:2d-energy},
the dashed and dotted lines indicate values of $E/E_{circ,200} = 1$
and $E/E_{circ,200} = 2$, respectively. For orbital energies less
negative than $E_{circ,200}$, the radius of a circular orbit lies
outside $r_{200}$. For each value of $0 < E/E_{circ,200} < 1$, there
is a maximum circularity for orbits with pericentres inside
$r_{200}$. This explains the ``forbidden'' region for high
circularities at $E/E_{circ,200} < 1$ on the E-L plane. Likewise, a
value of $E/E_{circ,200} = 2$ is equal to the potential energy at
$r_{200}$ and hence the maximum orbital energy for a subhalo on a
radial orbit with an apocentre inside of $r_{200}$. Subhaloes on
radial orbits with higher energies (values of $E/E_{circ,200} < 2$)
spend a fraction of their orbital period outside of $r_{200}$, raising
the average circularity measured inside of $r_{200}$.

By contrast, the nearly empty region at high values of
$E/E_{circ,200}$ and low values of $L/L_{circ}(E)$ in the E-L plane is
not a forbidden region. Instead, it reflects the fact that subhaloes
with low orbital energies and correspondingly short orbital periods
are more easily disrupted on radial orbits. As can be seen by the
solid black line on this panel, the median circularity increases for
more closely bound subhaloes above $E/E_{circ,200} = 2$.

While the subhaloes with the most negative energies thus typically
have high circularities and exist only near the halo centre, it does
{\it not} follow that subhaloes near the centre have high
circularities: instead, as can be seen on the L-R plane in the middle
column of Figure~\ref{fig:2d-energy}, the average circularity for
subhaloes is lowest near the halo centre. Among the subhalos on orbits
with highly negative energies and short orbital periods, subhalos on
more radial orbits get most easily disrupted. However, all subhalos
with short orbital periods are prone to tidal disruption, so the
central region of the halo is predominantly populated by high velocity
subhaloes with long orbital periods on highly radial orbits.

\begin{figure}
 \hspace{1.5mm}   \includegraphics*[trim = 0mm 19mm 0mm 0mm, clip, width = 0.975\columnwidth]{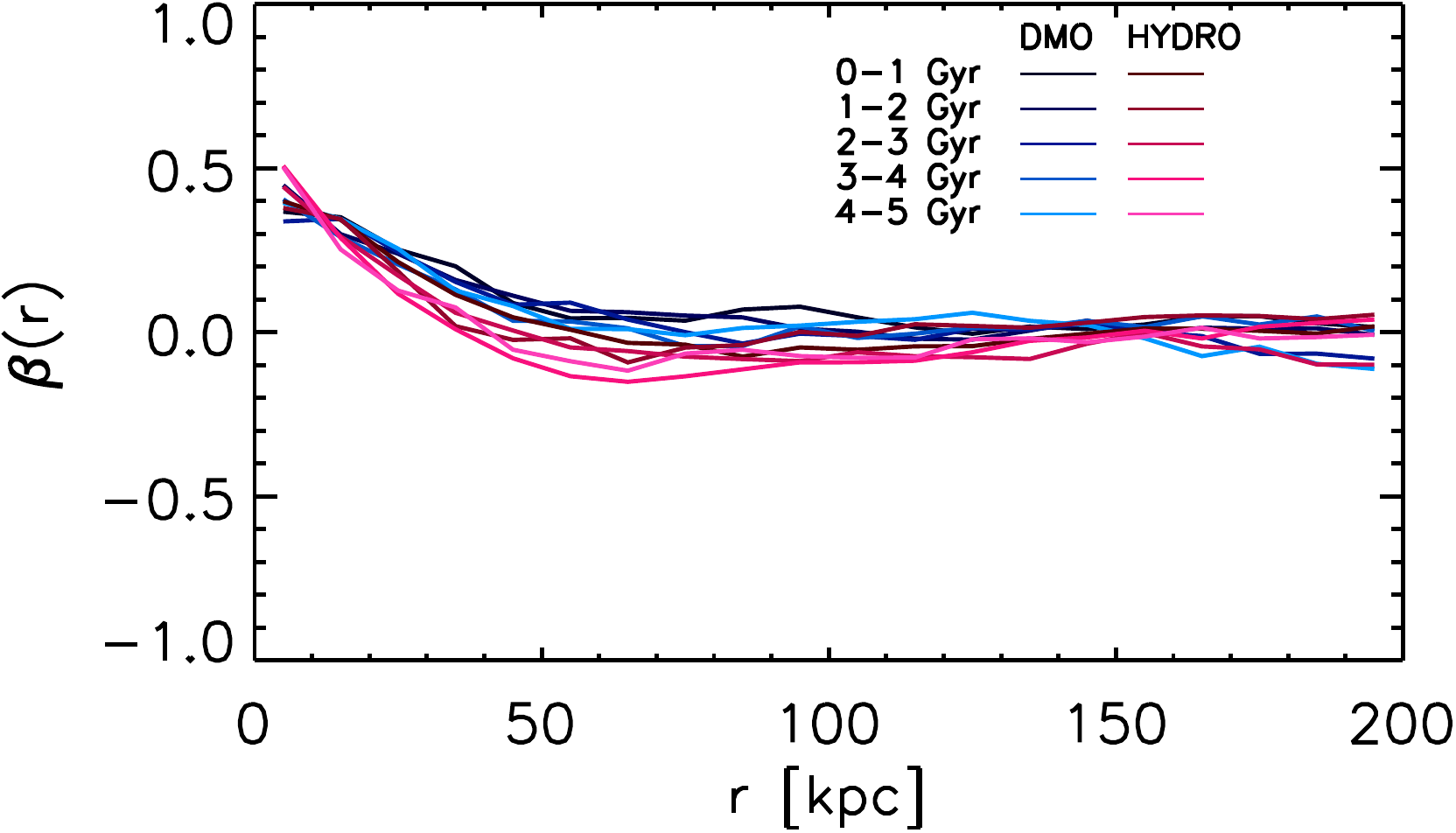} \\
    \includegraphics*[trim = 0mm 0mm 0mm 0mm, clip, width = \columnwidth]{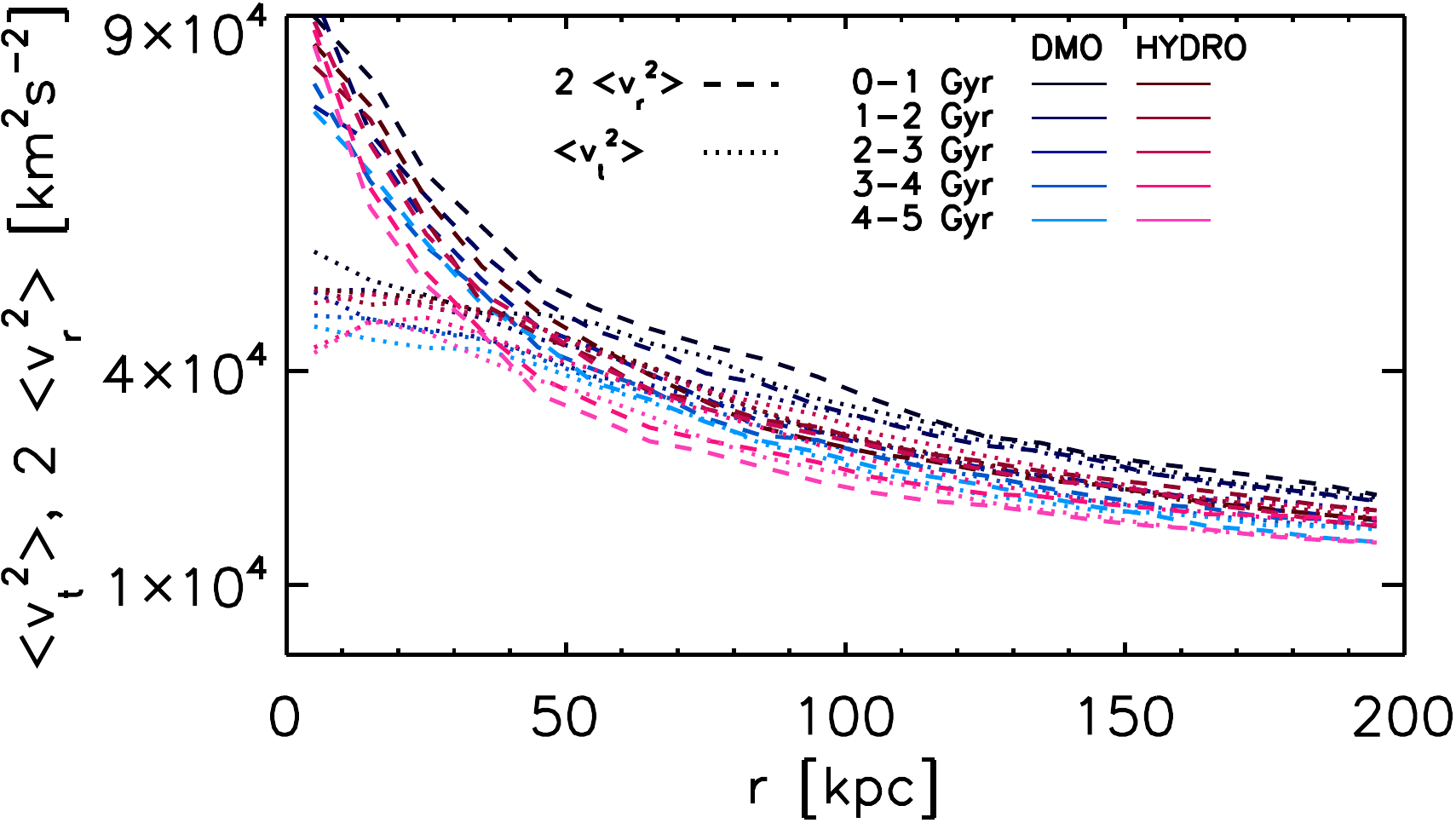}  
  \caption{Top: velocity anisotropy parameter, $\beta(r)$, profiles for
    subhaloes of mass $10^{6.5}-10^{8.5}\Ms$ in the DMO (blue) and
    hydrodynamic (red) simulations. Bottom: profiles of $2 \times
    \overline{v_r^2}$ (dashed) and $\overline{v_t^2}$ (dotted) in the
    same simulations. At large radii, the velocity dispersion in each
    dimension is similar, and the velocity anisotropy is close to
    zero. At small radii, there are fewer subhaloes with small radial
    velocities, and the velocity anisotropy increases.
    \label{fig:anisotropy}}
\end{figure}

\subsection{Velocity anisotropy profiles}\label{sec:anisotropy}
The velocity anisotropy, $\beta(r)$, quantifies the measured ratio
between the kinetic energy due to motions in the radial direction,
$v_r$, and in the tangential direction, $v_t$
\begin{equation}
\beta(r) = 1 - \frac{\overline{v_t^2(r)}}{2 \, \overline{v_r^2(r)}}, 
\end{equation}
The velocity anisotropy is zero for equal velocity dispersion in each
dimension, positive for more radial orbits, and negative for more
circular ones.

In the top panel of Figure~\ref{fig:anisotropy}, we show the velocity
anisotropy parameter of subhaloes as a function of radius in our
simulations. We find that the velocity anisotropy for subhaloes in the
mass range $10^{6.5}-10^{8.5}\Ms$ is close to zero at $r > 50$ kpc in
both the DMO and hydrodynamic simulations. At smaller radii, the
anisotropy rises to $\sim 0.$5 near the halo centre. In the bottom
panel of Figure~\ref{fig:anisotropy}, we show the mean of the square
of the tangential velocity components, $\overline{v_t^2}$, (dotted
lines), and twice the mean of the squares of the radial velocity
components, $2 \times \overline{v_r^2}$, (dashed lines), as a function
of radius. Both sets of lines rise towards the centre, and are nearly
equal at $r > 50$ kpc, corresponding to near zero velocity
anisotropy. At smaller radii, the average radial velocities rise much
more steeply, reflecting the prediction of a centrally rising velocity
bias described in Section~\ref{sec:velocity-bias}. However, the
increase in subhalo radial velocities is less than predicted by the
spherical equilibrium model, partly due to the fact that subhalo
disruption and infall are continuous processes, and the instantaneous
velocities of the existing subhaloes are not fully reflective of the
difference in the instantaneous substructure bias.

Interestingly, the centrally rising velocity anisotropy for subhaloes
is the opposite of that seen for spherical systems composed of
indissoluble bodies, such as stars in globular clusters, where orbits
become more isotropic near the centre \citep{Osipkov-1979,
  Merritt-1985}. This is easily understood: while interactions
isotropise the orbits near the centres of star clusters
\citep[e.g.][]{Baumgardt-2002}, tidal processes experienced by
subhaloes near the centre of a DM halo also lead to their disruption
over time. Hence, close to the halo centre, the subhalo population is
dominated by subhaloes with small pericentres but much larger
apocentres which limits the work done by tidal forces. As most
circular orbits with small pericentres are destroyed, and circular
orbits with large pericentres never enter the halo centre, the
innermost region contains predominantly subhaloes on highly eccentric
orbits, resulting in the increased central velocity anisotropy.

\section{Subhalo Velocity Distributions}\label{sec:velocities}
Due to the mass-velocity degeneracy inherent in gravitational
interactions of substructures with streams mentioned in the
introduction, the velocity probability density function (VPDF) of
substructures is an important prediction of any cosmological model. In
this section, we revisit the common assumption of locally Maxwellian
velocity distributions, and show that it is increasingly violated
towards the halo centre. We propose instead to parameterise the radial
velocity, $v_r$, by a bimodal Gaussian, and composites such as the
tangential velocity, $v_t$, and the total velocity norm, $|v|$, by
Rician distributions.

\begin{table}
\caption{Subhalo velocity PDF parameters}
\label{tab:fits-vel}
\setlength\tabcolsep{5pt}
\begin{tabularx}{\columnwidth}{c|cccc}
\hline \hline
& 0-10 kpc & 10-20 kpc & 20-50 kpc & 50-200 kpc\\[3pt]
\hline 
&\multicolumn{4}{c}{$v_r$ PDF parameters $\mu, \sigma$ [kms$^{-1}$]$^{[1]}$} \\[3pt]
\hline 
DMO & 173.3, 125.4 & 149.1, 131.0 & 91.3, 141.1 & 75.6, 88.7 \\
Hydro & 188.0, 120.7 & 151.6, 128.7 & 89.1, 134.2 & 71.7, 85.1 \\[3pt]
\hline
&\multicolumn{4}{c}{$v_t$ PDF parameters $\mu, \sigma$ [kms${-1}$]$^{[2]}$}  \\
\hline 
DMO & 159.3, 119.2 &165.3, 109.2 & 163.1, 95.1 &104.6, 86.2 \\
Hydro & 161.6, 118.4 & 181.4, 101.0 & 180.2, 82.4 & 110.3, 77.3  \\[3pt]
\hline
&\multicolumn{4}{c}{$|v|$ PDF parameters $\mu, \sigma$ [kms$^{-1}$]$^{[2]}$} \\
\hline 
DMO & 283.5, 78.0 & 266.6, 80.4 & 238.4, 85.4 & 165.0, 77.2 \\
Hydro & 290.3, 59.7 & 274.5, 67.9 & 244.1, 72.6 & 162.5, 70.8 \\[3pt]
\hline \hline
\end{tabularx}
\vspace{.3cm}\\
$^{[1]}$for a symmetric
double-Gaussian VPDF, as in \eqnref{eqn:double-gaussian-2}.
$^{[2]}$for a Rician VPDF, as in \eqnref{eqn:rice}.
\end{table}

\subsection{Non-Maxwellian distributions}
The velocity distribution of particles in haloes is commonly
characterised by a (locally) Maxwellian VPDF. A Maxwellian VPDF arises
under the assumption that particle velocities are isotropic, such that
all three velocity components are independent random variables whose
probability density functions (PDFs) are each given by normal
distributions,
\begin{equation}\label{eqn:normal}
P(v_i) = \frac{1}{\sigma \sqrt{2 \pi}} e^{-\frac{v_i^2}{2\sigma^2}},
\end{equation}
where $\sigma$ is the velocity dispersion in one dimension, and
isotropy implies a mean velocity of zero. If the three components are
independent and have identical distributions, integration over one or
two variables yields the 2D or 3D Maxwellian velocity PDFs,
\begin{equation}\label{eqn:maxwell-2}
P(|v_{2D}|)= \frac{v}{\sigma^2} e^{-v^2 / (2 \sigma^2)},
\end{equation}
also called the Rayleigh distribution, and
\begin{equation} \label{eqn:maxwell-3}
P(|v_{3D}|)=\sqrt{\frac{2}{\pi}} \frac{v^2}{\sigma^3} e^{-v^2 / (2 \sigma^2)},
\end{equation}
which is known as the Maxwell-Boltzmann distribution.

If $v_r$, $v_\theta$ and $v_\phi$ are independent degrees of freedom
with equal Gaussian distribution functions, the tangential velocity,
$v_t = \sqrt{v_\theta^2+v_\phi^2}$, should follow
\eqnref{eqn:maxwell-2}, and the norm of the total velocity, $|v| =
\sqrt{v_r^2+v_t^2}$, should follow \eqnref{eqn:maxwell-3}.

\begin{figure*}
  \begin{center}
    \includegraphics*[trim = 0mm 0mm 20mm 0mm, clip, height = 0.4\textwidth]{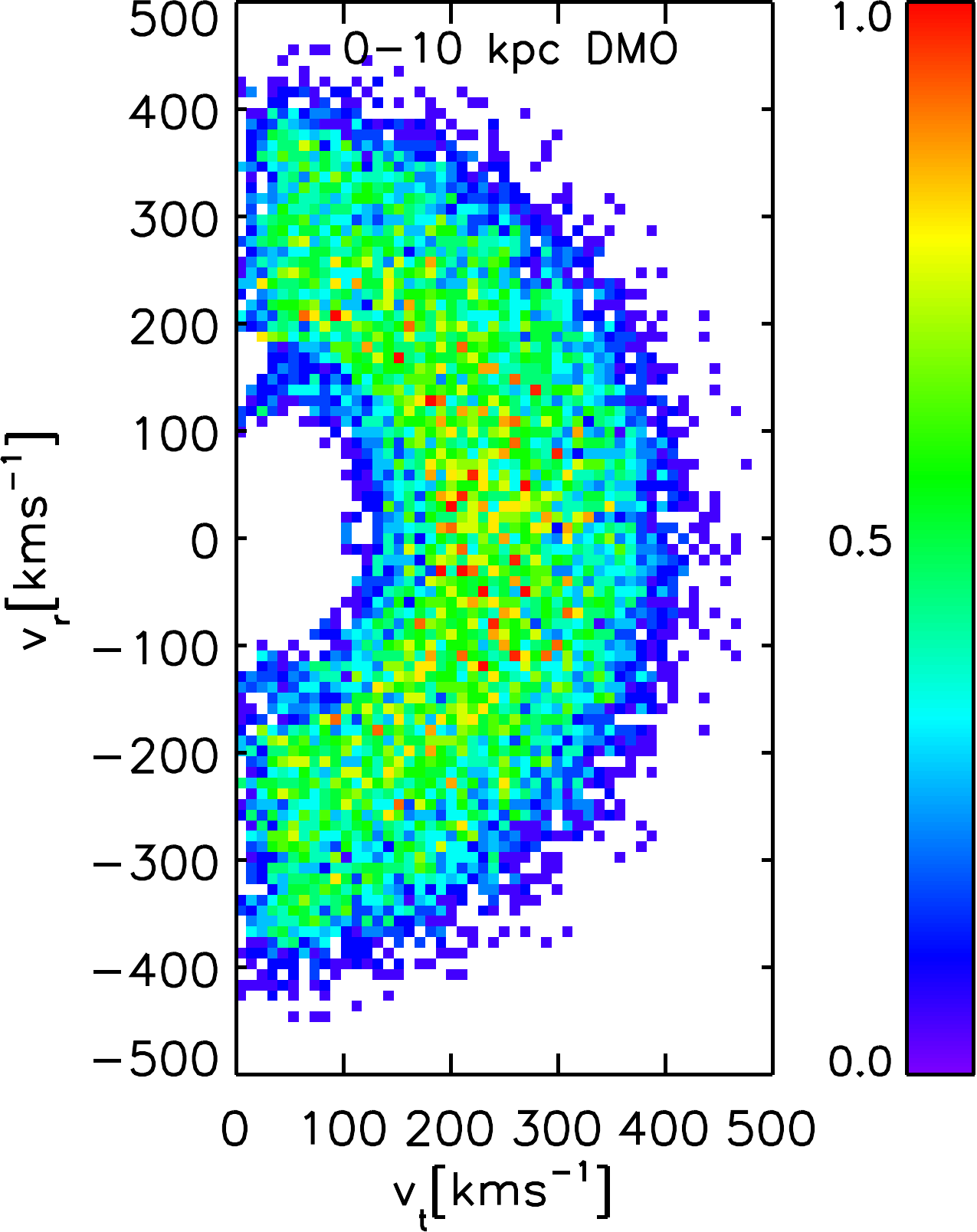}
    \includegraphics*[trim = 29mm 0mm 20mm 0mm, clip, height = 0.4\textwidth]{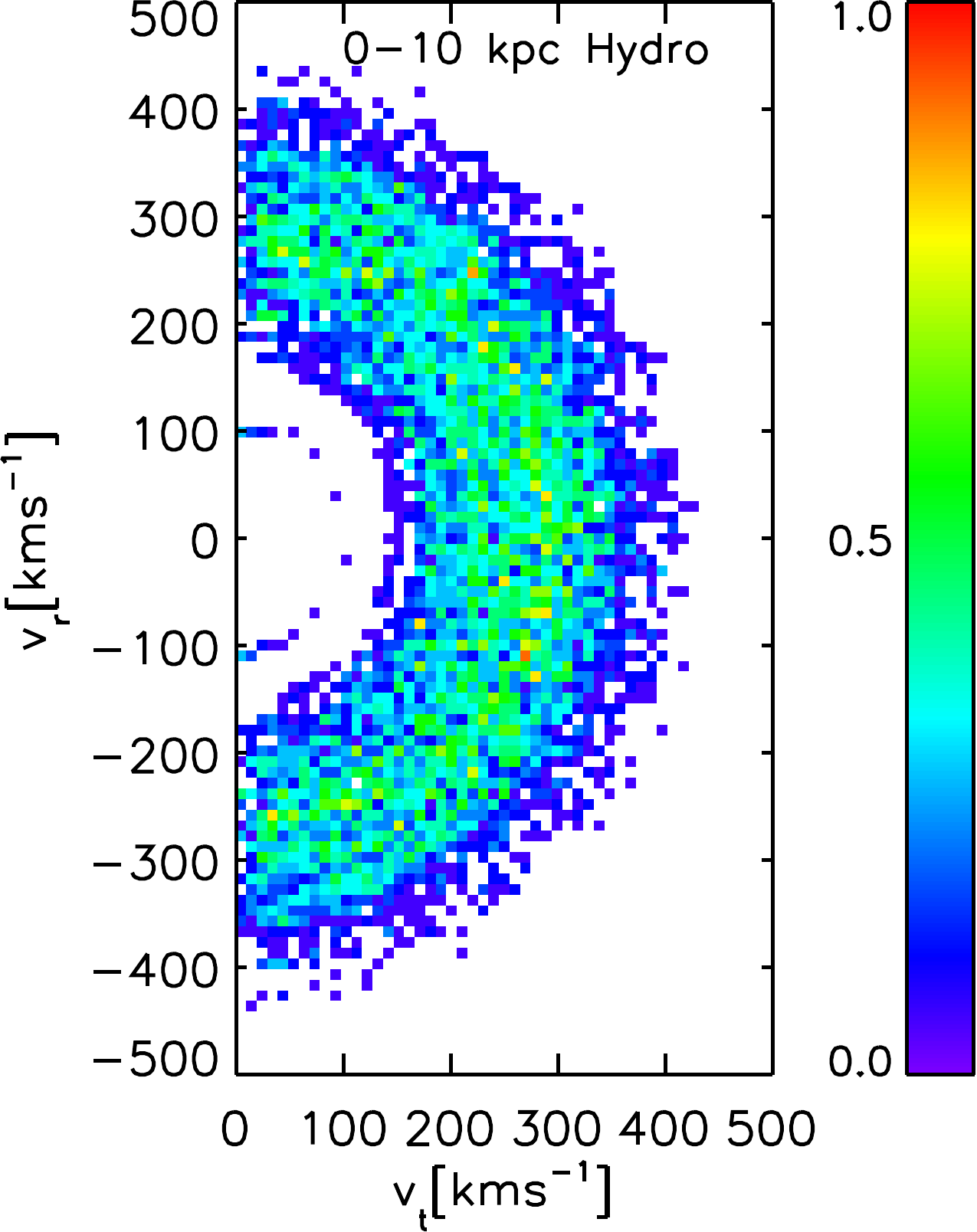}  
    \includegraphics*[trim = 0mm 0mm 20mm 0mm, clip, height = 0.4\textwidth]{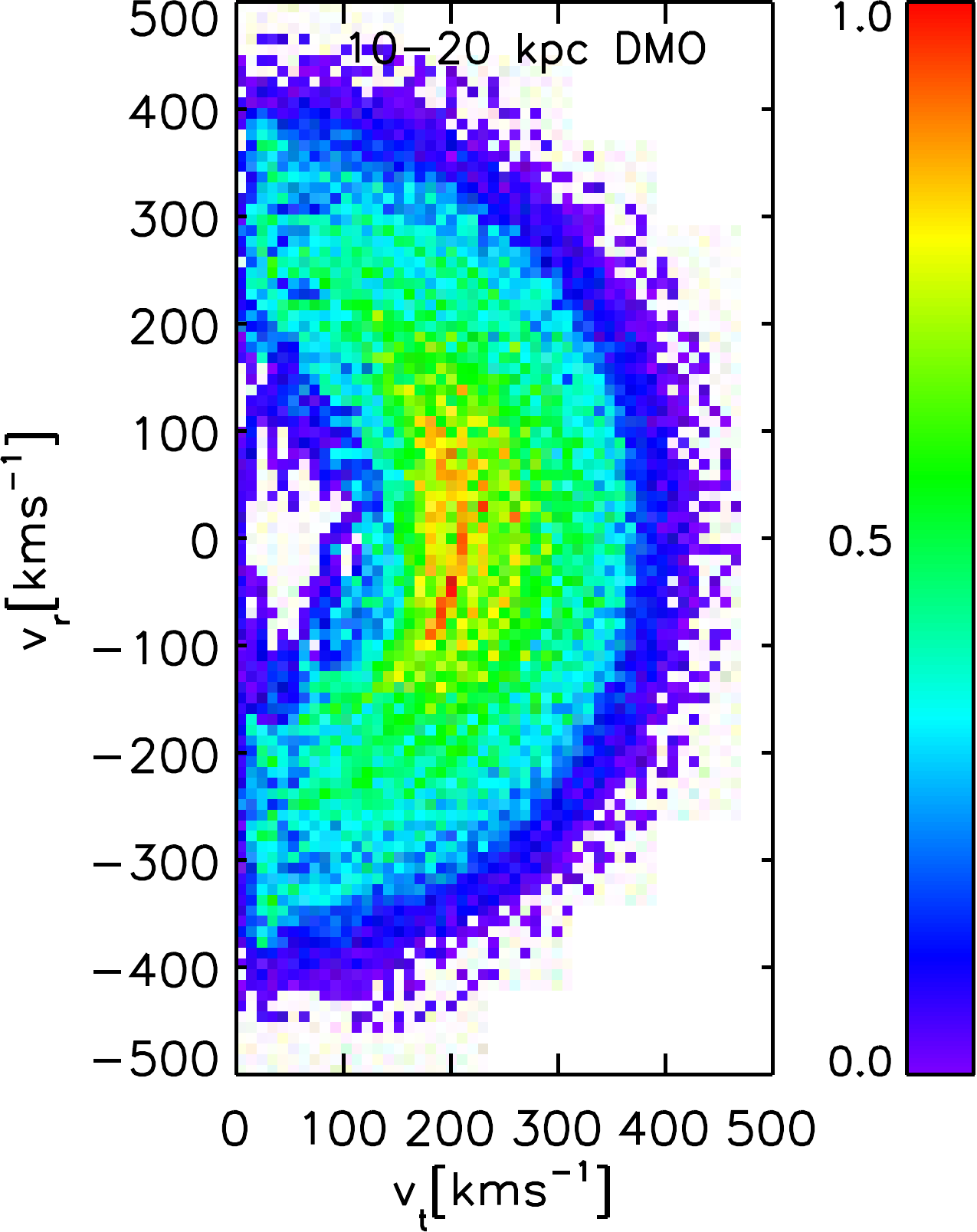}
    \includegraphics*[trim = 29mm 0mm 0mm 0mm, clip, height = 0.4\textwidth]{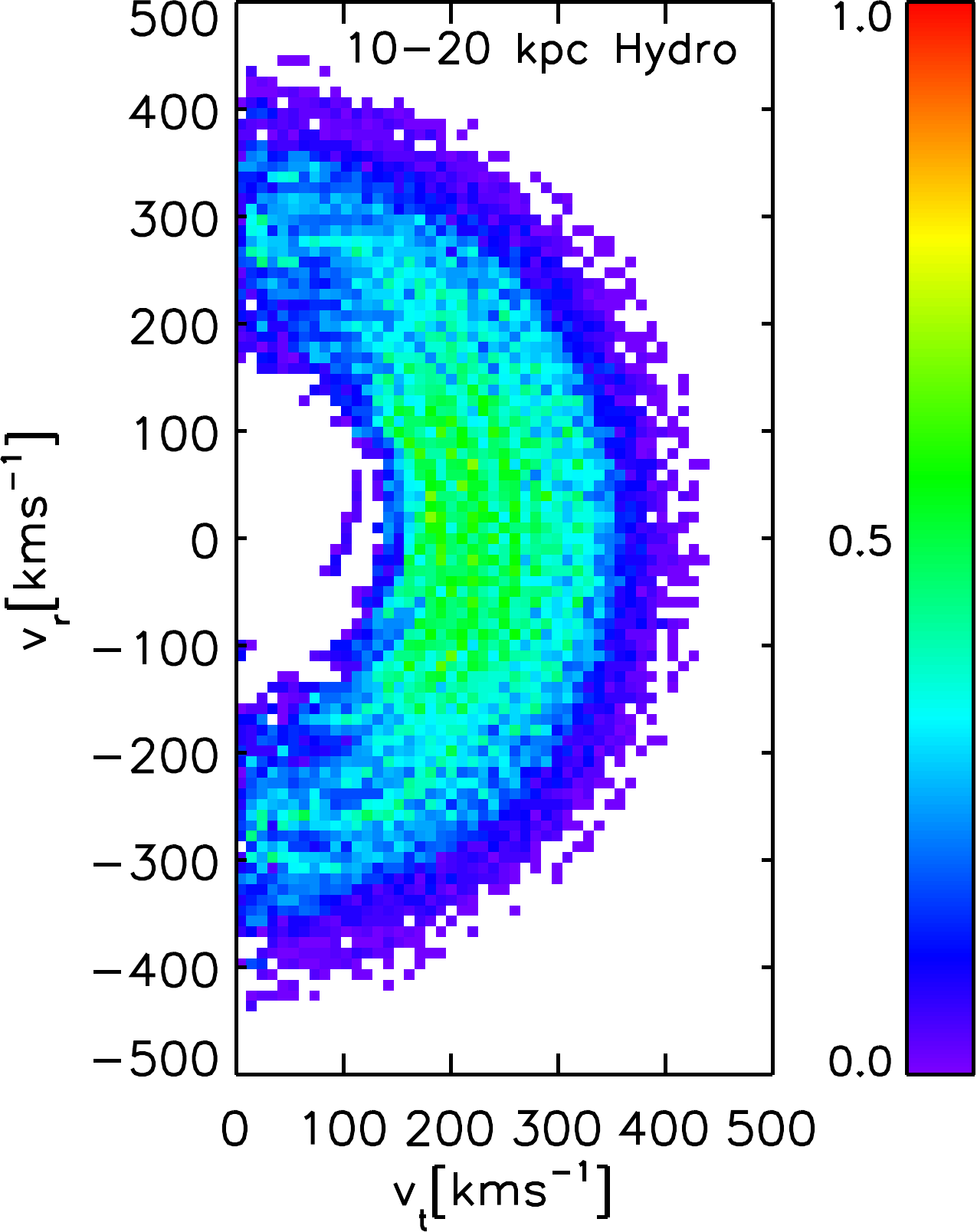} \\
\vspace{2mm}
    \includegraphics*[trim = 0mm 0mm 20mm 0mm, clip, height = 0.4\textwidth]{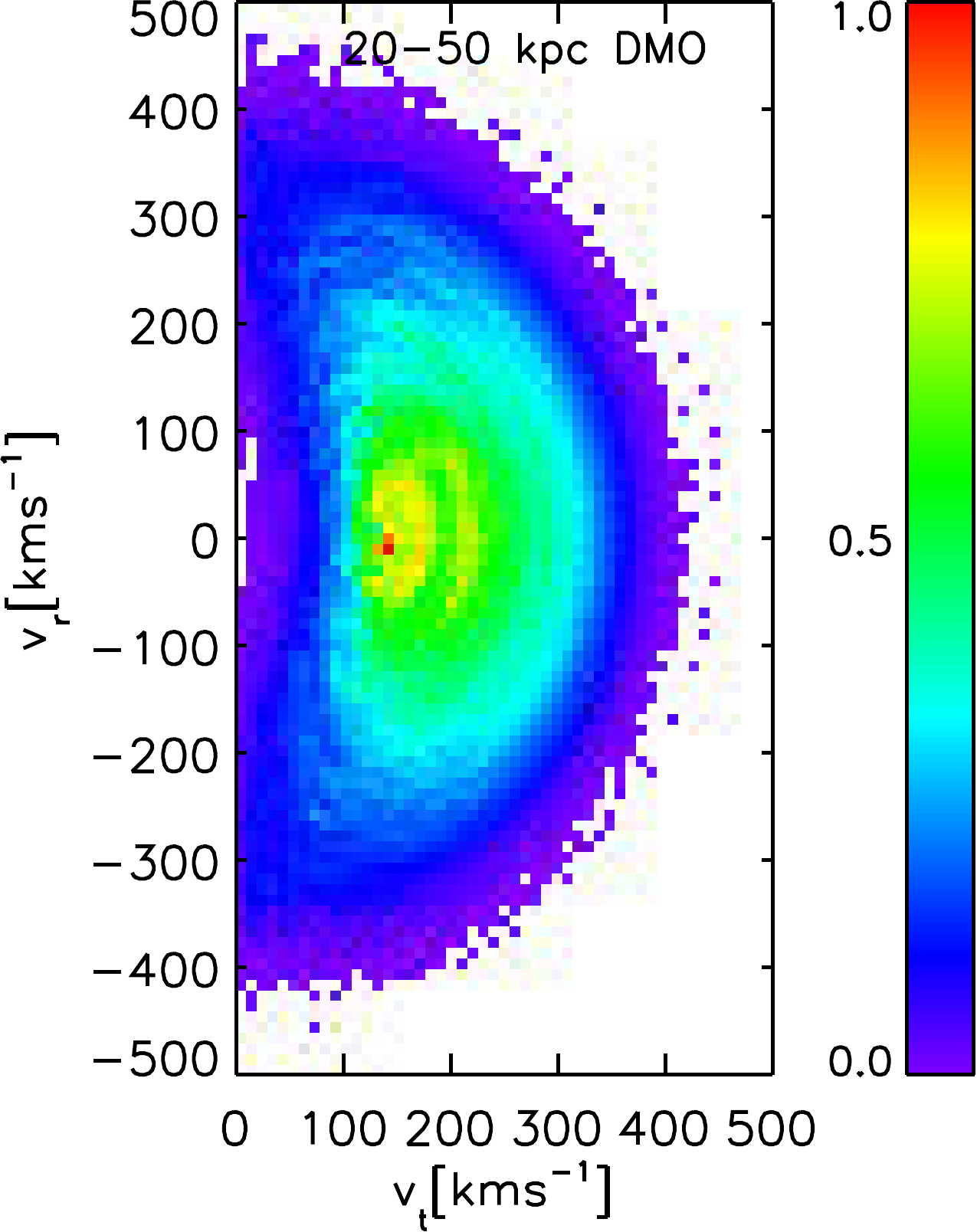}
    \includegraphics*[trim = 29mm 0mm 20mm 0mm, clip, height = 0.4\textwidth]{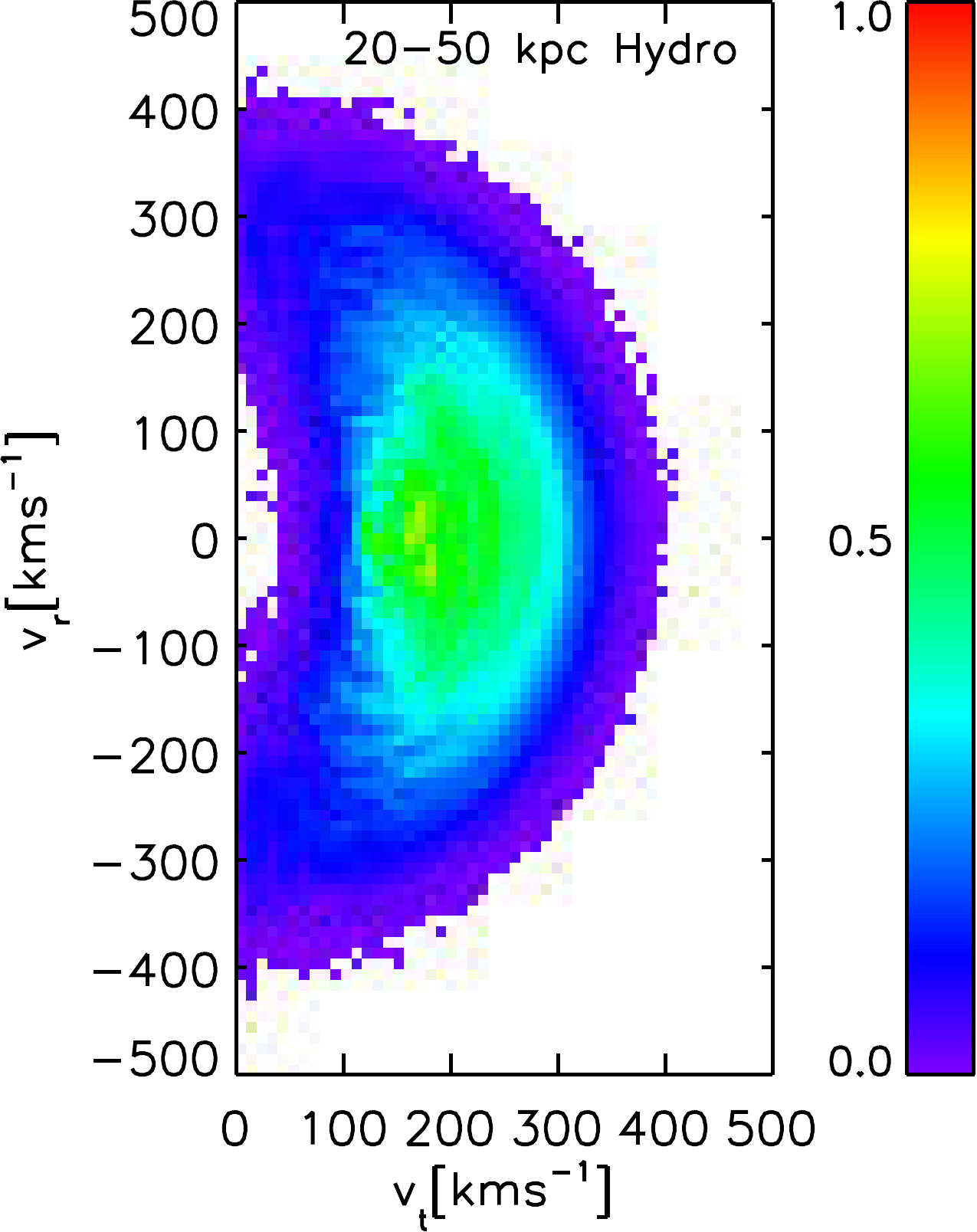}  
    \includegraphics*[trim = 0mm 0mm 20mm 0mm, clip, height = 0.4\textwidth]{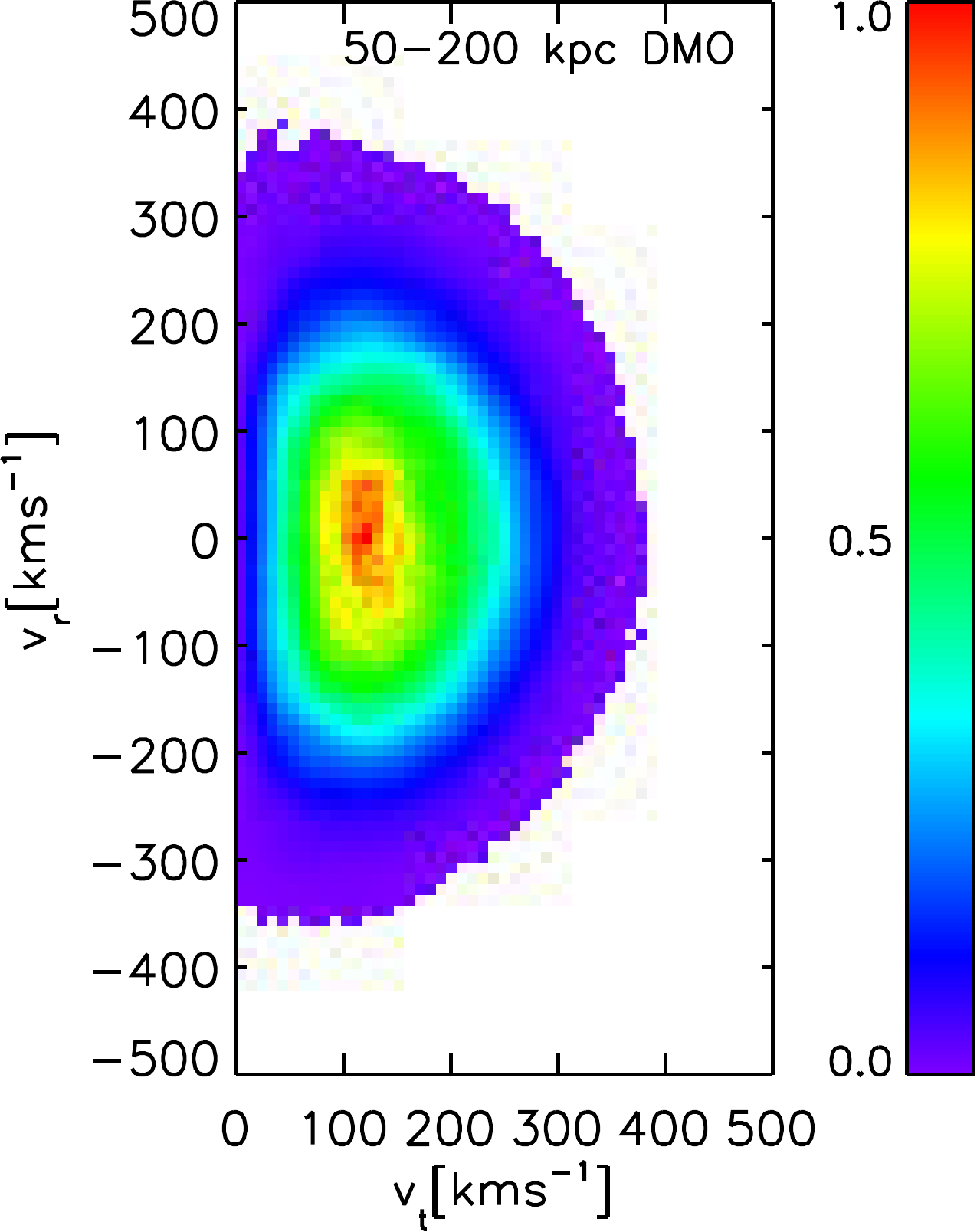}
    \includegraphics*[trim = 29mm 0mm 0mm 0mm, clip, height = 0.4\textwidth]{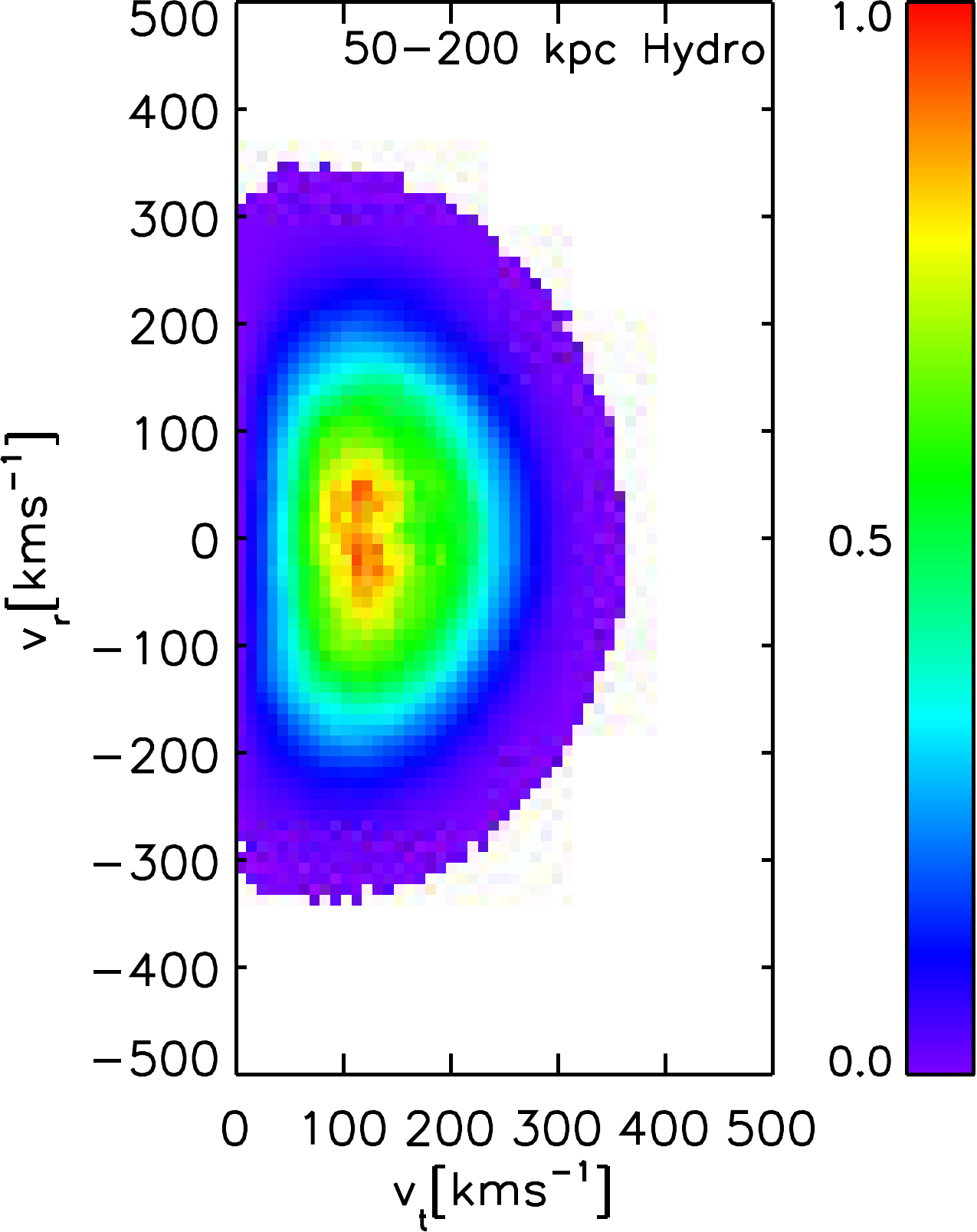} \\
  \end{center}

  \caption{Subhalo velocity distributions in the $v_r-v_t$ plane, in
    different radial shells, and for the DMO and hydrodynamic
    simulations, using all four haloes over 5 Gyr lookback time. At
    small radii, $v_r$ and $v_t$ are highly correlated, such that the
    mean velocity, $|v| = \sqrt{v_r^2+v_t^2}$ is approximately
    constant. At large radii, the mean velocity $|v|$ is smaller, so
    $v_t$ and $v_r$ are more independent, approximating a 2D Maxwell
    distribution. It can also be seen that, at all radii, the velocity
    distribution is slightly more concentrated in the hydrodynamic
    simulations, which is also evident from the projected velocity
    distributions shown in Figure~\ref{fig:velocities}. See
      footnote on page~\pageref{2d-explanation} for details of the 2D
      histograms.
      \label{fig:2d-velocities}}
\end{figure*}

While a local Maxwellian is a simple way to parameterise the total
velocity distribution, \cite{Kazantzidis-2004} have shown that it is
in fact not a steady-state solution to the velocity distribution
inside NFW haloes, as it leads to a quick dissolution of the cusp. It
has also been noted that a Maxwellian distribution is not a good fit
to the particle velocities measured in a high resolution numerical
simulations, and \cite{Vogelsberger-2009} have shown that DM particles
have prominent and long-lived, non-Gaussian velocity substructures,
which are relics of the assembly history of the
halo. \cite{Vergados-2008} argued that the particle velocity
distribution in an NFW-like halo should follow a Tsallis shape, based
on generalised Gaussian distributions that give better fits to the
high-velocity tails observed in the central regions of numerical
simulations.

Other attempts include truncating the Maxwellian at the escape
velocity (see e.g. \citealt{Fairbairn-2009}), while \cite{Kuhlen-2010}
opted empirically to fit more general distribution functions of the
form:
\begin{equation}
f(v_r) = \frac{1}{N_r}
e^{-\left(v_r^2/{2\sigma_r^2}\right)^{\alpha_r}}, \\
f(v_t) = \frac{v_t}{N_t}
e^{-\left(v_t^2/{2\sigma_t^2}\right)^{\alpha_t}}
\end{equation}
where $N_t$ and $N_r$ are normalisation constants, and $\alpha_r$,
$\alpha_t$ generalise the 1D and 2D Maxwellian distributions by including
additional free parameters.

\begin{figure*}
\vspace{-1mm}
  \begin{center}
    \includegraphics*[trim = 0mm 0mm 0mm 0mm, clip, height = 0.305\textwidth]{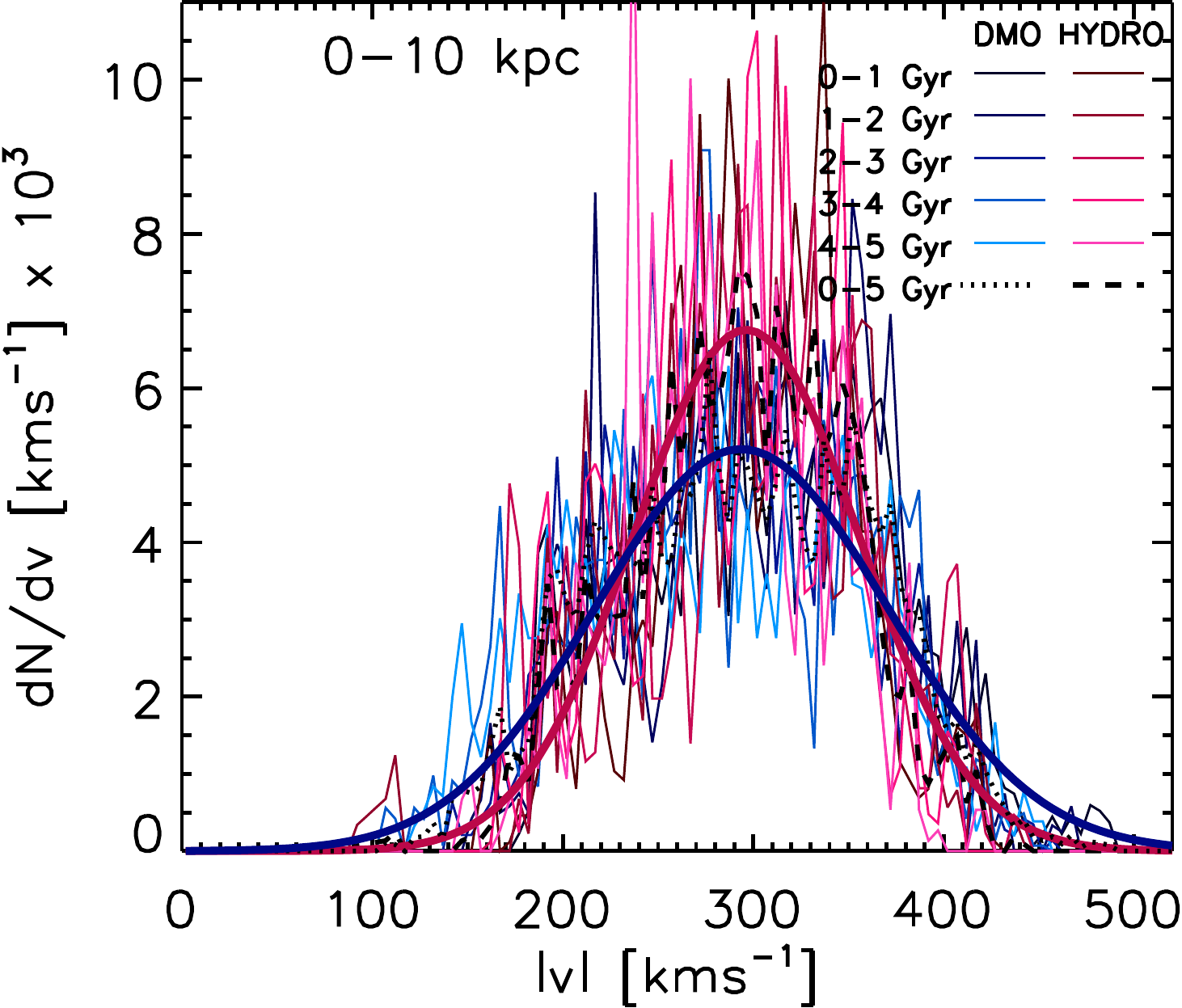} 
    \includegraphics*[trim = 0mm 0mm 0mm 0mm, clip, height = 0.305\textwidth]{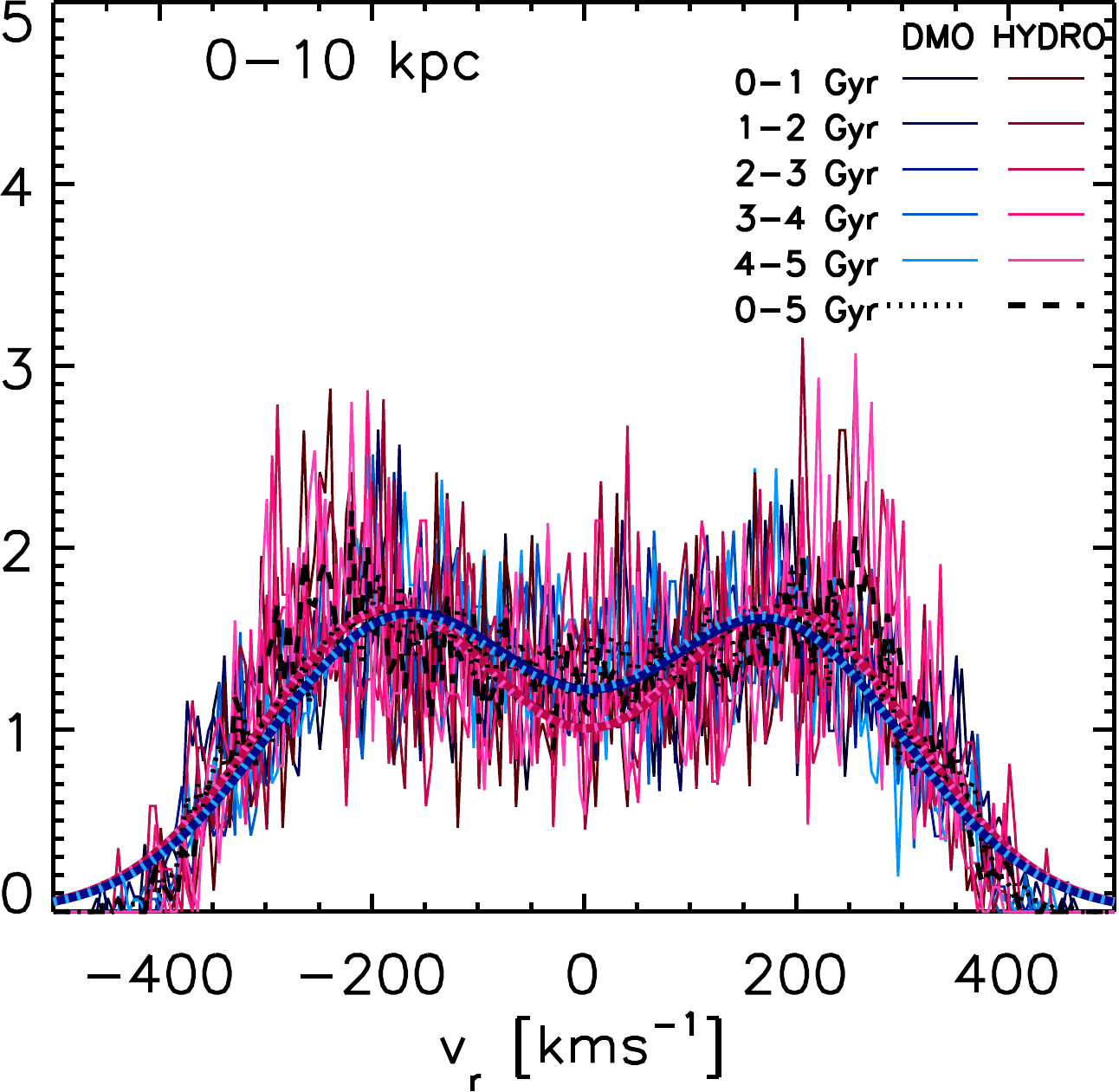} 
    \includegraphics*[trim = 0mm 0mm 0mm 0mm, clip, height = 0.305\textwidth]{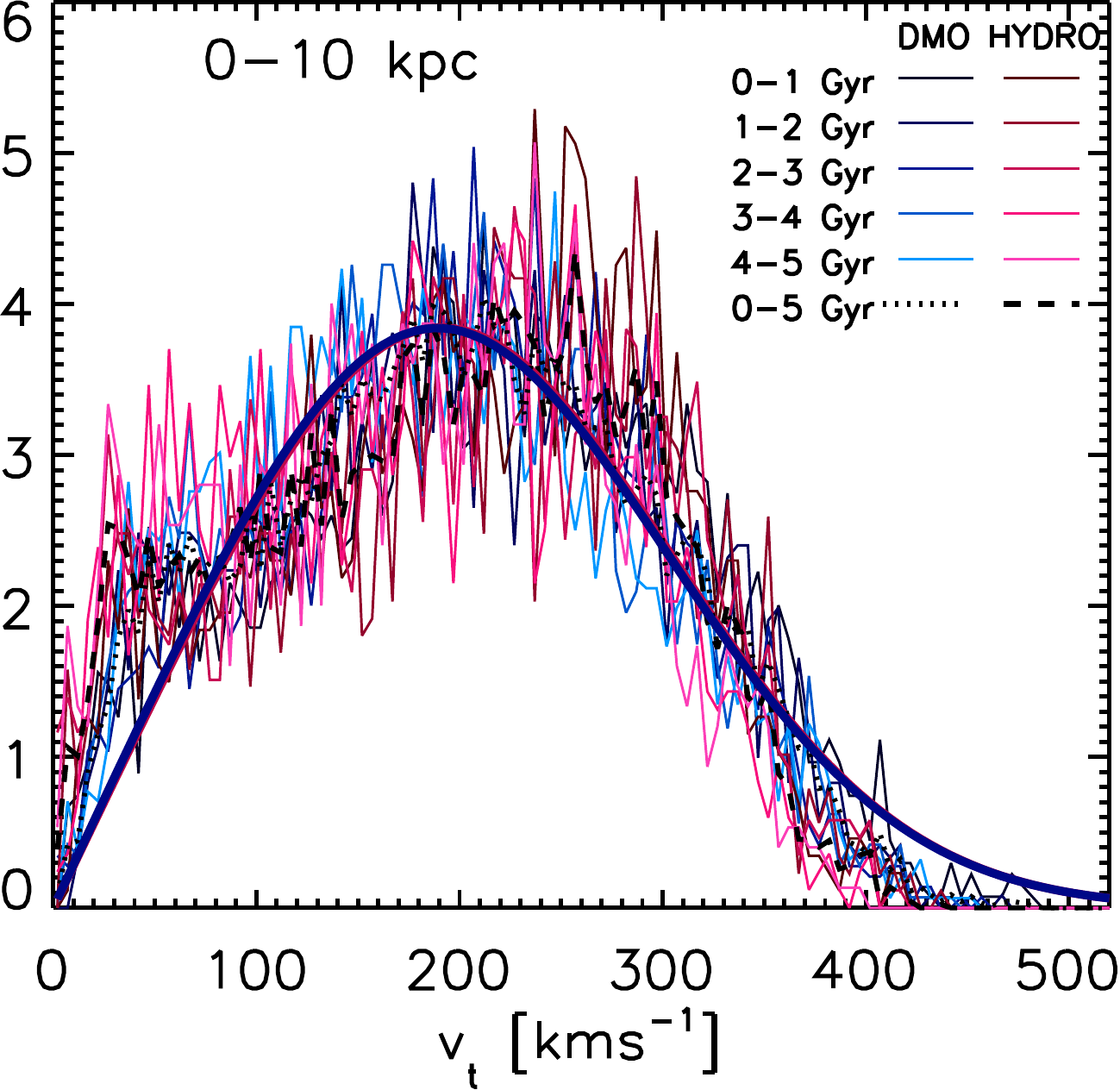} \\ 
    \includegraphics*[trim = 0mm 0mm 0mm 0mm, clip, height = 0.305\textwidth]{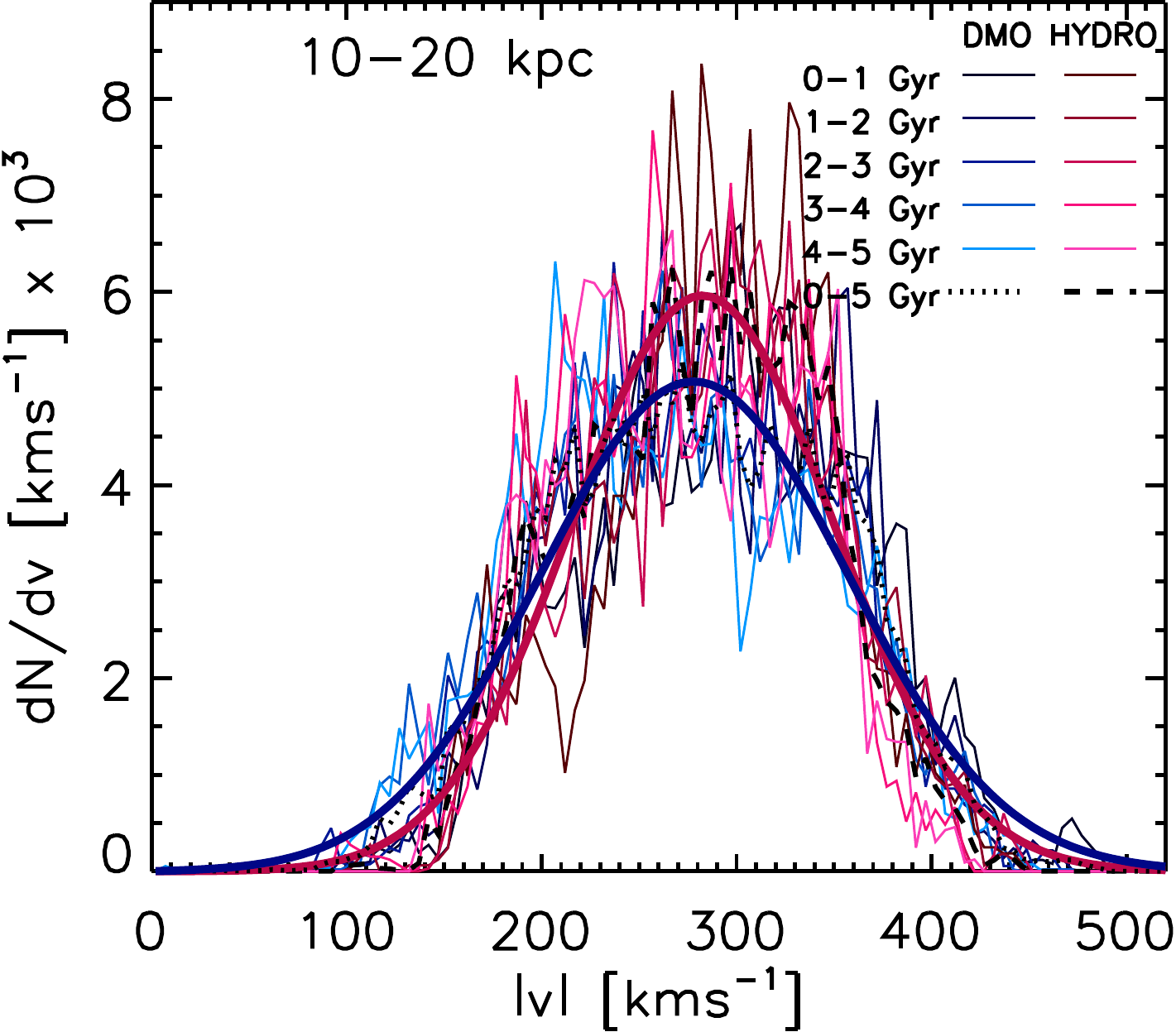} 
    \includegraphics*[trim = 0mm 0mm 0mm 0mm, clip, height = 0.305\textwidth]{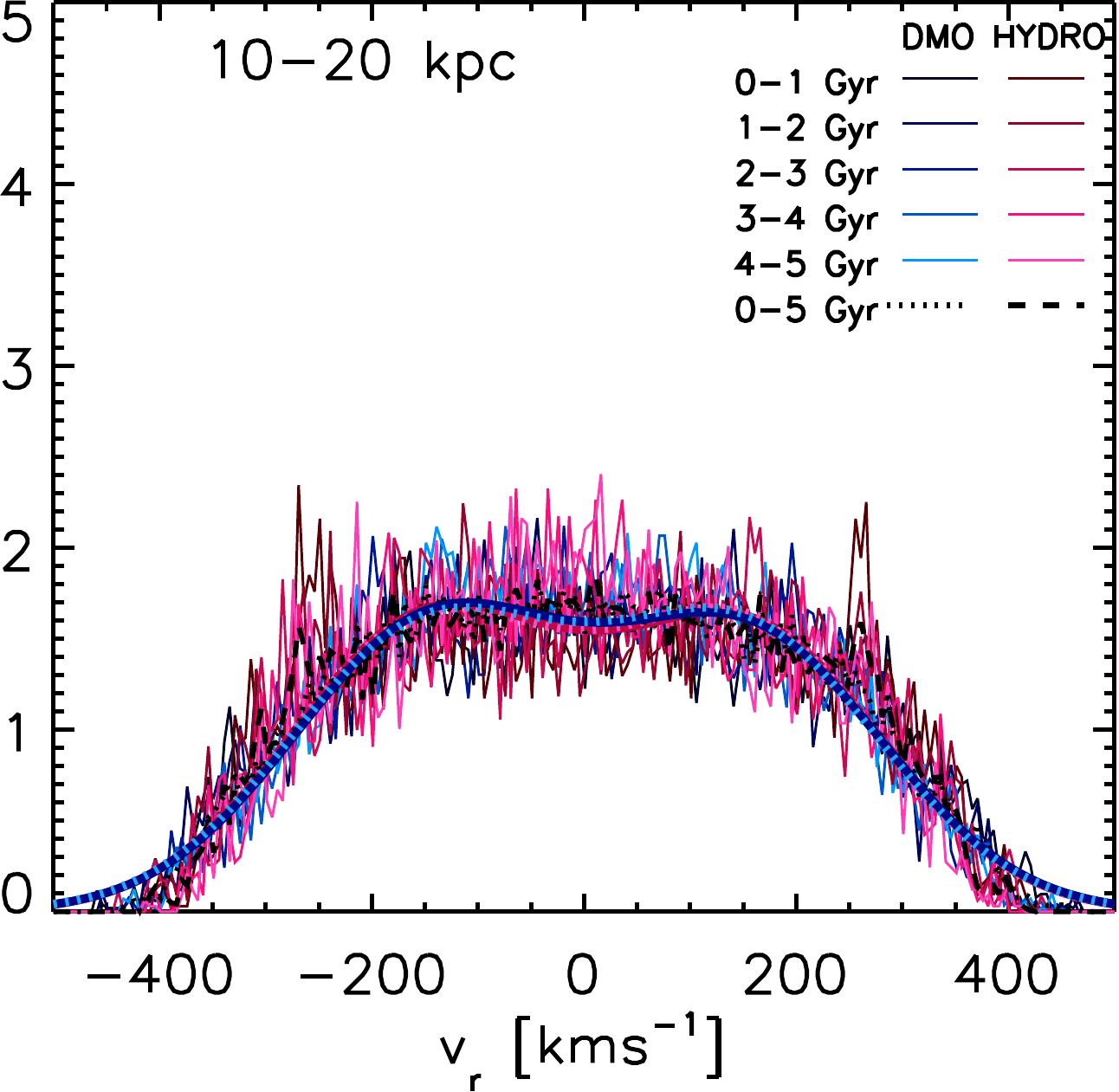} 
    \includegraphics*[trim = 0mm 0mm 0mm 0mm, clip, height = 0.305\textwidth]{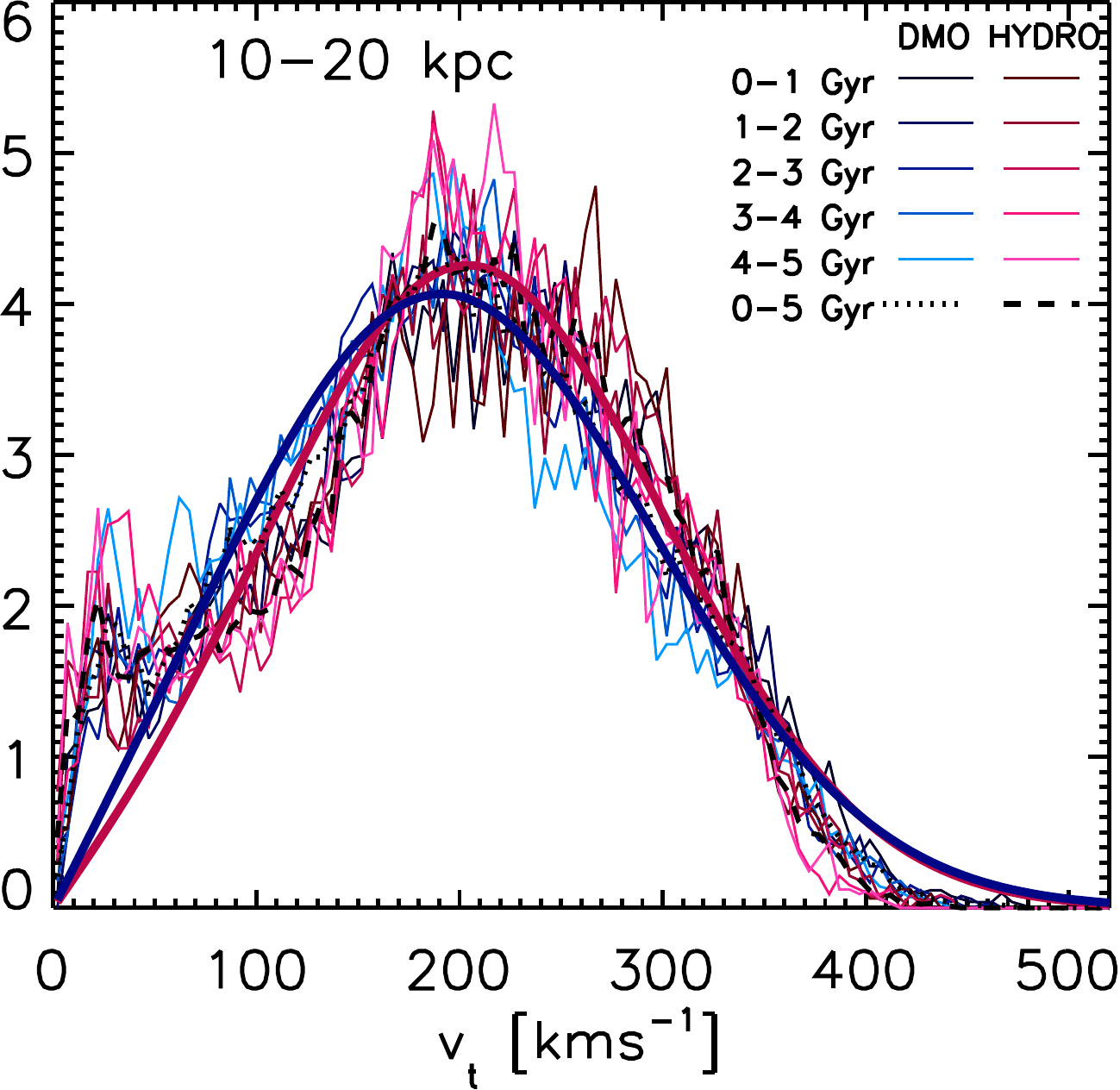} \\ 
    \includegraphics*[trim = 0mm 0mm 0mm 0mm, clip, height = 0.305\textwidth]{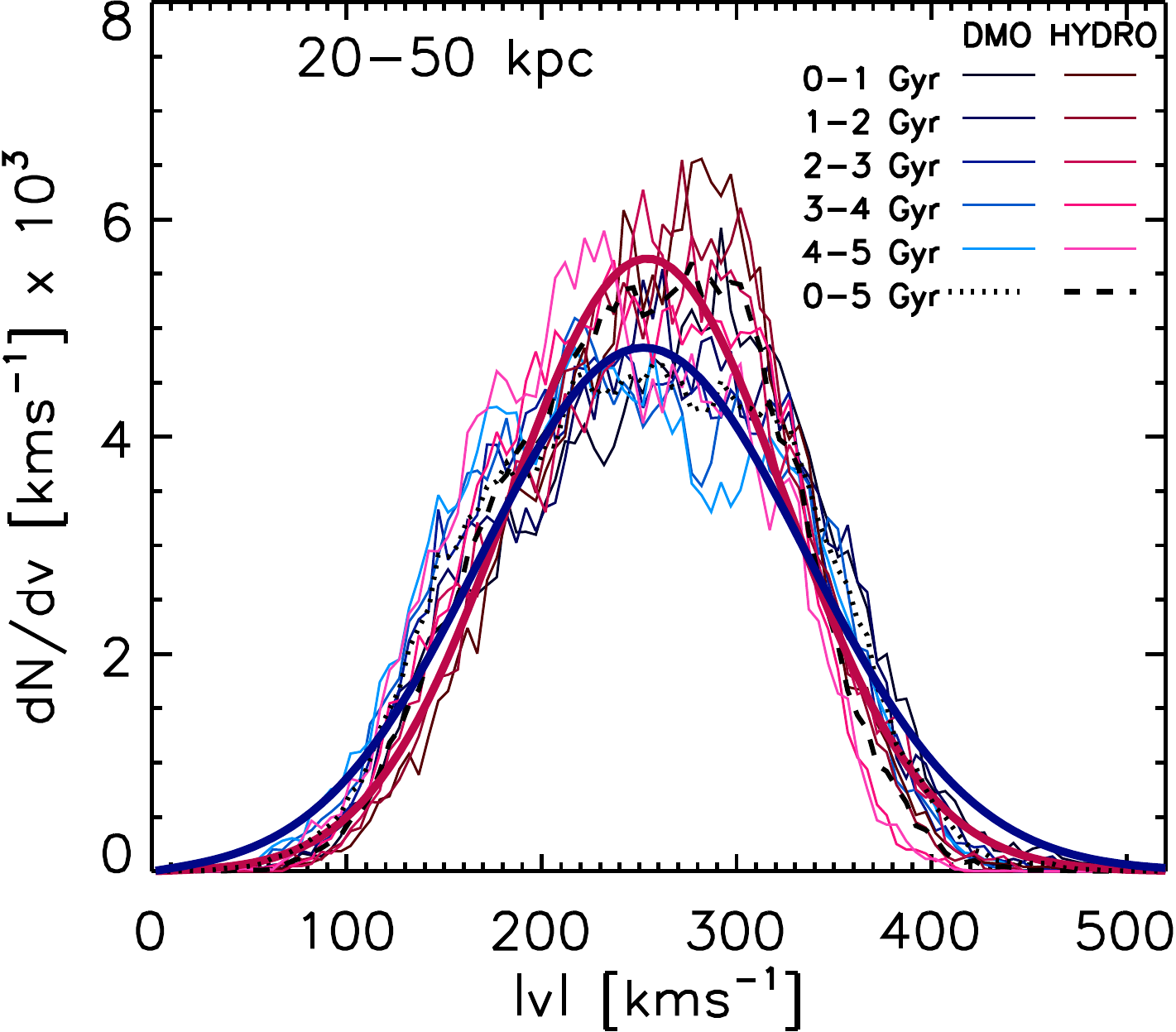} 
    \includegraphics*[trim = 0mm 0mm 0mm 0mm, clip, height = 0.305\textwidth]{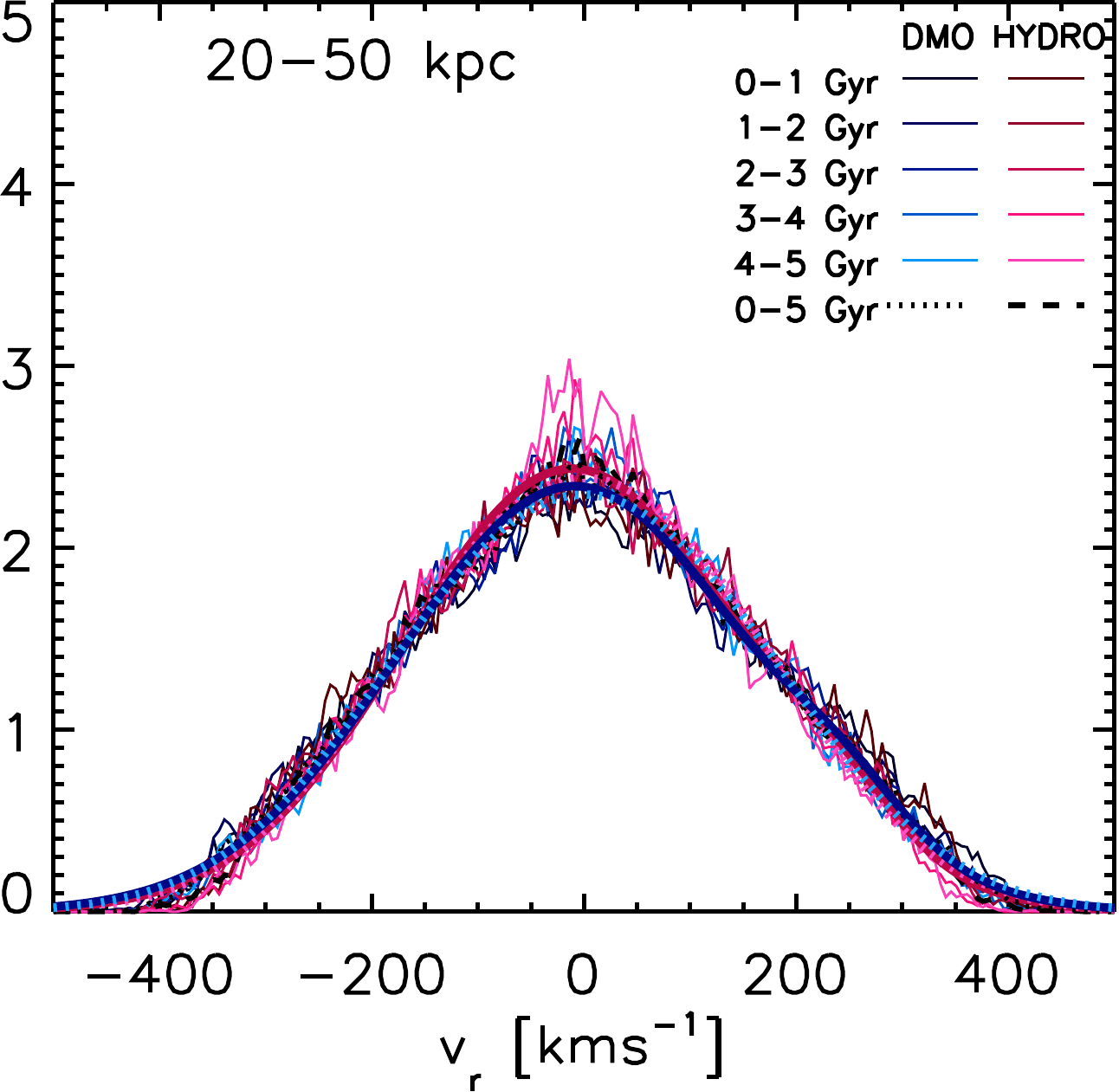} 
    \includegraphics*[trim = 0mm 0mm 0mm 0mm, clip, height = 0.305\textwidth]{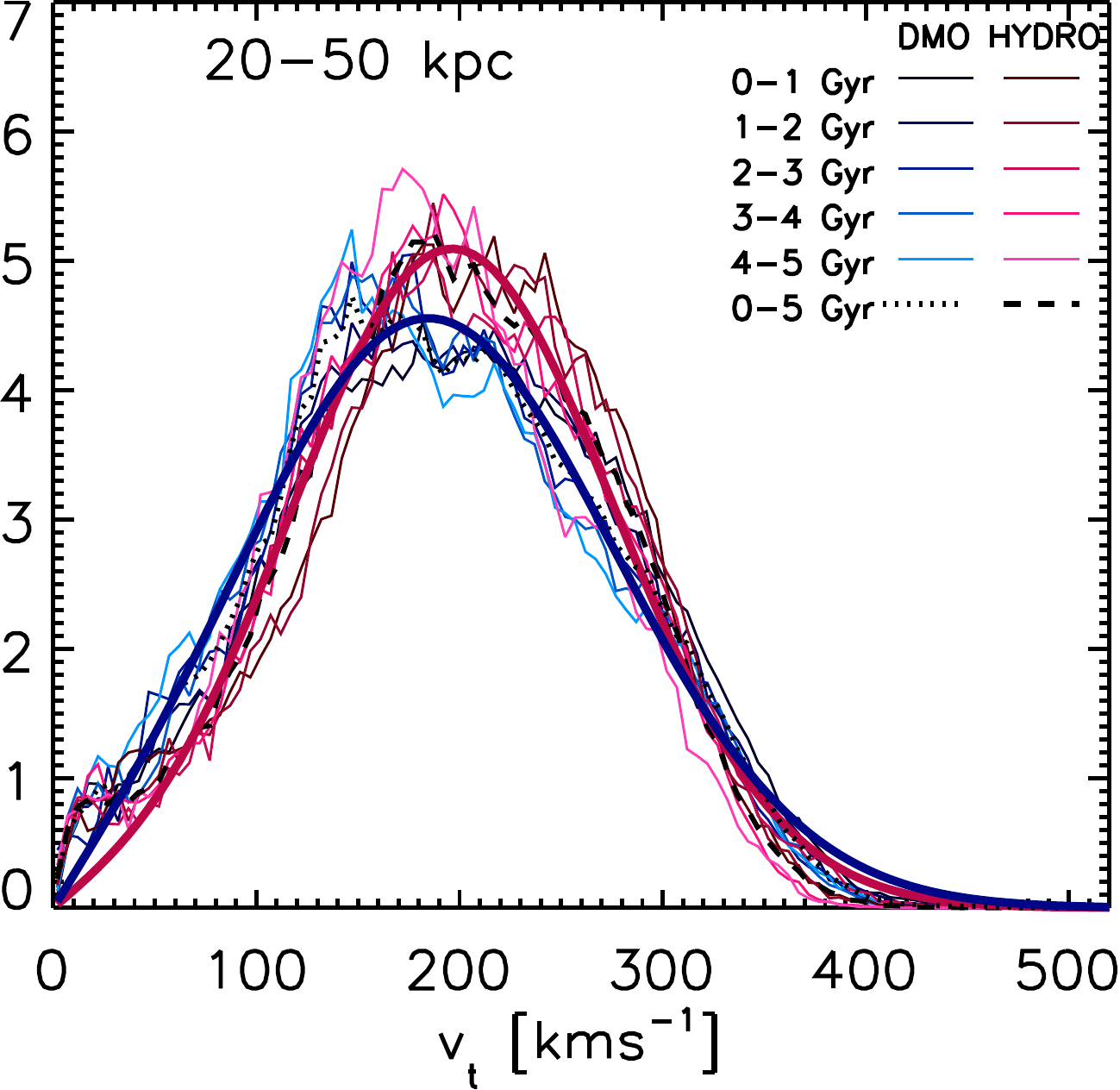} \\ 
    \includegraphics*[trim = 0mm 0mm 0mm 0mm, clip, height = 0.305\textwidth]{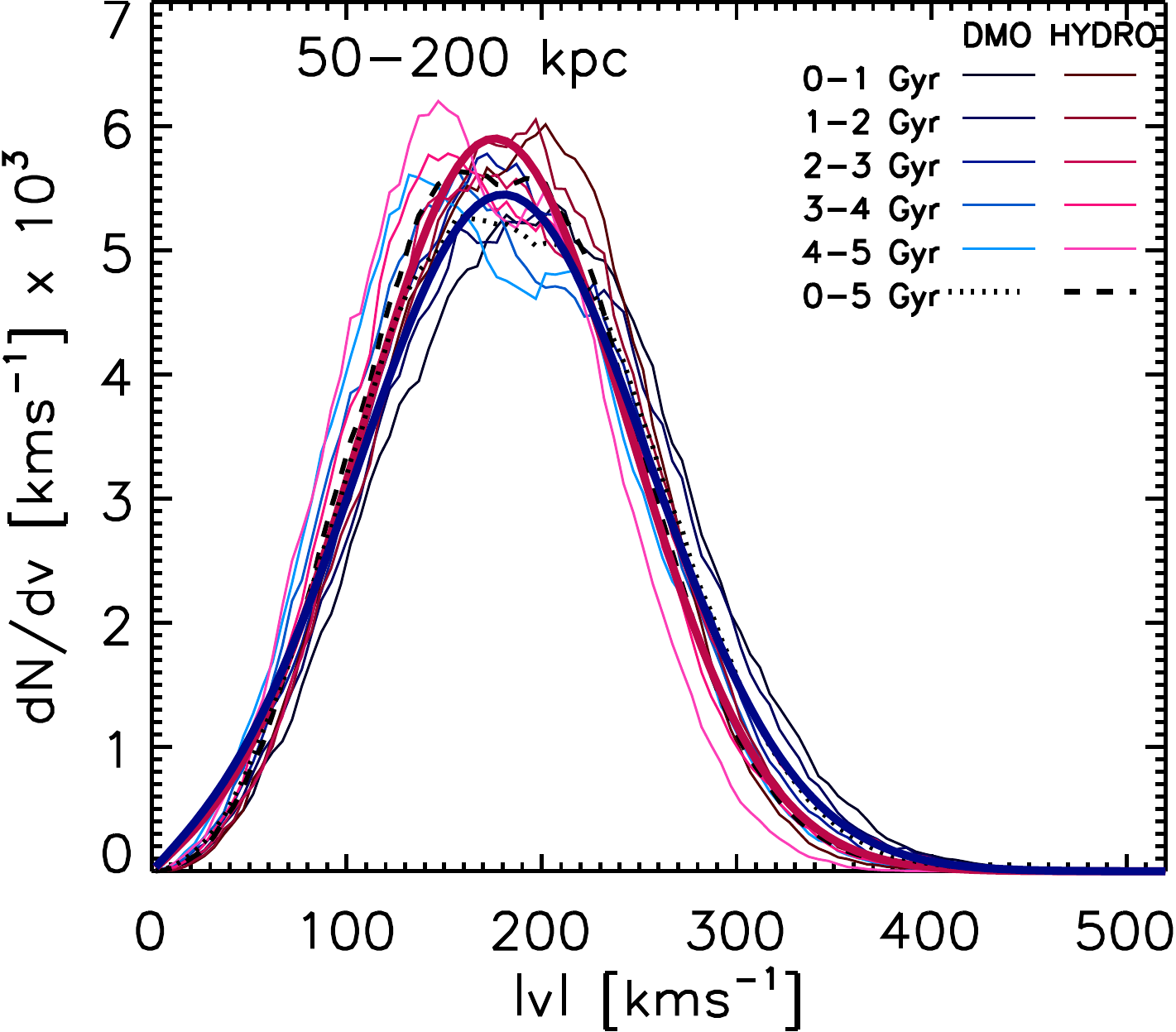} 
    \includegraphics*[trim = 0mm 0mm 0mm 0mm, clip, height = 0.305\textwidth]{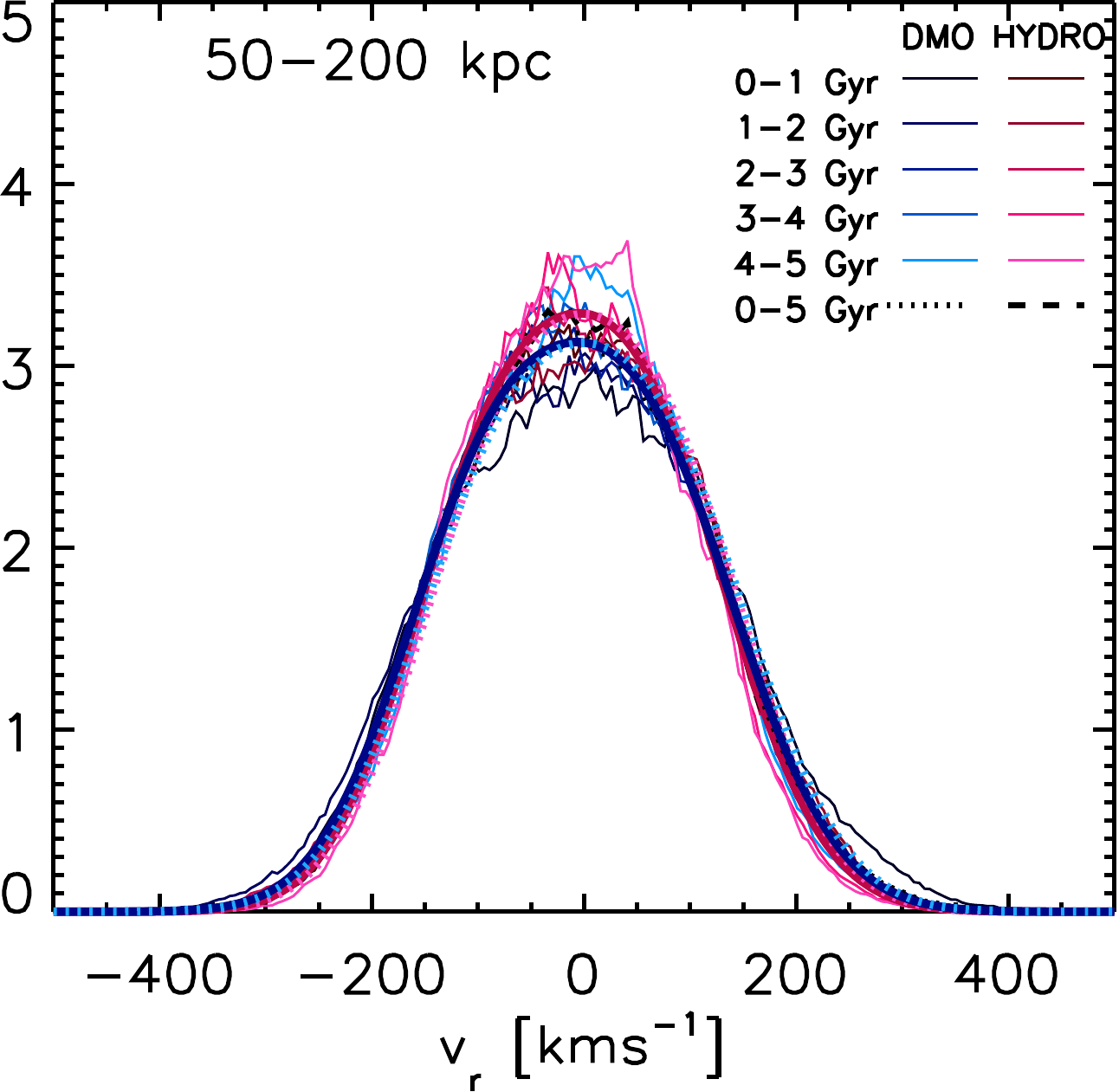} 
    \includegraphics*[trim = 0mm 0mm 0mm 0mm, clip, height = 0.305\textwidth]{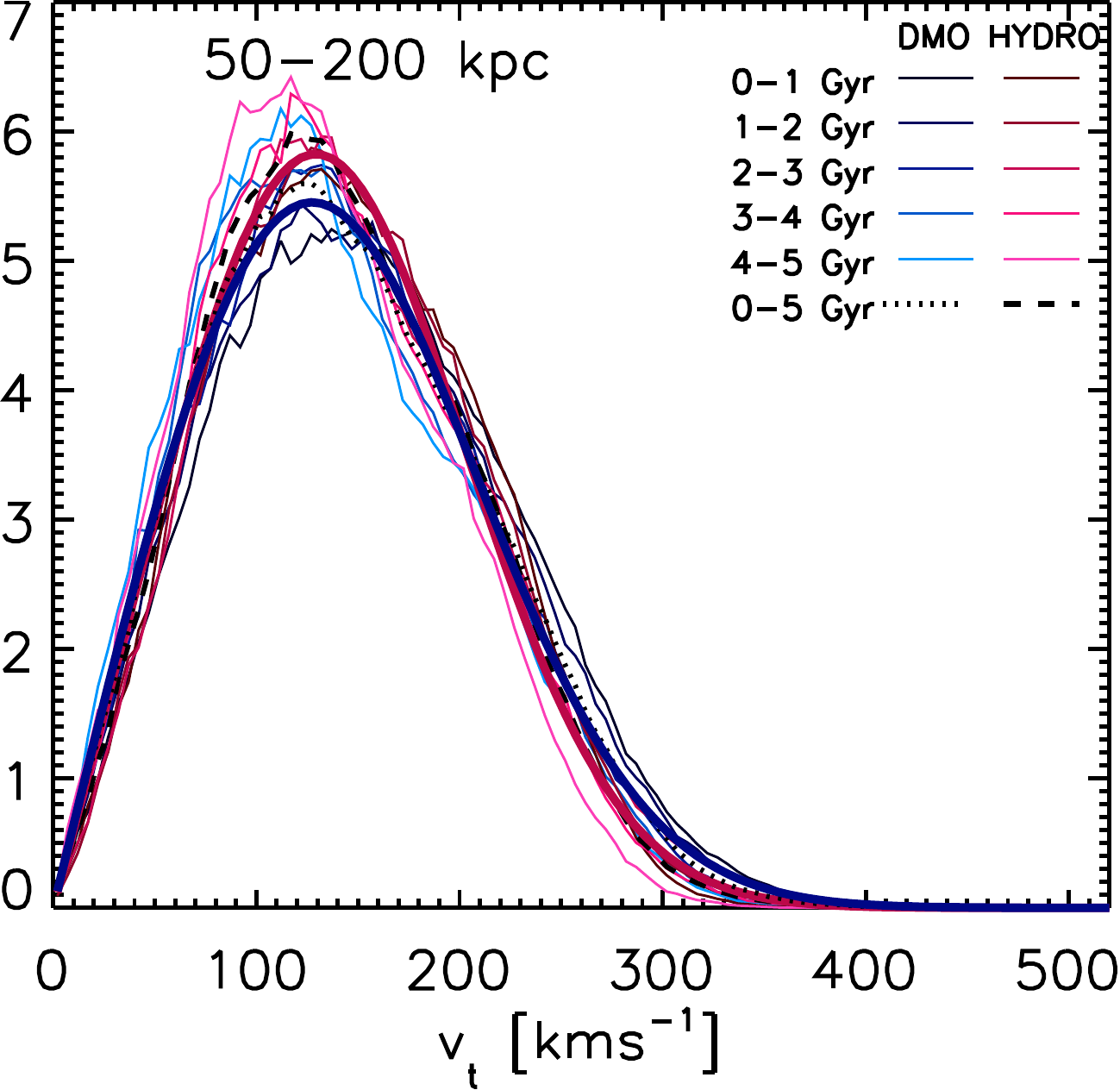} \\ 

  \end{center}
\vspace{-1.5mm}
  \caption{Probability density functions of subhalo velocities
    and velocity components in four radial shells for the same haloes
    and subhaloes shown in Figure~\ref{fig:radial-mf}. On all panels,
    thin lines show the results during different lookback time
    intervals; dotted and dashed black lines show the time-averaged
    results in the DMO and hydrodynamic simulations. Thick coloured
    lines show analytical fits, as described below. Left column: total
    velocity, $|v|$, with Rician fits (Eqn.~\ref{eqn:rice}, solid
    lines). Middle column: radial velocity, $v_r$ with fits to a
    general double-Gaussian with 5 free parameters
    (Eqn.~\ref{eqn:double-gaussian-5}, solid lines) and to a symmetric
    double-Gaussian with 2 free parameters
    (Eqn.~\ref{eqn:double-gaussian-2}, dotted lines). Right column:
    tangential velocity, $v_t$, with Rician fits (Eqn.~\ref{eqn:rice},
    solid lines). Note that the difference between individual time intervals is typically less than the
    scatter. A clear comparison between the time-averaged values and the
    fits is also shown in Figure~\ref{fig:velocities-appendix}.
    \label{fig:velocities}}
\end{figure*}

Independently of the velocity distributions for {\it particles}, it is
worth noting that the velocity profile of substructures may be
substantially different (see Section~\ref{sec:velocity-bias}). As we
discuss below, we also find the Maxwellian velocity distribution to be
merely a limiting case, only approximately true at large radii and low
velocities. It is strongly violated near the centre, where the
velocity anisotropy and the preferential disruption of low-velocity
subhaloes leads to highly non-Gaussian and non-Maxwellian VPDFs. 

\subsection{Total velocities} \label{sec:total-velocities} The
specific kinetic energy of a subhalo equals $\frac{1}{2} |v|^2 =
\frac{1}{2} (v_t^2 + v_r^2)$, where $v_r$ and $v_t$ are the radial and
tangential velocities. However, while there is considerable scatter in
the specific kinetic energies of different subhaloes at each radius,
$v_r$ and $v_t$ of a subhalo are clearly not independent.

Instead, near the halo centre, the radial and tangential velocities of
subhaloes have a bivariate velocity distribution, whose maximum occurs
at some distance $\mu > 0$ from the origin, with very few low-velocity
subhaloes. Instead of a Maxwellian, the probability density function
for $|v| = \sqrt{v_t^2 + v_r^2} $ may be described by a Rician
\citep{Rice-1945}:
\begin{equation}\label{eqn:rice}
  P\left(|v|\right) = \frac{|v|}{\sigma^2}
  e^{\frac{-\left(|v|^2+\mu^2\right)}{2\sigma^2}} I_0\left(\frac{|v| \mu}{\sigma^2}\right),
\end{equation}
where $I_0$ is the zeroth order modified Bessel function of the first
kind.

The discrepancy from a Maxwellian \eqnrefp{eqn:maxwell-3} is maximal
at small radii, where the mean specific kinetic energy is maximal, and
decreases as the mean specific kinetic energy decreases at large
radii. In the limit of $\mu = 0$, $v_t$ and $v_r$ become independent,
$I_0(0) = 1$, and the velocity distribution becomes Maxwellian.

Figure~\ref{fig:2d-velocities} demonstrates this behaviour in our
simulations. It shows the 2D velocity distribution function in the
$(v_r, v_t)$-plane measured over 5 Gyr in four radial bins, increasing
in radius from top left to bottom right. At $r<10$~ kpc, $\mu$ exceeds
the scatter, $\sigma$, and reflecting the near-absence of slow-moving
subhaloes with low values of both $v_r$ and $v_t$. At larger radii,
the average kinetic energy decreases and becomes comparable to the
scatter. Here, the velocity components $v_r$ and $v_t$ become more
independent, except for extreme values, where the orbital speed is
limited by the escape speed, $\sim 350-400$~kms$^{-1}$ at 50 kpc.

In the left column of Figure~\ref{fig:velocities}, we show the PDFs of
$|v|$ in the DMO and hydrodynamic simulations in the same four radial
bins shown in Figure~\ref{fig:2d-velocities}, together with fits to
the Rician PDFs \eqnrefp{eqn:rice}. We list the values of $\mu$ and
$\sigma$ in Table~\ref{tab:fits-vel}. As expected, we find $\mu$ to
increase towards the centre, from $165$ and $162$~kms$^{-1}$ at
$50-200$~kpc, to $284$ and $290$~kms$^{-1}$ at $<10$~kpc, for the DMO
and hydrodynamic simulations, respectively. The scatter $\sigma$ is
less dependent on radius, but it is $\sim 10-20\%$ lower in the
hydrodynamic simulations compared to the DMO simulations.

For comparison, in Appendix~\ref{appendix:maxwell}, we contrast Rician
and Maxwellian fits to the data shown in
Figure~\ref{fig:velocities}. We find that the latter are very poor
fits near the halo centre, but that the distributions become more
similar at the largest radii, as expected.

\subsection{Radial velocities}
As noted in Section~\ref{sec:energies}, the subhalo population near
the centre is dominated by subhaloes on radial orbits with long
orbital periods. Consequently, for small radii, the radial velocity
distribution of subhaloes is described by a double-peaked Gaussian of
the general form:
\begin{equation} \label{eqn:double-gaussian-5}
P(v_r)= \frac{a}{\sigma_1 \sqrt{2 \pi}}
e^{-\frac{\left(v_r-\mu_1\right)^2}{2\sigma_1^2}} + 
\frac{1-a}{\sigma_2 \sqrt{2 \pi}}
e^{-\frac{\left(v_r-\mu_2\right)^2}{2\sigma_2^2}} 
\end{equation}
where the 5 free parameters $\mu_1$, $\mu_2$, $\sigma_1$, $\sigma_2$,
and $a$ represent the mean and standard deviations of the first and
second Gaussian components, as well as the relative contribution of
the two components. A double Gaussian models the subhalo population at
each radius as a sum of an ``incoming'' and an ``outgoing''
population. In the full, five-parameter fit, we typically find a small
negative mean radial velocity, indicating either more incoming than
outgoing satellites as a result of recent infall and disruption, or
satellites losing orbital energy as a result of dynamical
friction. However, if orbital energy is exactly conserved,
\eqnref{eqn:double-gaussian-5} can be simplified to a symmetric
double-Gaussian, where we set $\sigma = \sigma_1 = \sigma_2$, $\mu =
\mu_1 = -\mu_2$, and $a=(1-a)=1/2$:

\begin{equation} \label{eqn:double-gaussian-2}
P(v_r)= \frac{1}{2\sigma \sqrt{2 \pi}}
e^{-\frac{\left(v_r-\mu\right)^2}{2\sigma^2}}+
\frac{1}{2\sigma \sqrt{2 \pi}}
e^{-\frac{\left(v_r+\mu\right)^2}{2\sigma^2}}
\end{equation} 

The middle column of Figure~\ref{fig:velocities} shows fits to our
simulation data using both \eqnref{eqn:double-gaussian-5} and
\eqnref{eqn:double-gaussian-2}, and we list the best-fit values for
$\mu$ and $\sigma$ for both the DMO and hydrodynamic simulations in
Table~\ref{tab:fits-vel}. It can be seen that, at large radii, $\sigma
> \mu$, resembling a (broadened) peak centred at $v_r = 0$,
approaching a single Gaussian in the limit $\mu = 0$. At smaller
radii, $\mu$ increases and the radial velocity distribution becomes
increasingly broad and, for $r< 20$ kpc, clearly
bimodal. Appendix~\ref{appendix:maxwell} compares the bimodal fit to
one with a single Gaussian and shows the convergence at large radii.

\subsection{Tangential velocities}
In principle, there are two orthogonal velocity components, $v_\theta$
and $v_\phi$, required in addition to the radial velocity, $v_r$, to
fully describe the velocity of a subhalo. Defining the tangential
velocity, $v_t = \sqrt{v_\theta^2 + v_\phi^2}$, if its two components
are independent Gaussian random variables with zero mean, the PDF for
$v_t$ may be expected to be a 2D-Maxwellian
\eqnrefp{eqn:maxwell-2}. At large radii, where the orbital anisotropy
is close to zero, we find a relatively good agreement, except for an
overprediction at the high-velocity tail, corresponding to subhaloes
above the escape velocity. However, as the anisotropy increases
towards the centre, the 2D-Maxwellian shape overpredicts the skewness
of the measured $v_t$ distribution. As shown in the right column of
Figure~\ref{fig:velocities}, we find that the tangential velocities in
each radial bin are quite well fit by Rician distributions
\eqnrefp{eqn:rice}. Appendix~\ref{appendix:maxwell} compares the
Rician fits to those of a 2D-Maxwellian, which severely overpredict
either the high- or low-velocity tails of the distributions at small
radii.

\section{Implications for substructure detection} \label{sec:observables}

\subsection{Substructure detection via lensing}
The detection of substructure around individual galaxies by strong
gravitational lensing depends not only on the mass of the
substructure, but in addition on its projected distance from the
Einstein radius.  Recently detections of dark substructures have been
made around massive elliptical galaxies that are typically embedded in
dark matter haloes, with total masses of $M\sim10^{13} \ M_{\odot}$
and typical Einstein radii of $r_{\rm E}\sim10 \ \rm kpc$
(e.g. \citealt{Vegetti-2012,Nierenberg-2014,Hezaveh-2016}). These
lensing haloes are an order of magnitude more massive than the
Milky-Way like host haloes we have studied in this paper.

The substructure abundance clearly depends on the host halo mass and
concentration. However, we believe that, when scaled by $r/r_{200}$,
the baryonic effects that suppress substructures found in the \apostle
simulations are likely to be a reasonable approximation to the effects
in host haloes of slightly larger mass, which are expected to have
slightly lower halo concentrations and stellar mass fractions
(e.g. \citealt{Moster-2010,Dutton-2014}).  Baryonic effects should not
be a major obstacle for detecting $\Lambda$CDM substructures through
lensing, or for ruling out $\Lambda$CDM in case of a significant
shortfall of detections relative to DMO predictions, at least in haloes
with similar central stellar densities and similar amounts of
adiabatic contraction.

\subsection{Substructure detection via stream gaps}
In order to detect dark matter substructures through the perturbations
they induce on globular cluster streams, both the mass function and
the velocity distribution of substructures are important, as the
interaction strength is proportional to the mass, and inversely
proportional to the relative velocity.

The Milky Way's two most prominent globular cluster streams are Pal-5
and GD-1, both discovered in the SDSS. Pal-5 \citep{Odenkirchen-2001}
extends over more than 20 degrees, with apogalactic and perigalactic
distances of $18.67$~kpc and $7.97$~kpc \citep{Kuepper-2015}, while
GD-1 \citep{Grillmair-2006} extends over 63 degrees, with apogalactic
and perigalactic distances of $28.75\pm2$~kpc and $14.43 \pm 0.5$~kpc
\citep{Willet-2009}. For the observable parts of the Pal-5 stream,
\cite{Kuepper-2015} estimate an age of $3.4^{+0.5}_ {-0.3}$ Gyr, while
\cite{Carlberg-2013} estimate a dynamical age of $2.3-4.6$ Gyr for
GD-1.

We expect the abundance of substructures inside the orbit of Pal-5 and
GD-1 to be reduced by $\sim 45-50\%$ relative to that inferred from
DMO simulations due to baryonic effects, with a slightly larger
reduction for Pal-5, due to its smaller mean galactocentric distance.

Compared to earlier work, we find two additional effects that will
need to be taken into account in future work. \cite{Erkal-2015b}
assume a prior for the substructure mass that is uniform in log$(M)$,
or a mass function with a slope of $-1$, we find steeper power laws,
with slopes between $-1.86$ and $-1.91$ in both the DMO and
hydrodynamic simulations.

More importantly, it has so far been assumed that the velocity PDF of
substructures is Maxwellian, with a mean velocity equal to
$v_{circ}/\sqrt{3}$ = 97 kms$^{-1}$ in the case of
\cite{Erkal-2015b}. However, as discussed in
Section~\ref{sec:total-velocities}, we find that this is a poor fit to
the subhalo velocities near the centre, where subhaloes are biased
towards much higher velocities, and where the Rician PDF contains far
fewer low-velocity subhaloes than a Maxwellian distribution fit to the
same data. Comparing the Maxwellian and Rician fits to the total
velocity within 10 kpc, shown in Appendix~\ref{appendix:maxwell}, we
find that the Maxwellians vastly overpredict the number of subhaloes
with low velocities, even considering that our Maxwellian fits have
mean velocities that are nearly twice as high as those assumed
previously. Given that low-velocity perturbers cause larger gaps and
are easier to detect, using accurate velocity priors is important for
the characterisation of perturbers, and any inferences derived from
it. An additional effect, particularly relevant for Pal-5, is the
potential confusion of perturbations by substructures with those
induced by giant molecular clouds. These are, of course, relatively
slow moving, and \cite{Amorisco-2016} point out that they induce
perturbations similar to those caused by dark matter subhaloes.

\subsection{Substructure detection via disk heating}

Similar to the perturbation of streams, perturbations of the Galactic
disk component by substructures are not only sensitive to the
substructure mass, but also to their impact velocity. Impacts of dark
substructures will heat the disk increasing the vertical velocity
dispersion, with the most pronounced effects typically seen in the
outer parts of the disk where the lower surface density results in a
correspondingly lower restoring force (e.g. \citealt{BT-2008}).  Based
on our analysis we find a reduction in the abundance of substructures
within 10~kpc of the halo centre by up to a factor of two in
hydrodynamic simulations as compared to DMO simulations.  As for
stream gaps, the velocity of perturbers determines their interaction
strength, and near the centre, we find a much lower number of
low-velocity substructures compared to the commonly assumed Maxwellian
velocity distribution function.

However, as a caveat it should be noted that the disk may not be such
a clean tracer of dark substructures, as other massive perturbers,
such as molecular clouds (e.g. \citealt{Lacey-1984,Hanninen-2002}),
and impacts by globular clusters \citep{VandePutte-2009} also result
in disk heating. In addition internal mechanisms, such as the growth
of a central bar component and spiral features in the Galactic disk
will also heat the disk (e.g. \citealt{Sellwood-2014,Grand-2016}).
Finally, if the disk itself is a major cause for the depletion of
substructures in the inner halo \citep{Donghia-2010} the substructures
that interact with the disk are likely to be a biased subset of the
entire substructure population. On the one hand, they are likely to be
the most strongly stripped after they have interacted with the
disk. On the other hand, if we measure the depletion of substructures
after one or more passages, we may overestimate the depletion factor
of the substructures at the time they interact with the disk.

\section{Summary}\label{sec:conclusion}
We have studied how baryonic effects can change the abundance of
substructures in the mass range M$=10^{6.5}-10^{8.5}\Ms$ inside Milky
Way mass haloes of M$_{200}\sim10^{12}\Ms$ over a lookback time of up
to 5 Gyr. We find that the abundance of subhaloes, independently of
subhalo mass, is reduced in hydrodynamic simulations of the same host
halo compared to their DMO counterpart. The depletion increases
towards the halo centre: at $r > 50$~kpc, the number of subhaloes in
the hydrodynamic simulations is above $3/4$ of that in the DMO
counterparts, dropping to $\sim 1/2$ at $r<10$~kpc. While baryonic
effects of this magnitude clearly need to be taken into account for
accurate predictions, they do not impede the detection of dark
substructures through stream gaps, disk heating, or lensing.

Purely in terms of substructure abundance, \cite{Donghia-2010} found a
stronger reduction, with the subhalo number reduced to $1/3$relative
to the original DMO simulation at $10^{7}\Ms$ by the effects of the
stellar disk alone. This is due in part to the much higher disk mass
($10\%$ of M$_{96}$, or $\sim 14\%$ of M$_{200}$) that they
assumed. They also reported a significant subhalo mass dependence,
with $1/2$ of subhaloes remaining at $10^9\Ms$, while we find a nearly
constant factor. One possible explanation for this may be numerical
resolution: while we limit our study to subhaloes with more than 50
particles, the lower resolution in \cite{Donghia-2010} means that
$10^7\Ms$ subhaloes only contain $\sim 20$ particles.

The central galaxies in our four simulations have stellar masses in
the range $(1.2-2.8)\times 10^{10}\Ms$, somewhat below the range of
$\sim5\ \pm 1 \times 10^{10} \Ms $ commonly assumed for the Milky Way
\citep[e.g.][]{Flynn-2006, Bovy-2013}. For a greater stellar mass, we
would expect some of the baryonic effects to increase, although we
note that the decline in subhalo abundance relative to DMO simulations
is due not only to the presence of the stellar component, but also to
the contraction of the halo itself, as well as to the almost complete
loss of baryons from low-mass haloes by reionisation and ram-pressure
stripping.

The processes that lead to a relative underdensity of subhaloes near
the centre also give rise to a positive velocity bias and rising
anisotropy of subhalo orbits, two effects we find enhanced in the
hydrodynamic simulation. Furthermore, we find that the velocity
distribution of substructures near the halo centre cannot be assumed
to be Maxwellian. The preferential disruption of strongly bound
subhaloes leads to velocity distributions with far fewer low-velocity
subhaloes than commonly assumed, and while the few surviving
low-velocity subhaloes near the halo centre have more circular orbits,
the overall subhalo population near the centre is dominated by
high-velocity subhaloes on highly radial orbits.  This impacts both
the total number and the strength of detectable substructure
interactions.

\section*{Acknowledgements}
T.~S., C.~S.~F. and S.~D.~M.~W. thank the organisers and fellow
participants of the Lorentz Centre Workshop ``Dark Matter on the
Smallest Scales'' for discussions that inspired this paper. T.~S.,
P.~P. and P.~H.~J.  acknowledge support of the Academy of Finland
grant 1274931. This work was supported by the Science and Technology
Facilities Council [grant number ST/F001166/1 and RF040218], the
European Research Council under the European Union's Seventh Framework
Programme (FP7/2007-2013) / ERC Grant agreement 278594
'GasAroundGalaxies', the National Science Foundation under Grant
No. PHYS-1066293. C.~S.~F. acknowledges ERC Advanced Grant 267291
'COSMIWAY'. This work used the DiRAC Data Centric system at Durham
University, operated by the Institute for Computational Cosmology on
behalf of the STFC DiRAC HPC Facility (www.dirac.ac.uk), facilities
hosted by the CSC-IT Center for Science in Espoo, Finland, which are
financed by the Finnish ministry of education, and resources provided
by WestGrid (www.westgrid.ca) and Compute Canada / Calcul Canada
(www.computecanada.ca). The DiRAC system is funded by BIS National
E-infrastructure capital grant ST/K00042X/1, STFC capital grant
ST/H008519/1, STFC DiRAC Operations grant ST/K003267/1, and Durham
University. DiRAC is part of the National E-Infrastructure. We have
used {\sc SciPy} \citep{scipy} and {\sc NumPy} \citep{numpy} and thank
their developers for making them freely available.

\bibliographystyle{mnras} \bibliography{paper}

\begin{thebibliography}{}
\makeatletter
\relax
\def\mn@urlcharsother{\let\do\@makeother \do\$\do\&\do\#\do\^\do\_\do\%\do\~}
\def\mn@doi{\begingroup\mn@urlcharsother \@ifnextchar [ {\mn@doi@}
  {\mn@doi@[]}}
\def\mn@doi@[#1]#2{\def\@tempa{#1}\ifx\@tempa\@empty \href
  {http://dx.doi.org/#2} {doi:#2}\else \href {http://dx.doi.org/#2} {#1}\fi
  \endgroup}
\def\mn@eprint#1#2{\mn@eprint@#1:#2::\@nil}
\def\mn@eprint@arXiv#1{\href {http://arxiv.org/abs/#1} {{\tt arXiv:#1}}}
\def\mn@eprint@dblp#1{\href {http://dblp.uni-trier.de/rec/bibtex/#1.xml}
  {dblp:#1}}
\def\mn@eprint@#1:#2:#3:#4\@nil{\def\@tempa {#1}\def\@tempb {#2}\def\@tempc
  {#3}\ifx \@tempc \@empty \let \@tempc \@tempb \let \@tempb \@tempa \fi \ifx
  \@tempb \@empty \def\@tempb {arXiv}\fi \@ifundefined
  {mn@eprint@\@tempb}{\@tempb:\@tempc}{\expandafter \expandafter \csname
  mn@eprint@\@tempb\endcsname \expandafter{\@tempc}}}

\bibitem[\protect\citeauthoryear{{Amorisco}, {G{\`o}mez}, {Vegetti}  \&
  {White}}{{Amorisco} et~al.}{2016}]{Amorisco-2016}
{Amorisco} N.~C.,  {G{\`o}mez} F.~A.,  {Vegetti} S.,   {White} S.~D.~M.,  2016,
  preprint, \href {http://adsabs.harvard.edu/abs/2016arXiv160602715A} {}
  (\mn@eprint {arXiv} {1606.02715})

\bibitem[\protect\citeauthoryear{{Arraki}, {Klypin}, {More}  \&
  {Trujillo-Gomez}}{{Arraki} et~al.}{2014}]{Arraki-2014}
{Arraki} K.~S.,  {Klypin} A.,  {More} S.,   {Trujillo-Gomez} S.,  2014, \mn@doi
  [\mnras] {10.1093/mnras/stt2279}, \href
  {http://adsabs.harvard.edu/abs/2014MNRAS.438.1466A} {438, 1466}

\bibitem[\protect\citeauthoryear{{Avila-Reese}, {Col{\'{\i}}n}, {Valenzuela},
  {D'Onghia}  \& {Firmani}}{{Avila-Reese} et~al.}{2001}]{Avila-Reese-2001}
{Avila-Reese} V.,  {Col{\'{\i}}n} P.,  {Valenzuela} O.,  {D'Onghia} E.,
  {Firmani} C.,  2001, \mn@doi [\apj] {10.1086/322411}, \href
  {http://adsabs.harvard.edu/abs/2001ApJ...559..516A} {559, 516}

\bibitem[\protect\citeauthoryear{{Baumgardt}, {Hut}  \& {Heggie}}{{Baumgardt}
  et~al.}{2002}]{Baumgardt-2002}
{Baumgardt} H.,  {Hut} P.,   {Heggie} D.~C.,  2002, \mn@doi [\mnras]
  {10.1046/j.1365-8711.2002.05736.x}, \href
  {http://adsabs.harvard.edu/abs/2002MNRAS.336.1069B} {336, 1069}

\bibitem[\protect\citeauthoryear{{Benson}, {Lacey}, {Baugh}, {Cole}  \&
  {Frenk}}{{Benson} et~al.}{2002}]{Benson-2002}
{Benson} A.~J.,  {Lacey} C.~G.,  {Baugh} C.~M.,  {Cole} S.,   {Frenk} C.~S.,
  2002, \mn@doi [\mnras] {10.1046/j.1365-8711.2002.05387.x}, \href
  {http://adsabs.harvard.edu/abs/2002MNRAS.333..156B} {333, 156}

\bibitem[\protect\citeauthoryear{{Benson}, {Lacey}, {Frenk}, {Baugh}  \&
  {Cole}}{{Benson} et~al.}{2004}]{Benson-2004}
{Benson} A.~J.,  {Lacey} C.~G.,  {Frenk} C.~S.,  {Baugh} C.~M.,   {Cole} S.,
  2004, \mn@doi [\mnras] {10.1111/j.1365-2966.2004.07870.x}, \href
  {http://adsabs.harvard.edu/abs/2004MNRAS.351.1215B} {351, 1215}

\bibitem[\protect\citeauthoryear{{Binney} \& {Tremaine}}{{Binney} \&
  {Tremaine}}{2008}]{BT-2008}
{Binney} J.,  {Tremaine} S.,  2008, {Galactic Dynamics: Second Edition}.
Princeton University Press

\bibitem[\protect\citeauthoryear{{Booth} \& {Schaye}}{{Booth} \&
  {Schaye}}{2009}]{Booth-2009}
{Booth} C.~M.,  {Schaye} J.,  2009, \mn@doi [\mnras]
  {10.1111/j.1365-2966.2009.15043.x}, \href
  {http://adsabs.harvard.edu/abs/2009MNRAS.398...53B} {398, 53}

\bibitem[\protect\citeauthoryear{{Bose}, {Hellwing}, {Frenk}, {Jenkins},
  {Lovell}, {Helly}  \& {Li}}{{Bose} et~al.}{2016}]{Bose-2016}
{Bose} S.,  {Hellwing} W.~A.,  {Frenk} C.~S.,  {Jenkins} A.,  {Lovell} M.~R.,
  {Helly} J.~C.,   {Li} B.,  2016, \mn@doi [\mnras] {10.1093/mnras/stv2294},
  \href {http://adsabs.harvard.edu/abs/2016MNRAS.455..318B} {455, 318}

\bibitem[\protect\citeauthoryear{{Bovy} \& {Rix}}{{Bovy} \&
  {Rix}}{2013}]{Bovy-2013}
{Bovy} J.,  {Rix} H.-W.,  2013, \mn@doi [\apj] {10.1088/0004-637X/779/2/115},
  \href {http://adsabs.harvard.edu/abs/2013ApJ...779..115B} {779, 115}

\bibitem[\protect\citeauthoryear{{Bovy}, {Erkal}  \& {Sanders}}{{Bovy}
  et~al.}{2016}]{Bovy-2016}
{Bovy} J.,  {Erkal} D.,   {Sanders} J.~L.,  2016, preprint, \href
  {http://adsabs.harvard.edu/abs/2016arXiv160603470B} {} (\mn@eprint {arXiv}
  {1606.03470})

\bibitem[\protect\citeauthoryear{{Boylan-Kolchin}, {Bullock}  \&
  {Kaplinghat}}{{Boylan-Kolchin} et~al.}{2011}]{Boylan-Kolchin-2011}
{Boylan-Kolchin} M.,  {Bullock} J.~S.,   {Kaplinghat} M.,  2011, \mn@doi
  [\mnras] {10.1111/j.1745-3933.2011.01074.x}, \href
  {http://adsabs.harvard.edu/abs/2011MNRAS.415L..40B} {415, L40}

\bibitem[\protect\citeauthoryear{{Brooks}, {Kuhlen}, {Zolotov}  \&
  {Hooper}}{{Brooks} et~al.}{2013}]{Brooks-2013}
{Brooks} A.~M.,  {Kuhlen} M.,  {Zolotov} A.,   {Hooper} D.,  2013, \mn@doi
  [\apj] {10.1088/0004-637X/765/1/22}, \href
  {http://adsabs.harvard.edu/abs/2013ApJ...765...22B} {765, 22}

\bibitem[\protect\citeauthoryear{{Bullock}, {Kravtsov}  \&
  {Weinberg}}{{Bullock} et~al.}{2000}]{Bullock-2000}
{Bullock} J.~S.,  {Kravtsov} A.~V.,   {Weinberg} D.~H.,  2000, \mn@doi [\apj]
  {10.1086/309279}, \href {http://adsabs.harvard.edu/abs/2000ApJ...539..517B}
  {539, 517}

\bibitem[\protect\citeauthoryear{{Carlberg} \& {Grillmair}}{{Carlberg} \&
  {Grillmair}}{2013}]{Carlberg-2013}
{Carlberg} R.~G.,  {Grillmair} C.~J.,  2013, \mn@doi [\apj]
  {10.1088/0004-637X/768/2/171}, \href
  {http://adsabs.harvard.edu/abs/2013ApJ...768..171C} {768, 171}

\bibitem[\protect\citeauthoryear{{Chan}, {Kere{\v s}}, {O{\~n}orbe}, {Hopkins},
  {Muratov}, {Faucher-Gigu{\`e}re}  \& {Quataert}}{{Chan}
  et~al.}{2015}]{Chan-2015}
{Chan} T.~K.,  {Kere{\v s}} D.,  {O{\~n}orbe} J.,  {Hopkins} P.~F.,  {Muratov}
  A.~L.,  {Faucher-Gigu{\`e}re} C.-A.,   {Quataert} E.,  2015, \mn@doi [\mnras]
  {10.1093/mnras/stv2165}, \href
  {http://adsabs.harvard.edu/abs/2015MNRAS.454.2981C} {454, 2981}

\bibitem[\protect\citeauthoryear{{Crain} et~al.,}{{Crain}
  et~al.}{2015}]{Crain-2015}
{Crain} R.~A.,  et~al., 2015, \mn@doi [\mnras] {10.1093/mnras/stv725}, \href
  {http://adsabs.harvard.edu/abs/2015MNRAS.450.1937C} {450, 1937}

\bibitem[\protect\citeauthoryear{{Croft}, {Weinberg}, {Bolte}, {Burles},
  {Hernquist}, {Katz}, {Kirkman}  \& {Tytler}}{{Croft}
  et~al.}{2002}]{Croft-2002}
{Croft} R.~A.~C.,  {Weinberg} D.~H.,  {Bolte} M.,  {Burles} S.,  {Hernquist}
  L.,  {Katz} N.,  {Kirkman} D.,   {Tytler} D.,  2002, \mn@doi [\apj]
  {10.1086/344099}, \href {http://adsabs.harvard.edu/abs/2002ApJ...581...20C}
  {581, 20}

\bibitem[\protect\citeauthoryear{{D'Onghia}, {Springel}, {Hernquist}  \&
  {Keres}}{{D'Onghia} et~al.}{2010}]{Donghia-2010}
{D'Onghia} E.,  {Springel} V.,  {Hernquist} L.,   {Keres} D.,  2010, \mn@doi
  [\apj] {10.1088/0004-637X/709/2/1138}, \href
  {http://adsabs.harvard.edu/abs/2010ApJ...709.1138D} {709, 1138}

\bibitem[\protect\citeauthoryear{{Dalal} \& {Kochanek}}{{Dalal} \&
  {Kochanek}}{2002}]{Dalal-2002}
{Dalal} N.,  {Kochanek} C.~S.,  2002, \mn@doi [\apj] {10.1086/340303}, \href
  {http://adsabs.harvard.edu/abs/2002ApJ...572...25D} {572, 25}

\bibitem[\protect\citeauthoryear{{Dalla Vecchia} \& {Schaye}}{{Dalla Vecchia}
  \& {Schaye}}{2012}]{DallaVecchia-2012}
{Dalla Vecchia} C.,  {Schaye} J.,  2012, \mn@doi [\mnras]
  {10.1111/j.1365-2966.2012.21704.x}, \href
  {http://adsabs.harvard.edu/abs/2012MNRAS.426..140D} {426, 140}

\bibitem[\protect\citeauthoryear{{Davis}, {Efstathiou}, {Frenk}  \&
  {White}}{{Davis} et~al.}{1985}]{Davis-1985}
{Davis} M.,  {Efstathiou} G.,  {Frenk} C.~S.,   {White} S.~D.~M.,  1985,
  \mn@doi [\apj] {10.1086/163168}, \href
  {http://adsabs.harvard.edu/abs/1985ApJ...292..371D} {292, 371}

\bibitem[\protect\citeauthoryear{{Diemand}, {Moore}  \& {Stadel}}{{Diemand}
  et~al.}{2004}]{Diemand-2004}
{Diemand} J.,  {Moore} B.,   {Stadel} J.,  2004, \mn@doi [\mnras]
  {10.1111/j.1365-2966.2004.07940.x}, \href
  {http://adsabs.harvard.edu/abs/2004MNRAS.352..535D} {352, 535}

\bibitem[\protect\citeauthoryear{{Diemand}, {Kuhlen}  \& {Madau}}{{Diemand}
  et~al.}{2007}]{Diemand-2007}
{Diemand} J.,  {Kuhlen} M.,   {Madau} P.,  2007, \mn@doi [\apj]
  {10.1086/520573}, \href {http://adsabs.harvard.edu/abs/2007ApJ...667..859D}
  {667, 859}

\bibitem[\protect\citeauthoryear{{Dolag}, {Borgani}, {Murante}  \&
  {Springel}}{{Dolag} et~al.}{2009}]{Dolag-2009}
{Dolag} K.,  {Borgani} S.,  {Murante} G.,   {Springel} V.,  2009, \mn@doi
  [\mnras] {10.1111/j.1365-2966.2009.15034.x}, \href
  {http://adsabs.harvard.edu/abs/2009MNRAS.399..497D} {399, 497}

\bibitem[\protect\citeauthoryear{{Dutton} \& {Treu}}{{Dutton} \&
  {Treu}}{2014}]{Dutton-2014}
{Dutton} A.~A.,  {Treu} T.,  2014, \mn@doi [\mnras] {10.1093/mnras/stt2489},
  \href {http://adsabs.harvard.edu/abs/2014MNRAS.438.3594D} {438, 3594}

\bibitem[\protect\citeauthoryear{{Dutton}, {Macci{\`o}}, {Frings}, {Wang},
  {Stinson}, {Penzo}  \& {Kang}}{{Dutton} et~al.}{2016}]{Dutton-2016}
{Dutton} A.~A.,  {Macci{\`o}} A.~V.,  {Frings} J.,  {Wang} L.,  {Stinson}
  G.~S.,  {Penzo} C.,   {Kang} X.,  2016, \mn@doi [\mnras]
  {10.1093/mnrasl/slv193}, \href
  {http://adsabs.harvard.edu/abs/2016MNRAS.457L..74D} {457, L74}

\bibitem[\protect\citeauthoryear{{Erkal} \& {Belokurov}}{{Erkal} \&
  {Belokurov}}{2015a}]{Erkal-2015a}
{Erkal} D.,  {Belokurov} V.,  2015a, \mn@doi [\mnras] {10.1093/mnras/stv655},
  \href {http://adsabs.harvard.edu/abs/2015MNRAS.450.1136E} {450, 1136}

\bibitem[\protect\citeauthoryear{{Erkal} \& {Belokurov}}{{Erkal} \&
  {Belokurov}}{2015b}]{Erkal-2015b}
{Erkal} D.,  {Belokurov} V.,  2015b, \mn@doi [\mnras] {10.1093/mnras/stv2122},
  \href {http://adsabs.harvard.edu/abs/2015MNRAS.454.3542E} {454, 3542}

\bibitem[\protect\citeauthoryear{{Fairbairn} \& {Schwetz}}{{Fairbairn} \&
  {Schwetz}}{2009}]{Fairbairn-2009}
{Fairbairn} M.,  {Schwetz} T.,  2009, \mn@doi [\jcap]
  {10.1088/1475-7516/2009/01/037}, \href
  {http://adsabs.harvard.edu/abs/2009JCAP...01..037F} {1, 037}

\bibitem[\protect\citeauthoryear{{Fattahi} et~al.,}{{Fattahi}
  et~al.}{2016}]{Fattahi-2015}
{Fattahi} A.,  et~al., 2016, \mn@doi [\mnras] {10.1093/mnras/stv2970}, \href
  {http://adsabs.harvard.edu/abs/2016MNRAS.457..844F} {457, 844}

\bibitem[\protect\citeauthoryear{{Feldmann} \& {Spolyar}}{{Feldmann} \&
  {Spolyar}}{2015}]{Feldmann-2015}
{Feldmann} R.,  {Spolyar} D.,  2015, \mn@doi [\mnras] {10.1093/mnras/stu2147},
  \href {http://adsabs.harvard.edu/abs/2015MNRAS.446.1000F} {446, 1000}

\bibitem[\protect\citeauthoryear{{Flynn}, {Holmberg}, {Portinari}, {Fuchs}  \&
  {Jahrei{\ss}}}{{Flynn} et~al.}{2006}]{Flynn-2006}
{Flynn} C.,  {Holmberg} J.,  {Portinari} L.,  {Fuchs} B.,   {Jahrei{\ss}} H.,
  2006, \mn@doi [\mnras] {10.1111/j.1365-2966.2006.10911.x}, \href
  {http://adsabs.harvard.edu/abs/2006MNRAS.372.1149F} {372, 1149}

\bibitem[\protect\citeauthoryear{{Ghigna}, {Moore}, {Governato}, {Lake},
  {Quinn}  \& {Stadel}}{{Ghigna} et~al.}{2000}]{Ghigna-2000}
{Ghigna} S.,  {Moore} B.,  {Governato} F.,  {Lake} G.,  {Quinn} T.,   {Stadel}
  J.,  2000, \mn@doi [\apj] {10.1086/317221}, \href
  {http://adsabs.harvard.edu/abs/2000ApJ...544..616G} {544, 616}

\bibitem[\protect\citeauthoryear{{Gilmore} et~al.,}{{Gilmore}
  et~al.}{2012}]{Gaia-2012}
{Gilmore} G.,  et~al., 2012, The Messenger, \href
  {http://adsabs.harvard.edu/abs/2012Msngr.147...25G} {147, 25}

\bibitem[\protect\citeauthoryear{{Governato} et~al.,}{{Governato}
  et~al.}{2010}]{Governato-2010}
{Governato} F.,  et~al., 2010, \mn@doi [\nat] {10.1038/nature08640}, \href
  {http://adsabs.harvard.edu/abs/2010Natur.463..203G} {463, 203}

\bibitem[\protect\citeauthoryear{{Grand}, {Springel}, {G{\'o}mez}, {Marinacci},
  {Pakmor}, {Campbell}  \& {Jenkins}}{{Grand} et~al.}{2016}]{Grand-2016}
{Grand} R.~J.~J.,  {Springel} V.,  {G{\'o}mez} F.~A.,  {Marinacci} F.,
  {Pakmor} R.,  {Campbell} D.~J.~R.,   {Jenkins} A.,  2016, \mn@doi [\mnras]
  {10.1093/mnras/stw601}, \href
  {http://adsabs.harvard.edu/abs/2016MNRAS.459..199G} {459, 199}

\bibitem[\protect\citeauthoryear{{Grillmair} \& {Dionatos}}{{Grillmair} \&
  {Dionatos}}{2006}]{Grillmair-2006}
{Grillmair} C.~J.,  {Dionatos} O.,  2006, \mn@doi [\apjl] {10.1086/505111},
  \href {http://adsabs.harvard.edu/abs/2006ApJ...643L..17G} {643, L17}

\bibitem[\protect\citeauthoryear{{H{\"a}nninen} \& {Flynn}}{{H{\"a}nninen} \&
  {Flynn}}{2002}]{Hanninen-2002}
{H{\"a}nninen} J.,  {Flynn} C.,  2002, \mn@doi [\mnras]
  {10.1046/j.1365-8711.2002.05956.x}, \href
  {http://adsabs.harvard.edu/abs/2002MNRAS.337..731H} {337, 731}

\bibitem[\protect\citeauthoryear{{Hezaveh} et~al.,}{{Hezaveh}
  et~al.}{2016}]{Hezaveh-2016}
{Hezaveh} Y.~D.,  et~al., 2016, \mn@doi [\apj] {10.3847/0004-637X/823/1/37},
  \href {http://adsabs.harvard.edu/abs/2016ApJ...823...37H} {823, 37}

\bibitem[\protect\citeauthoryear{{Hopkins}}{{Hopkins}}{2013}]{Hopkins-2013}
{Hopkins} P.~F.,  2013, \mn@doi [\mnras] {10.1093/mnras/sts210}, \href
  {http://adsabs.harvard.edu/abs/2013MNRAS.428.2840H} {428, 2840}

\bibitem[\protect\citeauthoryear{{Hopkins}, {Cox}, {Younger}  \&
  {Hernquist}}{{Hopkins} et~al.}{2009}]{Hopkins-2009}
{Hopkins} P.~F.,  {Cox} T.~J.,  {Younger} J.~D.,   {Hernquist} L.,  2009,
  \mn@doi [\apj] {10.1088/0004-637X/691/2/1168}, \href
  {http://adsabs.harvard.edu/abs/2009ApJ...691.1168H} {691, 1168}

\bibitem[\protect\citeauthoryear{{Ibata}, {Lewis}, {Irwin}  \& {Quinn}}{{Ibata}
  et~al.}{2002}]{Ibata-2002}
{Ibata} R.~A.,  {Lewis} G.~F.,  {Irwin} M.~J.,   {Quinn} T.,  2002, \mn@doi
  [\mnras] {10.1046/j.1365-8711.2002.05358.x}, \href
  {http://adsabs.harvard.edu/abs/2002MNRAS.332..915I} {332, 915}

\bibitem[\protect\citeauthoryear{{Jenkins}}{{Jenkins}}{2010}]{Jenkins-2010}
{Jenkins} A.,  2010, \mn@doi [\mnras] {10.1111/j.1365-2966.2010.16259.x}, \href
  {http://adsabs.harvard.edu/abs/2010MNRAS.403.1859J} {403, 1859}

\bibitem[\protect\citeauthoryear{{Jenkins}}{{Jenkins}}{2013}]{Jenkins-2013}
{Jenkins} A.,  2013, \mn@doi [\mnras] {10.1093/mnras/stt1154}, \href
  {http://adsabs.harvard.edu/abs/2013MNRAS.434.2094J} {434, 2094}

\bibitem[\protect\citeauthoryear{{Jiang}, {Helly}, {Cole}  \& {Frenk}}{{Jiang}
  et~al.}{2014}]{Jiang-2014}
{Jiang} L.,  {Helly} J.~C.,  {Cole} S.,   {Frenk} C.~S.,  2014, \mn@doi
  [\mnras] {10.1093/mnras/stu390}, \href
  {http://adsabs.harvard.edu/abs/2014MNRAS.440.2115J} {440, 2115}

\bibitem[\protect\citeauthoryear{{Johansson}, {Naab}  \& {Burkert}}{{Johansson}
  et~al.}{2009}]{Johansson-2009}
{Johansson} P.~H.,  {Naab} T.,   {Burkert} A.,  2009, \mn@doi [\apj]
  {10.1088/0004-637X/690/1/802}, \href
  {http://adsabs.harvard.edu/abs/2009ApJ...690..802J} {690, 802}

\bibitem[\protect\citeauthoryear{Jones, Oliphant, Peterson  et~al.}{Jones
  et~al.}{01  }]{scipy}
Jones E.,  Oliphant T.,  Peterson P.,   et~al., 2001--, {SciPy}: Open source
  scientific tools for {Python}, \url {http://www.scipy.org/}

\bibitem[\protect\citeauthoryear{{Kazantzidis}, {Magorrian}  \&
  {Moore}}{{Kazantzidis} et~al.}{2004}]{Kazantzidis-2004}
{Kazantzidis} S.,  {Magorrian} J.,   {Moore} B.,  2004, \mn@doi [\apj]
  {10.1086/380192}, \href {http://adsabs.harvard.edu/abs/2004ApJ...601...37K}
  {601, 37}

\bibitem[\protect\citeauthoryear{{Kazantzidis}, {Bullock}, {Zentner},
  {Kravtsov}  \& {Moustakas}}{{Kazantzidis} et~al.}{2008}]{Kazantzidis-2008}
{Kazantzidis} S.,  {Bullock} J.~S.,  {Zentner} A.~R.,  {Kravtsov} A.~V.,
  {Moustakas} L.~A.,  2008, \mn@doi [\apj] {10.1086/591958}, \href
  {http://adsabs.harvard.edu/abs/2008ApJ...688..254K} {688, 254}

\bibitem[\protect\citeauthoryear{{Kennedy}, {Frenk}, {Cole}  \&
  {Benson}}{{Kennedy} et~al.}{2014}]{Kennedy-2014}
{Kennedy} R.,  {Frenk} C.,  {Cole} S.,   {Benson} A.,  2014, \mn@doi [\mnras]
  {10.1093/mnras/stu719}, \href
  {http://adsabs.harvard.edu/abs/2014MNRAS.442.2487K} {442, 2487}

\bibitem[\protect\citeauthoryear{{Klypin}, {Kravtsov}, {Valenzuela}  \&
  {Prada}}{{Klypin} et~al.}{1999}]{Klypin-1999}
{Klypin} A.,  {Kravtsov} A.~V.,  {Valenzuela} O.,   {Prada} F.,  1999, \mn@doi
  [\apj] {10.1086/307643}, \href
  {http://adsabs.harvard.edu/abs/1999ApJ...522...82K} {522, 82}

\bibitem[\protect\citeauthoryear{{Komatsu}}{{Komatsu}}{2011}]{Komatsu-2011}
{Komatsu} E. e.~a.,  2011, \mn@doi [\apjs] {10.1088/0067-0049/192/2/18}, \href
  {http://adsabs.harvard.edu/abs/2011ApJS..192...18K} {192, 18}

\bibitem[\protect\citeauthoryear{{Kuhlen}, {Weiner}, {Diemand}, {Madau},
  {Moore}, {Potter}, {Stadel}  \& {Zemp}}{{Kuhlen} et~al.}{2010}]{Kuhlen-2010}
{Kuhlen} M.,  {Weiner} N.,  {Diemand} J.,  {Madau} P.,  {Moore} B.,  {Potter}
  D.,  {Stadel} J.,   {Zemp} M.,  2010, \mn@doi [\jcap]
  {10.1088/1475-7516/2010/02/030}, \href
  {http://adsabs.harvard.edu/abs/2010JCAP...02..030K} {2, 030}

\bibitem[\protect\citeauthoryear{{K{\"u}pper}, {Balbinot}, {Bonaca},
  {Johnston}, {Hogg}, {Kroupa}  \& {Santiago}}{{K{\"u}pper}
  et~al.}{2015}]{Kuepper-2015}
{K{\"u}pper} A.~H.~W.,  {Balbinot} E.,  {Bonaca} A.,  {Johnston} K.~V.,  {Hogg}
  D.~W.,  {Kroupa} P.,   {Santiago} B.~X.,  2015, \mn@doi [\apj]
  {10.1088/0004-637X/803/2/80}, \href
  {http://adsabs.harvard.edu/abs/2015ApJ...803...80K} {803, 80}

\bibitem[\protect\citeauthoryear{{LSST Science Collaboration} et~al.,}{{LSST
  Science Collaboration} et~al.}{2009}]{LSST-2009}
{LSST Science Collaboration} et~al., 2009, preprint, \href
  {http://adsabs.harvard.edu/abs/2009arXiv0912.0201L} {} (\mn@eprint {arXiv}
  {0912.0201})

\bibitem[\protect\citeauthoryear{{Lacey}}{{Lacey}}{1984}]{Lacey-1984}
{Lacey} C.~G.,  1984, \mn@doi [\mnras] {10.1093/mnras/208.4.687}, \href
  {http://adsabs.harvard.edu/abs/1984MNRAS.208..687L} {208, 687}

\bibitem[\protect\citeauthoryear{{Lovell} et~al.,}{{Lovell}
  et~al.}{2012}]{Lovell-2012}
{Lovell} M.~R.,  et~al., 2012, \mn@doi [\mnras]
  {10.1111/j.1365-2966.2011.20200.x}, \href
  {http://adsabs.harvard.edu/abs/2012MNRAS.420.2318L} {420, 2318}

\bibitem[\protect\citeauthoryear{{Mandelbaum}, {Seljak}, {Cool}, {Blanton},
  {Hirata}  \& {Brinkmann}}{{Mandelbaum} et~al.}{2006}]{Mandelbaum-2006}
{Mandelbaum} R.,  {Seljak} U.,  {Cool} R.~J.,  {Blanton} M.,  {Hirata} C.~M.,
  {Brinkmann} J.,  2006, \mn@doi [\mnras] {10.1111/j.1365-2966.2006.10906.x},
  \href {http://adsabs.harvard.edu/abs/2006MNRAS.372..758M} {372, 758}

\bibitem[\protect\citeauthoryear{{Mao} \& {Schneider}}{{Mao} \&
  {Schneider}}{1998}]{Mao-1998}
{Mao} S.,  {Schneider} P.,  1998, \mn@doi [\mnras]
  {10.1046/j.1365-8711.1998.01319.x}, \href
  {http://adsabs.harvard.edu/abs/1998MNRAS.295..587M} {295, 587}

\bibitem[\protect\citeauthoryear{{Merritt}}{{Merritt}}{1985}]{Merritt-1985}
{Merritt} D.,  1985, \mn@doi [\aj] {10.1086/113810}, \href
  {http://adsabs.harvard.edu/abs/1985AJ.....90.1027M} {90, 1027}

\bibitem[\protect\citeauthoryear{{Metcalf} \& {Madau}}{{Metcalf} \&
  {Madau}}{2001}]{Metcalf-2001}
{Metcalf} R.~B.,  {Madau} P.,  2001, \mn@doi [\apj] {10.1086/323695}, \href
  {http://adsabs.harvard.edu/abs/2001ApJ...563....9M} {563, 9}

\bibitem[\protect\citeauthoryear{{Moster}, {Macci{\`o}}, {Somerville},
  {Johansson}  \& {Naab}}{{Moster} et~al.}{2010}]{Moster-2010}
{Moster} B.~P.,  {Macci{\`o}} A.~V.,  {Somerville} R.~S.,  {Johansson} P.~H.,
  {Naab} T.,  2010, \mn@doi [\mnras] {10.1111/j.1365-2966.2009.16190.x}, \href
  {http://adsabs.harvard.edu/abs/2010MNRAS.403.1009M} {403, 1009}

\bibitem[\protect\citeauthoryear{{Navarro} \& {White}}{{Navarro} \&
  {White}}{1994}]{Navarro-1994}
{Navarro} J.~F.,  {White} S.~D.~M.,  1994, \mn@doi [\mnras]
  {10.1093/mnras/267.2.401}, \href
  {http://adsabs.harvard.edu/abs/1994MNRAS.267..401N} {267, 401}

\bibitem[\protect\citeauthoryear{{Navarro}, {Eke}  \& {Frenk}}{{Navarro}
  et~al.}{1996}]{Navarro-1996}
{Navarro} J.~F.,  {Eke} V.~R.,   {Frenk} C.~S.,  1996, \mnras, \href
  {http://adsabs.harvard.edu/abs/1996MNRAS.283L..72N} {283, L72}

\bibitem[\protect\citeauthoryear{{Nierenberg}, {Treu}, {Wright}, {Fassnacht}
  \& {Auger}}{{Nierenberg} et~al.}{2014}]{Nierenberg-2014}
{Nierenberg} A.~M.,  {Treu} T.,  {Wright} S.~A.,  {Fassnacht} C.~D.,   {Auger}
  M.~W.,  2014, \mn@doi [\mnras] {10.1093/mnras/stu862}, \href
  {http://adsabs.harvard.edu/abs/2014MNRAS.442.2434N} {442, 2434}

\bibitem[\protect\citeauthoryear{{Ocvirk} et~al.,}{{Ocvirk}
  et~al.}{2015}]{Ocvirk-2015}
{Ocvirk} P.,  et~al., 2015, preprint, \href
  {http://adsabs.harvard.edu/abs/2015arXiv151100011O} {} (\mn@eprint {arXiv}
  {1511.00011})

\bibitem[\protect\citeauthoryear{{Odenkirchen} et~al.,}{{Odenkirchen}
  et~al.}{2001}]{Odenkirchen-2001}
{Odenkirchen} M.,  et~al., 2001, \mn@doi [\apjl] {10.1086/319095}, \href
  {http://adsabs.harvard.edu/abs/2001ApJ...548L.165O} {548, L165}

\bibitem[\protect\citeauthoryear{{Okamoto}, {Gao}  \& {Theuns}}{{Okamoto}
  et~al.}{2008}]{Okamoto-2008}
{Okamoto} T.,  {Gao} L.,   {Theuns} T.,  2008, \mn@doi [\mnras]
  {10.1111/j.1365-2966.2008.13830.x}, \href
  {http://adsabs.harvard.edu/abs/2008MNRAS.390..920O} {390, 920}

\bibitem[\protect\citeauthoryear{{Onions} et~al.,}{{Onions}
  et~al.}{2012}]{Onions-2012}
{Onions} J.,  et~al., 2012, \mn@doi [\mnras]
  {10.1111/j.1365-2966.2012.20947.x}, \href
  {http://adsabs.harvard.edu/abs/2012MNRAS.423.1200O} {423, 1200}

\bibitem[\protect\citeauthoryear{{Osipkov}}{{Osipkov}}{1979}]{Osipkov-1979}
{Osipkov} L.~P.,  1979, Pisma v Astronomicheskii Zhurnal, \href
  {http://adsabs.harvard.edu/abs/1979PAZh....5...77O} {5, 77}

\bibitem[\protect\citeauthoryear{{Papastergis}, {Martin}, {Giovanelli}  \&
  {Haynes}}{{Papastergis} et~al.}{2011}]{Papastergis-2011}
{Papastergis} E.,  {Martin} A.~M.,  {Giovanelli} R.,   {Haynes} M.~P.,  2011,
  \mn@doi [\apj] {10.1088/0004-637X/739/1/38}, \href
  {http://adsabs.harvard.edu/abs/2011ApJ...739...38P} {739, 38}

\bibitem[\protect\citeauthoryear{{Perryman} et~al.,}{{Perryman}
  et~al.}{2001}]{Perryman-2001}
{Perryman} M.~A.~C.,  et~al., 2001, \mn@doi [\aap]
  {10.1051/0004-6361:20010085}, \href
  {http://adsabs.harvard.edu/abs/2001A%26A...369..339P} {369, 339}

\bibitem[\protect\citeauthoryear{{Quinn}, {Hernquist}  \& {Fullagar}}{{Quinn}
  et~al.}{1993}]{Quinn-1993}
{Quinn} P.~J.,  {Hernquist} L.,   {Fullagar} D.~P.,  1993, \mn@doi [\apj]
  {10.1086/172184}, \href {http://adsabs.harvard.edu/abs/1993ApJ...403...74Q}
  {403, 74}

\bibitem[\protect\citeauthoryear{Rice}{Rice}{1945}]{Rice-1945}
Rice S.~O.,  1945, \mn@doi [Bell System Technical Journal]
  {10.1002/j.1538-7305.1945.tb00453.x}, 24, 46

\bibitem[\protect\citeauthoryear{{Rosas-Guevara} et~al.,}{{Rosas-Guevara}
  et~al.}{2015}]{Rosas-Guevara-2013}
{Rosas-Guevara} Y.~M.,  et~al., 2015, \mn@doi [\mnras] {10.1093/mnras/stv2056},
  \href {http://adsabs.harvard.edu/abs/2015MNRAS.454.1038R} {454, 1038}

\bibitem[\protect\citeauthoryear{{Sawala}, {Frenk}, {Crain}, {Jenkins},
  {Schaye}, {Theuns}  \& {Zavala}}{{Sawala} et~al.}{2013}]{Sawala-2013}
{Sawala} T.,  {Frenk} C.~S.,  {Crain} R.~A.,  {Jenkins} A.,  {Schaye} J.,
  {Theuns} T.,   {Zavala} J.,  2013, \mn@doi [\mnras] {10.1093/mnras/stt259},
  \href {http://adsabs.harvard.edu/abs/2013MNRAS.431.1366S} {431, 1366}

\bibitem[\protect\citeauthoryear{{Sawala} et~al.,}{{Sawala}
  et~al.}{2015}]{Sawala-2015}
{Sawala} T.,  et~al., 2015, \mn@doi [\mnras] {10.1093/mnras/stu2753}, \href
  {http://adsabs.harvard.edu/abs/2015MNRAS.448.2941S} {448, 2941}

\bibitem[\protect\citeauthoryear{{Sawala} et~al.,}{{Sawala}
  et~al.}{2016a}]{Sawala-2016a}
{Sawala} T.,  et~al., 2016a, \mn@doi [\mnras] {10.1093/mnras/stv2597}, \href
  {http://adsabs.harvard.edu/abs/2014arXiv1406.6362S} {456, 85}

\bibitem[\protect\citeauthoryear{{Sawala} et~al.,}{{Sawala}
  et~al.}{2016b}]{Sawala-2016b}
{Sawala} T.,  et~al., 2016b, \mn@doi [\mnras] {10.1093/mnras/stw145}, \href
  {http://adsabs.harvard.edu/abs/2016MNRAS.457.1931S} {457, 1931}

\bibitem[\protect\citeauthoryear{{Schaller} et~al.,}{{Schaller}
  et~al.}{2015a}]{Schaller-2015}
{Schaller} M.,  et~al., 2015a, \mn@doi [\mnras] {10.1093/mnras/stv1067}, \href
  {http://adsabs.harvard.edu/abs/2015MNRAS.451.1247S} {451, 1247}

\bibitem[\protect\citeauthoryear{{Schaller}, {Dalla Vecchia}, {Schaye},
  {Bower}, {Theuns}, {Crain}, {Furlong}  \& {McCarthy}}{{Schaller}
  et~al.}{2015b}]{Schaller-2015b}
{Schaller} M.,  {Dalla Vecchia} C.,  {Schaye} J.,  {Bower} R.~G.,  {Theuns} T.,
   {Crain} R.~A.,  {Furlong} M.,   {McCarthy} I.~G.,  2015b, \mn@doi [\mnras]
  {10.1093/mnras/stv2169}, \href
  {http://adsabs.harvard.edu/abs/2015MNRAS.454.2277S} {454, 2277}

\bibitem[\protect\citeauthoryear{{Schaller}, {Frenk}, {Fattahi}, {Navarro},
  {Oman}  \& {Sawala}}{{Schaller} et~al.}{2016}]{Schaller-2016}
{Schaller} M.,  {Frenk} C.~S.,  {Fattahi} A.,  {Navarro} J.~F.,  {Oman} K.~A.,
   {Sawala} T.,  2016, \mn@doi [\mnras] {10.1093/mnrasl/slw101}, \href
  {http://adsabs.harvard.edu/abs/2016MNRAS.461L..56S} {461, L56}

\bibitem[\protect\citeauthoryear{{Schaye} \& {Dalla Vecchia}}{{Schaye} \&
  {Dalla Vecchia}}{2008}]{Schaye-2008}
{Schaye} J.,  {Dalla Vecchia} C.,  2008, \mn@doi [\mnras]
  {10.1111/j.1365-2966.2007.12639.x}, \href
  {http://adsabs.harvard.edu/abs/2008MNRAS.383.1210S} {383, 1210}

\bibitem[\protect\citeauthoryear{{Schaye}, {Theuns}, {Rauch}, {Efstathiou}  \&
  {Sargent}}{{Schaye} et~al.}{2000}]{Schaye-2000}
{Schaye} J.,  {Theuns} T.,  {Rauch} M.,  {Efstathiou} G.,   {Sargent} W.~L.~W.,
   2000, \mn@doi [\mnras] {10.1046/j.1365-8711.2000.03815.x}, \href
  {http://adsabs.harvard.edu/abs/2000MNRAS.318..817S} {318, 817}

\bibitem[\protect\citeauthoryear{{Schaye} et~al.}{{Schaye}
  et~al.}{2015}]{Schaye-2014}
{Schaye} J.,  et~al., 2015, \mn@doi [\mnras] {10.1093/mnras/stu2058}, \href
  {http://adsabs.harvard.edu/abs/2015MNRAS.446..521S} {446, 521}

\bibitem[\protect\citeauthoryear{{Sellwood}}{{Sellwood}}{2014}]{Sellwood-2014}
{Sellwood} J.~A.,  2014, \mn@doi [Reviews of Modern Physics]
  {10.1103/RevModPhys.86.1}, \href
  {http://adsabs.harvard.edu/abs/2014RvMP...86....1S} {86, 1}

\bibitem[\protect\citeauthoryear{{Sellwood}, {Nelson}  \&
  {Tremaine}}{{Sellwood} et~al.}{1998}]{Sellwood-1998}
{Sellwood} J.~A.,  {Nelson} R.~W.,   {Tremaine} S.,  1998, \mn@doi [\apj]
  {10.1086/306280}, \href {http://adsabs.harvard.edu/abs/1998ApJ...506..590S}
  {506, 590}

\bibitem[\protect\citeauthoryear{{Spergel} \& {Steinhardt}}{{Spergel} \&
  {Steinhardt}}{2000}]{Spergel-2000}
{Spergel} D.~N.,  {Steinhardt} P.~J.,  2000, \mn@doi [Physical Review Letters]
  {10.1103/PhysRevLett.84.3760}, \href
  {http://adsabs.harvard.edu/abs/2000PhRvL..84.3760S} {84, 3760}

\bibitem[\protect\citeauthoryear{{Springel}}{{Springel}}{2005}]{Springel-2005}
{Springel} V.,  2005, \mn@doi [\mnras] {10.1111/j.1365-2966.2005.09655.x},
  \href {http://adsabs.harvard.edu/abs/2005MNRAS.364.1105S} {364, 1105}

\bibitem[\protect\citeauthoryear{{Springel}, {White}, {Tormen}  \&
  {Kauffmann}}{{Springel} et~al.}{2001}]{Springel-2001}
{Springel} V.,  {White} S.~D.~M.,  {Tormen} G.,   {Kauffmann} G.,  2001,
  \mn@doi [\mnras] {10.1046/j.1365-8711.2001.04912.x}, \href
  {http://adsabs.harvard.edu/abs/2001MNRAS.328..726S} {328, 726}

\bibitem[\protect\citeauthoryear{{Springel} et~al.,}{{Springel}
  et~al.}{2008}]{Springel-2008}
{Springel} V.,  et~al., 2008, \mn@doi [\mnras]
  {10.1111/j.1365-2966.2008.14066.x}, \href
  {http://adsabs.harvard.edu/abs/2008MNRAS.391.1685S} {391, 1685}

\bibitem[\protect\citeauthoryear{{Stewart}, {Bullock}, {Wechsler}  \&
  {Maller}}{{Stewart} et~al.}{2009}]{Stewart-2009}
{Stewart} K.~R.,  {Bullock} J.~S.,  {Wechsler} R.~H.,   {Maller} A.~H.,  2009,
  \mn@doi [\apj] {10.1088/0004-637X/702/1/307}, \href
  {http://adsabs.harvard.edu/abs/2009ApJ...702..307S} {702, 307}

\bibitem[\protect\citeauthoryear{{Strigari}, {Frenk}  \& {White}}{{Strigari}
  et~al.}{2014}]{Strigari-2014}
{Strigari} L.~E.,  {Frenk} C.~S.,   {White} S.~D.~M.,  2014, preprint, \href
  {http://adsabs.harvard.edu/abs/2014arXiv1406.6079S} {} (\mn@eprint {arXiv}
  {1406.6079})

\bibitem[\protect\citeauthoryear{{The Dark Energy Survey Collaboration}}{{The
  Dark Energy Survey Collaboration}}{2005}]{DES-2005}
{The Dark Energy Survey Collaboration} 2005, preprint, \href
  {http://adsabs.harvard.edu/abs/2005astro.ph.10346T} {} (\mn@eprint {arXiv}
  {astro-ph/0510346})

\bibitem[\protect\citeauthoryear{{Tikhonov} \& {Klypin}}{{Tikhonov} \&
  {Klypin}}{2009}]{Tikhonov-2009}
{Tikhonov} A.~V.,  {Klypin} A.,  2009, \mn@doi [\mnras]
  {10.1111/j.1365-2966.2009.14686.x}, \href
  {http://adsabs.harvard.edu/abs/2009MNRAS.395.1915T} {395, 1915}

\bibitem[\protect\citeauthoryear{{Toth} \& {Ostriker}}{{Toth} \&
  {Ostriker}}{1992}]{Toth-1992}
{Toth} G.,  {Ostriker} J.~P.,  1992, \mn@doi [\apj] {10.1086/171185}, \href
  {http://adsabs.harvard.edu/abs/1992ApJ...389....5T} {389, 5}

\bibitem[\protect\citeauthoryear{{Vande Putte}, {Cropper}  \&
  {Ferreras}}{{Vande Putte} et~al.}{2009}]{VandePutte-2009}
{Vande Putte} D.,  {Cropper} M.,   {Ferreras} I.,  2009, \mn@doi [\mnras]
  {10.1111/j.1365-2966.2009.15044.x}, \href
  {http://adsabs.harvard.edu/abs/2009MNRAS.397.1587V} {397, 1587}

\bibitem[\protect\citeauthoryear{{Vegetti}, {Lagattuta}, {McKean}, {Auger},
  {Fassnacht}  \& {Koopmans}}{{Vegetti} et~al.}{2012}]{Vegetti-2012}
{Vegetti} S.,  {Lagattuta} D.~J.,  {McKean} J.~P.,  {Auger} M.~W.,  {Fassnacht}
  C.~D.,   {Koopmans} L.~V.~E.,  2012, \mn@doi [\nat] {10.1038/nature10669},
  \href {http://adsabs.harvard.edu/abs/2012Natur.481..341V} {481, 341}

\bibitem[\protect\citeauthoryear{{Vegetti}, {Koopmans}, {Auger}, {Treu}  \&
  {Bolton}}{{Vegetti} et~al.}{2014}]{Vegetti-2014}
{Vegetti} S.,  {Koopmans} L.~V.~E.,  {Auger} M.~W.,  {Treu} T.,   {Bolton}
  A.~S.,  2014, \mn@doi [\mnras] {10.1093/mnras/stu943}, \href
  {http://adsabs.harvard.edu/abs/2014MNRAS.442.2017V} {442, 2017}

\bibitem[\protect\citeauthoryear{{Vergados}, {Hansen}  \& {Host}}{{Vergados}
  et~al.}{2008}]{Vergados-2008}
{Vergados} J.~D.,  {Hansen} S.~H.,   {Host} O.,  2008, \mn@doi [\prd]
  {10.1103/PhysRevD.77.023509}, \href
  {http://adsabs.harvard.edu/abs/2008PhRvD..77b3509V} {77, 023509}

\bibitem[\protect\citeauthoryear{{Viel}, {Becker}, {Bolton}  \&
  {Haehnelt}}{{Viel} et~al.}{2013}]{Viel-2013}
{Viel} M.,  {Becker} G.~D.,  {Bolton} J.~S.,   {Haehnelt} M.~G.,  2013, \mn@doi
  [\prd] {10.1103/PhysRevD.88.043502}, \href
  {http://adsabs.harvard.edu/abs/2013PhRvD..88d3502V} {88, 043502}

\bibitem[\protect\citeauthoryear{{Vogelsberger} et~al.,}{{Vogelsberger}
  et~al.}{2009}]{Vogelsberger-2009}
{Vogelsberger} M.,  et~al., 2009, \mn@doi [\mnras]
  {10.1111/j.1365-2966.2009.14630.x}, \href
  {http://adsabs.harvard.edu/abs/2009MNRAS.395..797V} {395, 797}

\bibitem[\protect\citeauthoryear{{Walker} \& {Pe{\~n}arrubia}}{{Walker} \&
  {Pe{\~n}arrubia}}{2011}]{Walker-2011}
{Walker} M.~G.,  {Pe{\~n}arrubia} J.,  2011, \mn@doi [\apj]
  {10.1088/0004-637X/742/1/20}, \href
  {http://adsabs.harvard.edu/abs/2011ApJ...742...20W} {742, 20}

\bibitem[\protect\citeauthoryear{{Walker}, {Mihos}  \& {Hernquist}}{{Walker}
  et~al.}{1996}]{Walker-1996}
{Walker} I.~R.,  {Mihos} J.~C.,   {Hernquist} L.,  1996, \mn@doi [\apj]
  {10.1086/176956}, \href {http://adsabs.harvard.edu/abs/1996ApJ...460..121W}
  {460, 121}

\bibitem[\protect\citeauthoryear{{Wiersma}, {Schaye}  \& {Smith}}{{Wiersma}
  et~al.}{2009a}]{Wiersma-2009}
{Wiersma} R.~P.~C.,  {Schaye} J.,   {Smith} B.~D.,  2009a, \mn@doi [\mnras]
  {10.1111/j.1365-2966.2008.14191.x}, \href
  {http://adsabs.harvard.edu/abs/2009MNRAS.393...99W} {393, 99}

\bibitem[\protect\citeauthoryear{{Wiersma}, {Schaye}, {Theuns}, {Dalla Vecchia}
   \& {Tornatore}}{{Wiersma} et~al.}{2009b}]{Wiersma-2009b}
{Wiersma} R.~P.~C.,  {Schaye} J.,  {Theuns} T.,  {Dalla Vecchia} C.,
  {Tornatore} L.,  2009b, \mn@doi [\mnras] {10.1111/j.1365-2966.2009.15331.x},
  \href {http://adsabs.harvard.edu/abs/2009MNRAS.399..574W} {399, 574}

\bibitem[\protect\citeauthoryear{{Willett}, {Newberg}, {Zhang}, {Yanny}  \&
  {Beers}}{{Willett} et~al.}{2009}]{Willet-2009}
{Willett} B.~A.,  {Newberg} H.~J.,  {Zhang} H.,  {Yanny} B.,   {Beers} T.~C.,
  2009, \mn@doi [\apj] {10.1088/0004-637X/697/1/207}, \href
  {http://adsabs.harvard.edu/abs/2009ApJ...697..207W} {697, 207}

\bibitem[\protect\citeauthoryear{{Wright} et~al.,}{{Wright}
  et~al.}{1992}]{Wright-1992}
{Wright} E.~L.,  et~al., 1992, \mn@doi [\apjl] {10.1086/186506}, \href
  {http://adsabs.harvard.edu/abs/1992ApJ...396L..13W} {396, L13}

\bibitem[\protect\citeauthoryear{{Xu} et~al.,}{{Xu} et~al.}{2009}]{Xu-2009}
{Xu} D.~D.,  et~al., 2009, \mn@doi [\mnras] {10.1111/j.1365-2966.2009.15230.x},
  \href {http://adsabs.harvard.edu/abs/2009MNRAS.398.1235X} {398, 1235}

\bibitem[\protect\citeauthoryear{{Xu}, {Sluse}, {Gao}, {Wang}, {Frenk}, {Mao},
  {Schneider}  \& {Springel}}{{Xu} et~al.}{2015}]{Xu-2015}
{Xu} D.,  {Sluse} D.,  {Gao} L.,  {Wang} J.,  {Frenk} C.,  {Mao} S.,
  {Schneider} P.,   {Springel} V.,  2015, \mn@doi [\mnras]
  {10.1093/mnras/stu2673}, \href
  {http://adsabs.harvard.edu/abs/2015MNRAS.447.3189X} {447, 3189}

\bibitem[\protect\citeauthoryear{{Yoon}, {Johnston}  \& {Hogg}}{{Yoon}
  et~al.}{2011}]{Yoon-2011}
{Yoon} J.~H.,  {Johnston} K.~V.,   {Hogg} D.~W.,  2011, \mn@doi [\apj]
  {10.1088/0004-637X/731/1/58}, \href
  {http://adsabs.harvard.edu/abs/2011ApJ...731...58Y} {731, 58}

\bibitem[\protect\citeauthoryear{{Yurin} \& {Springel}}{{Yurin} \&
  {Springel}}{2015}]{Yurin-2015}
{Yurin} D.,  {Springel} V.,  2015, \mn@doi [\mnras] {10.1093/mnras/stv1454},
  \href {http://adsabs.harvard.edu/abs/2015MNRAS.452.2367Y} {452, 2367}

\bibitem[\protect\citeauthoryear{{Zhao}}{{Zhao}}{1996}]{Zhao-1996}
{Zhao} H.,  1996, \mn@doi [\mnras] {10.1093/mnras/278.2.488}, \href
  {http://adsabs.harvard.edu/abs/1996MNRAS.278..488Z} {278, 488}

\bibitem[\protect\citeauthoryear{{Zolotov} et~al.,}{{Zolotov}
  et~al.}{2012}]{Zolotov-2012}
{Zolotov} A.,  et~al., 2012, \mn@doi [\apj] {10.1088/0004-637X/761/1/71}, \href
  {http://adsabs.harvard.edu/abs/2012ApJ...761...71Z} {761, 71}

\bibitem[\protect\citeauthoryear{van~der Walt, Colbert  \& Varoquaux}{van~der
  Walt et~al.}{2011}]{numpy}
van~der Walt S.,  Colbert S.~C.,   Varoquaux G.,  2011, \mn@doi [Computing in
  Science Engineering] {10.1109/MCSE.2011.37}, 13, 22

\makeatother
\end{thebibliography}

\appendix

\section{Halo Reference Frame and Orbital Interpolation}
\label{appendix:orbits}

\subsection{Host halo reference frame}\label{section:reference}
Whereas satellite subhaloes are typically tidally truncated at small
radi, the fact that their host haloes are extended structures
complicates the choice of reference frame. Because the centre of mass
(CM) of a halo depends on material in the loosely bound outskirts, far
away from the pericentres of satellites orbits, a more physical and
more common definition of the host halo's position is the minimum of
its gravitational potential, or more specifically, the position of the
particle with the lowest potential energy, which we denote as CP.

Considering that the centre of mass and centre of potential of a halo
can differ by $\sim 10$~kpc, for subhaloes that come much closer to
the centre, the combination of CP positions and CM velocities is
unsuitable, and can result in significant errors in the estimated
orbital parameters. For this reason, in this work, we use the
positions and velocities for both the main halo and subhaloes relative
to those of the CP $({\bf x}_{CP}, \dot{{\bf x}}_{CP})$ where the time
derivative $\dot{{\bf x}}_{CP}$ is obtained through higher order
interpolation.

\begin{figure}
  \begin{center}
    \includegraphics*[width = \columnwidth]{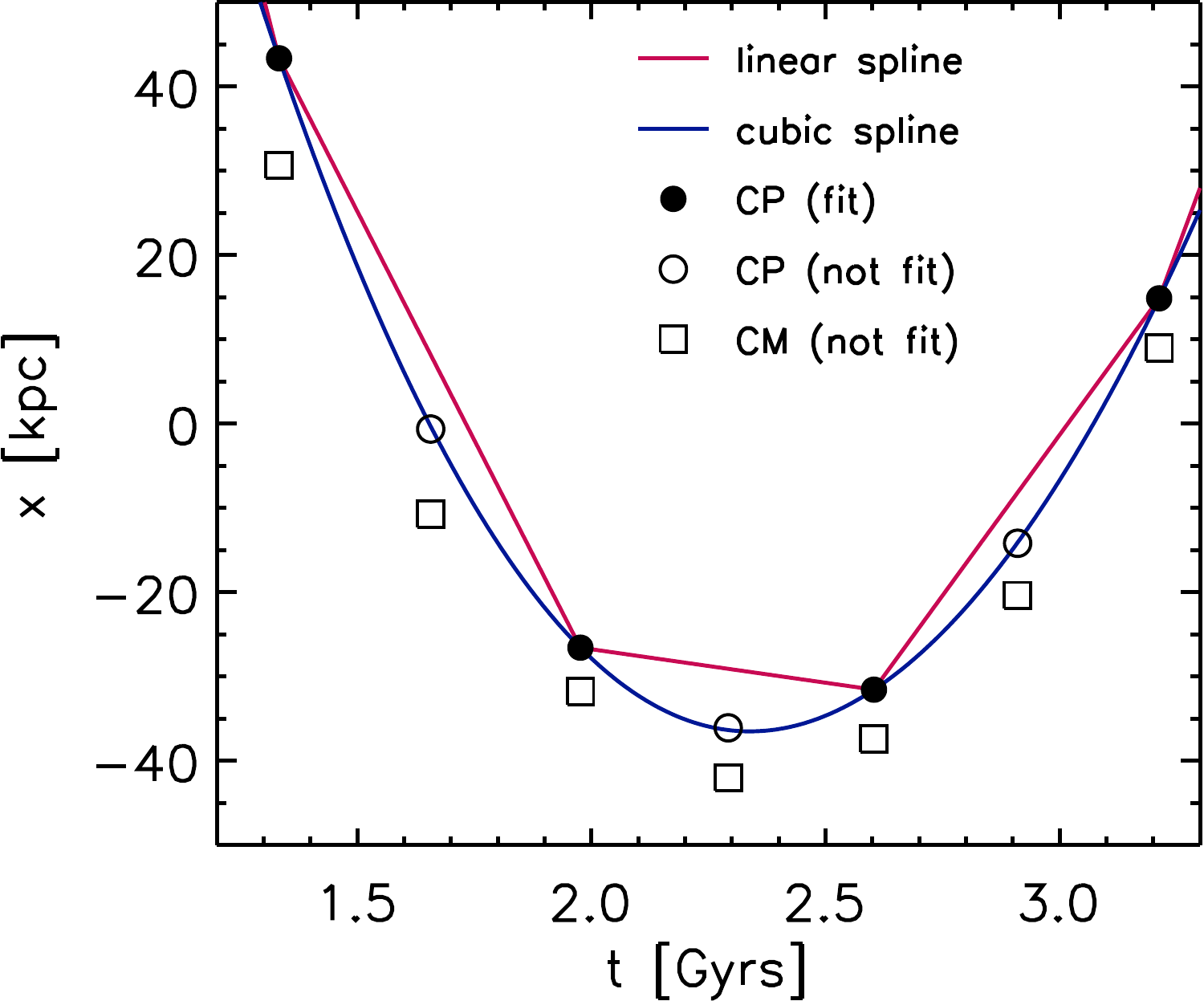}
  \end{center}
  \caption{Evolution of the centre of potential (CP, circles) and
    centre of mass (CM, squares) of one of the host haloes, as a
    function of lookback time. For illustration, a linear
    least-squared fit to the centre of potential has been
    subtracted. The red and blue lines show a linear and a cubic fit
    to the CP at the times indicated by filled circles, the open
    symbols show intermediate times not used in the fit. The cubic
    spline accurately predicts the CP at the intermediate points to
    less than 1~kpc, while the distance between the CP and the CM can
    exceed 10~kpc. \label{fig:interpolation-central}}
\end{figure}

In Figure~\ref{fig:interpolation-central}, we show the evolution of
the CP and CM of one of the host haloes during a time interval of
$\sim 2$ Gyr, with symbols indicating the values at individual
snapshots, and lines showing the intermediate values obtained via
interpolation. For illustration purposes, a linear least-squared fit
to the CP has been subtracted from the reference frame. Red and blue
lines show linear and cubic spline interpolations to those CP
coordinates which are represented by filled circles. Open circles
denote intermediate CP coordinates used only for validation of the
interpolation. Using only half of the snapshots and cubic splines, the
difference between the true and interpolated values of CP is under
1~kpc. As noted above, Figure~\ref{fig:interpolation-central} also
shows that the separation between the CM and CP can be $\sim10$~kpc,
making the CM frame a poor choice for the motion of satellites in the
inner tens of kpc of a halo.

\begin{figure}
  \begin{center}
    \includegraphics*[width = \columnwidth]{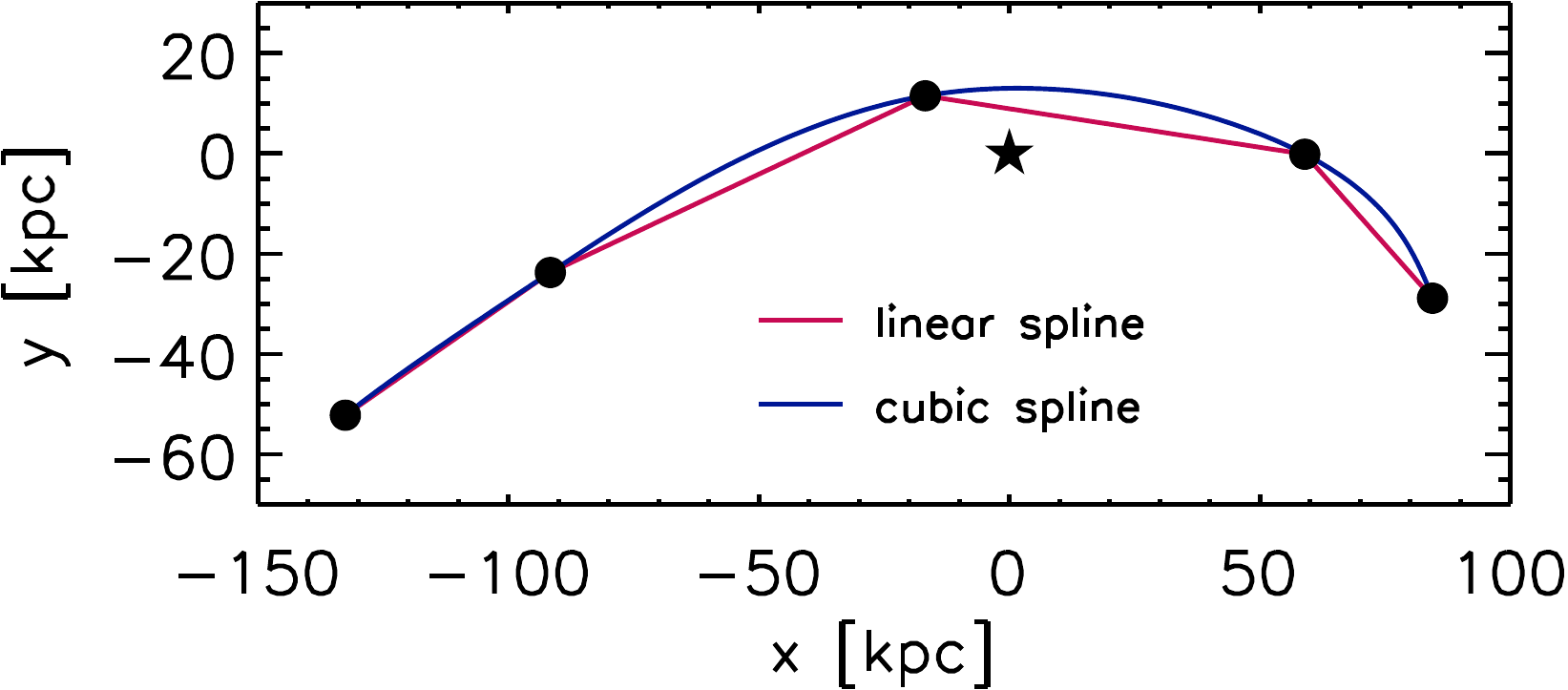} \\
    \includegraphics*[width = \columnwidth]{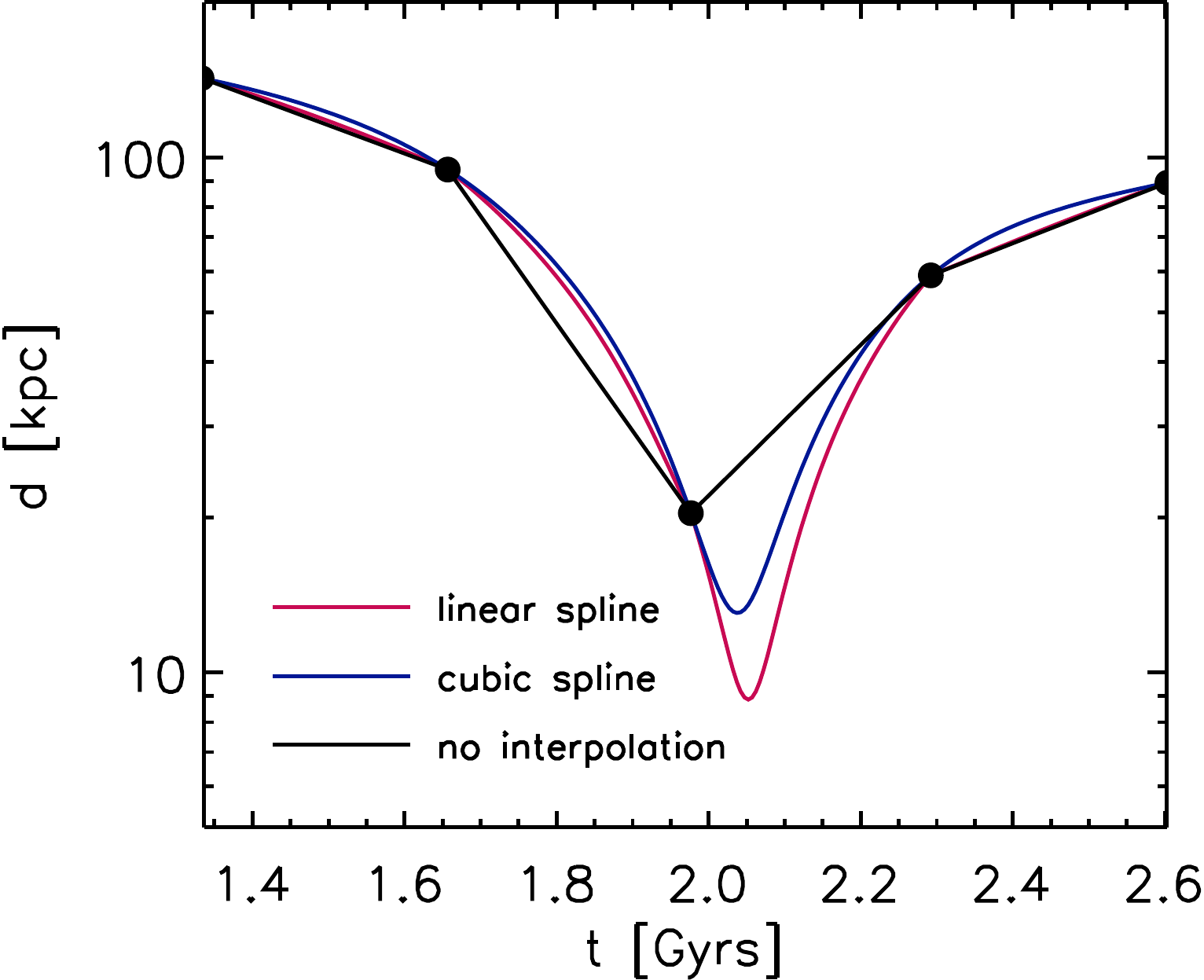} 
  \end{center}
  \caption{Top: circles denote positions at five snapshots of a
    subhalo near pericentre relative to the host halo CP marked by the
    star symbol. Positions in between snapshots are interpolated using
    linear (red) and third oder (blue) splines. Bottom: distance of
    the satellite to the host as a function of time, assuming no
    interpolation (black), or using the above interpolations with
    corresponding colours. Accurate estimates of the orbit near
    pericentre requires higher order interpolation; not interpolating
    overestimates the true pericentre distance, while linear
    interpolation underestimates
    it. \label{fig:interpolation-satellites}}
\end{figure}

\subsection{Orbit interpolations}
In Figure~\ref{fig:interpolation-satellites} we illustrate the
importance of accurate orbital interpolation for measuring the orbital
evolution, and hence the abundance and velocities of subhaloes near the
halo centre. In the top panel, we show the positions of a subhalo near
pericentre, relative to the host halo CP at five snapshots. Connecting
lines show reconstructions of the orbit using linear (assuming
constant velocity), and cubic (assuming acceleration that changes at
most linearly) interpolations. 

The bottom panel of Figure~\ref{fig:interpolation-satellites} shows
the distance of the satellite to centre as a function of time,
resulting from the different interpolation schemes, and also assuming
no interpolation. Without interpolation, the measured pericentre
distance is only an upper bound to the true value, so the abundance of
subhaloes near the centre is almost always systematically
underestimated. Using linear interpolation, the pericentre of a
parabolic orbit is underestimated, unless the time intervals are so
long that the entire pericentre passage is missed (consider, in the
top panel of Figure~\ref{fig:interpolation-satellites}, a straight
line between the first and final data points). As a result, with
sufficiently small but finite timesteps, linear interpolation
systematically underestimates the distance, and hence overestimates
the abundance of substructures near the centre. Naturally, we have
assumed that the reference frame, i.e. the host CP itself, is known
accurately at all times; otherwise additional errors arise.

\section{Comparison to Maxwellian and Gaussian Velocity Distributions}
\label{appendix:maxwell}

\begin{figure*}
\vspace{-1mm}
  \begin{center}
    \includegraphics*[trim = 0mm 0mm 0mm 0mm, clip, height = 0.305\textwidth]{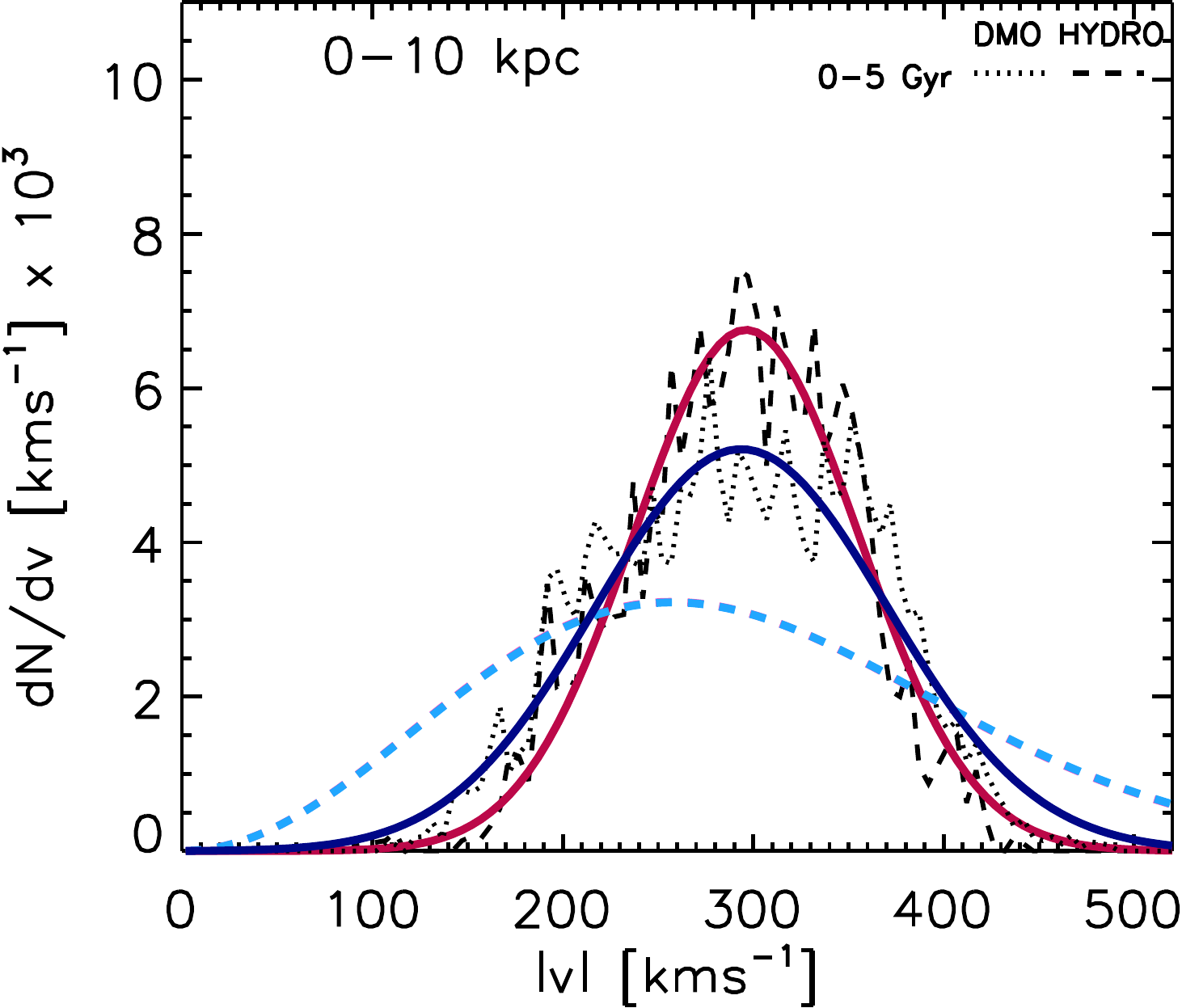} 
    \includegraphics*[trim = 0mm 0mm 0mm 0mm, clip, height = 0.305\textwidth]{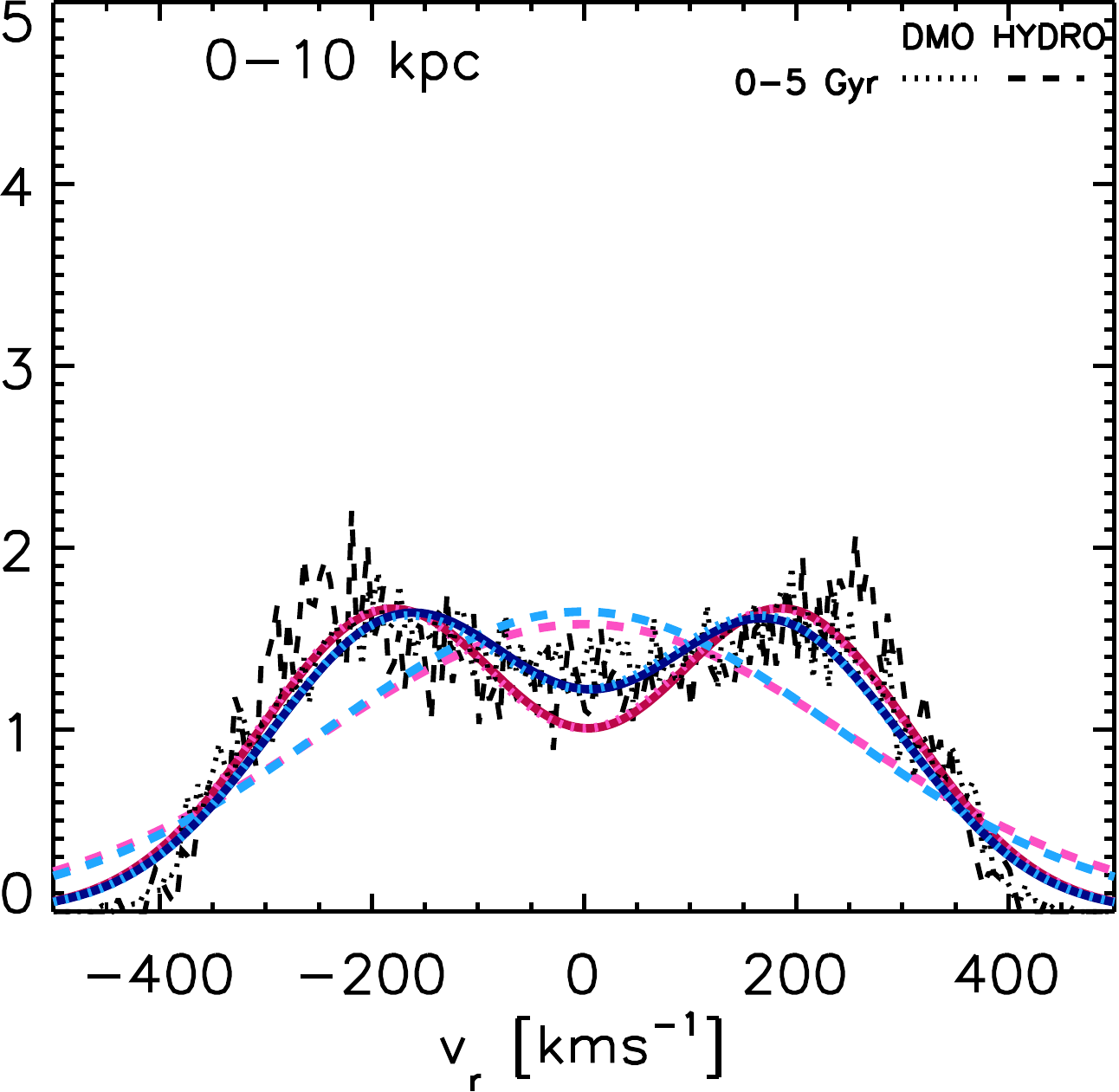} 
    \includegraphics*[trim = 0mm 0mm 0mm 0mm, clip, height = 0.305\textwidth]{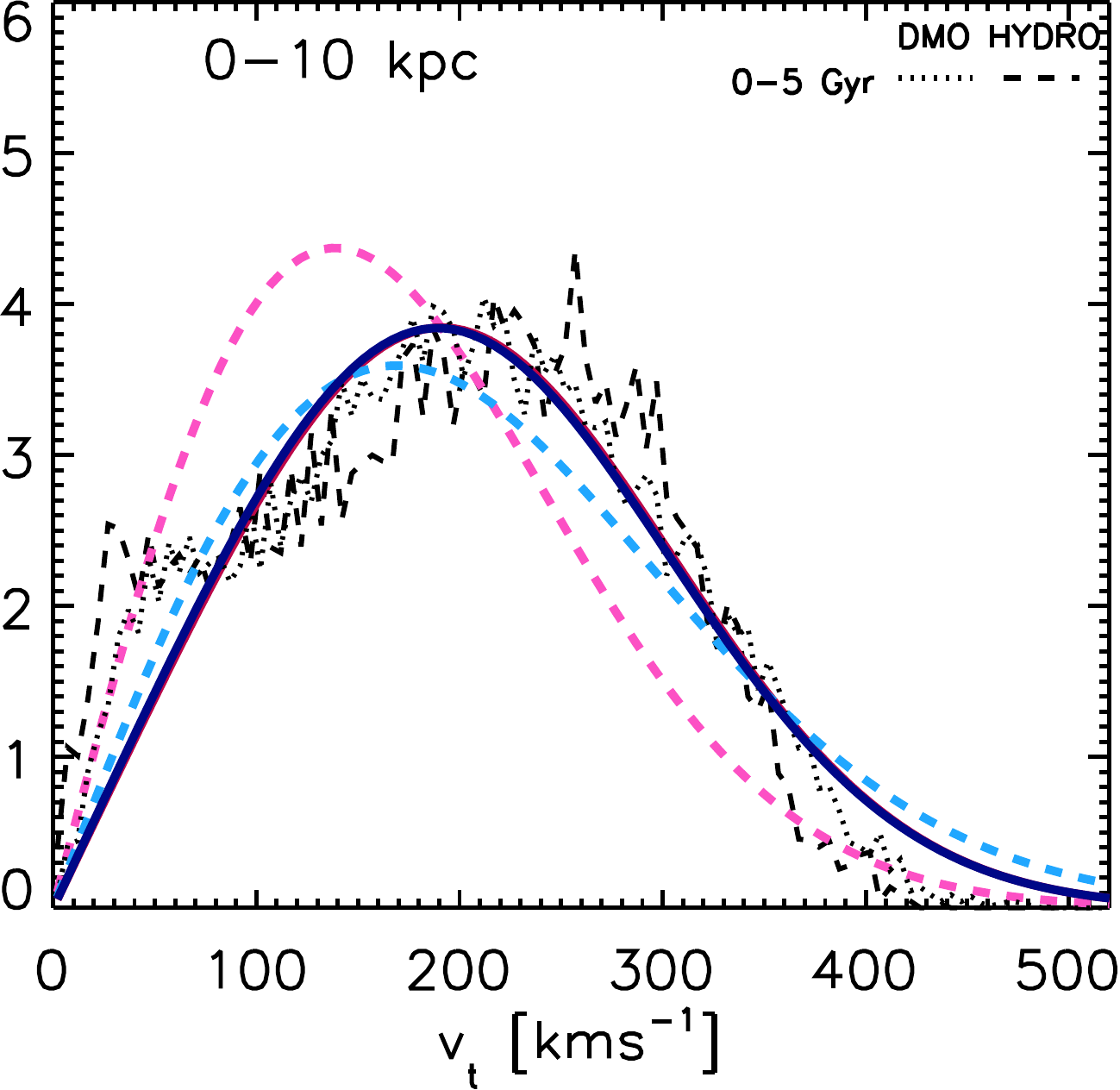} \\ 
    \includegraphics*[trim = 0mm 0mm 0mm 0mm, clip, height = 0.305\textwidth]{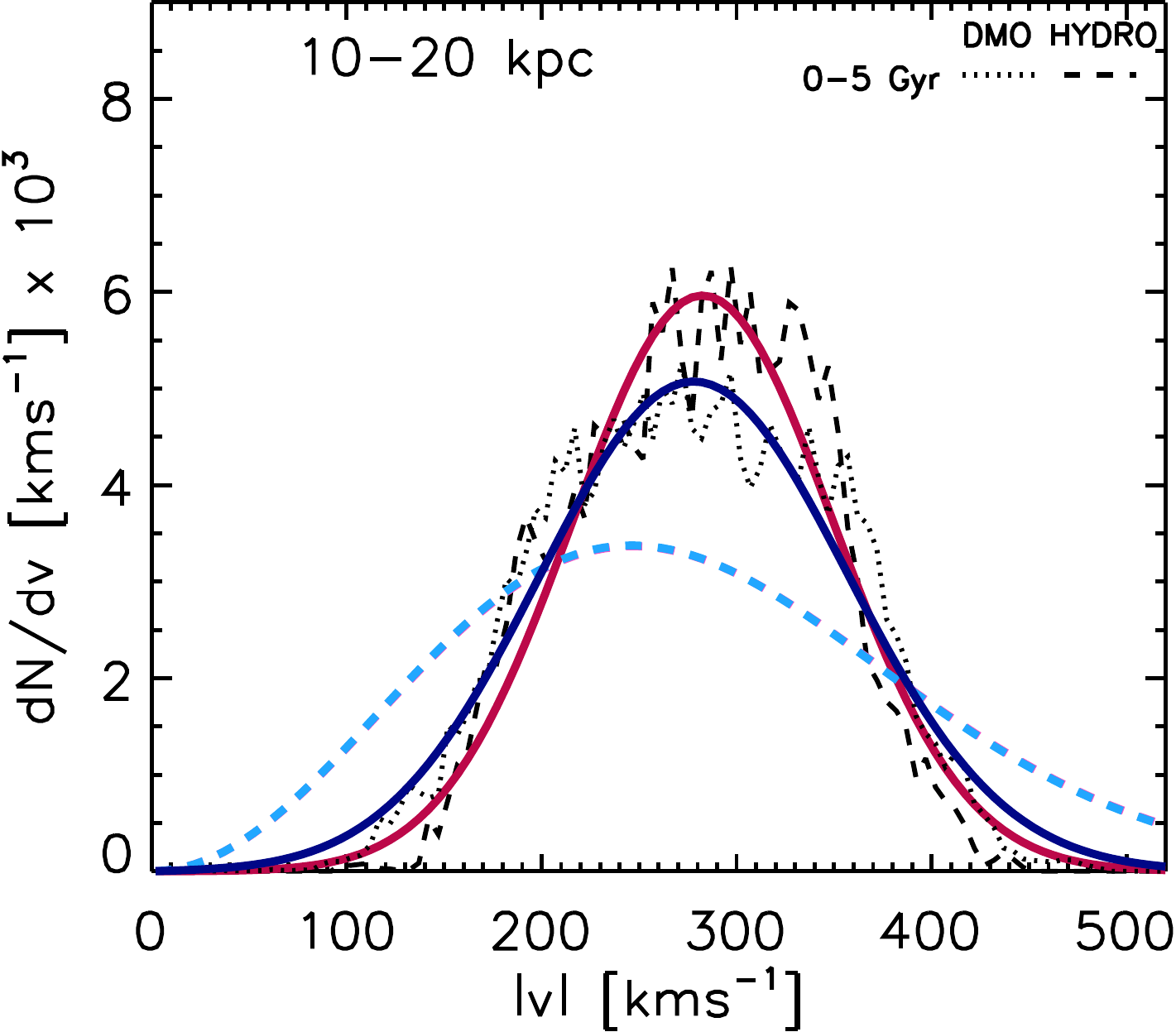} 
    \includegraphics*[trim = 0mm 0mm 0mm 0mm, clip, height = 0.305\textwidth]{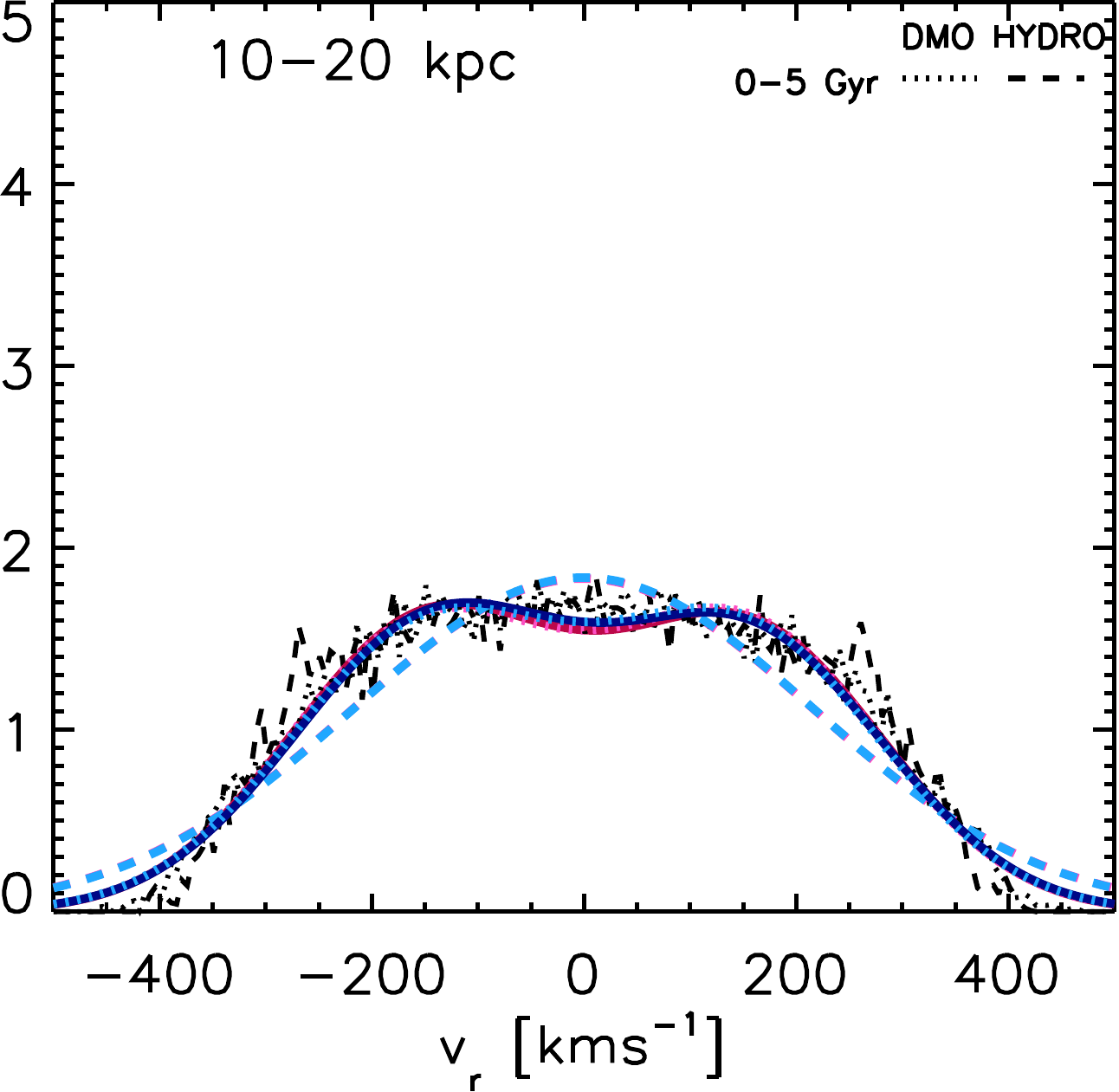} 
    \includegraphics*[trim = 0mm 0mm 0mm 0mm, clip, height = 0.305\textwidth]{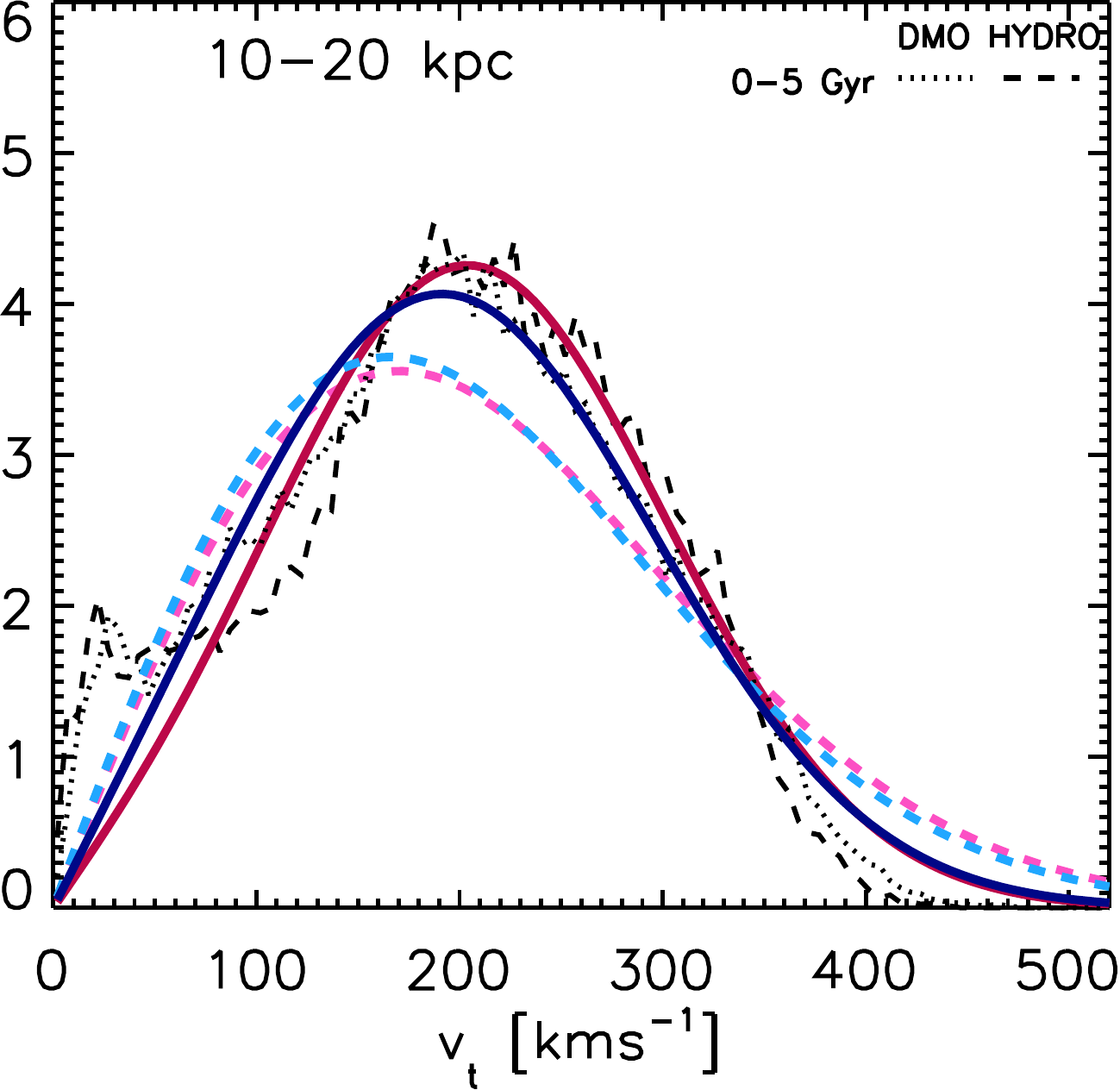} \\ 
    \includegraphics*[trim = 0mm 0mm 0mm 0mm, clip, height = 0.305\textwidth]{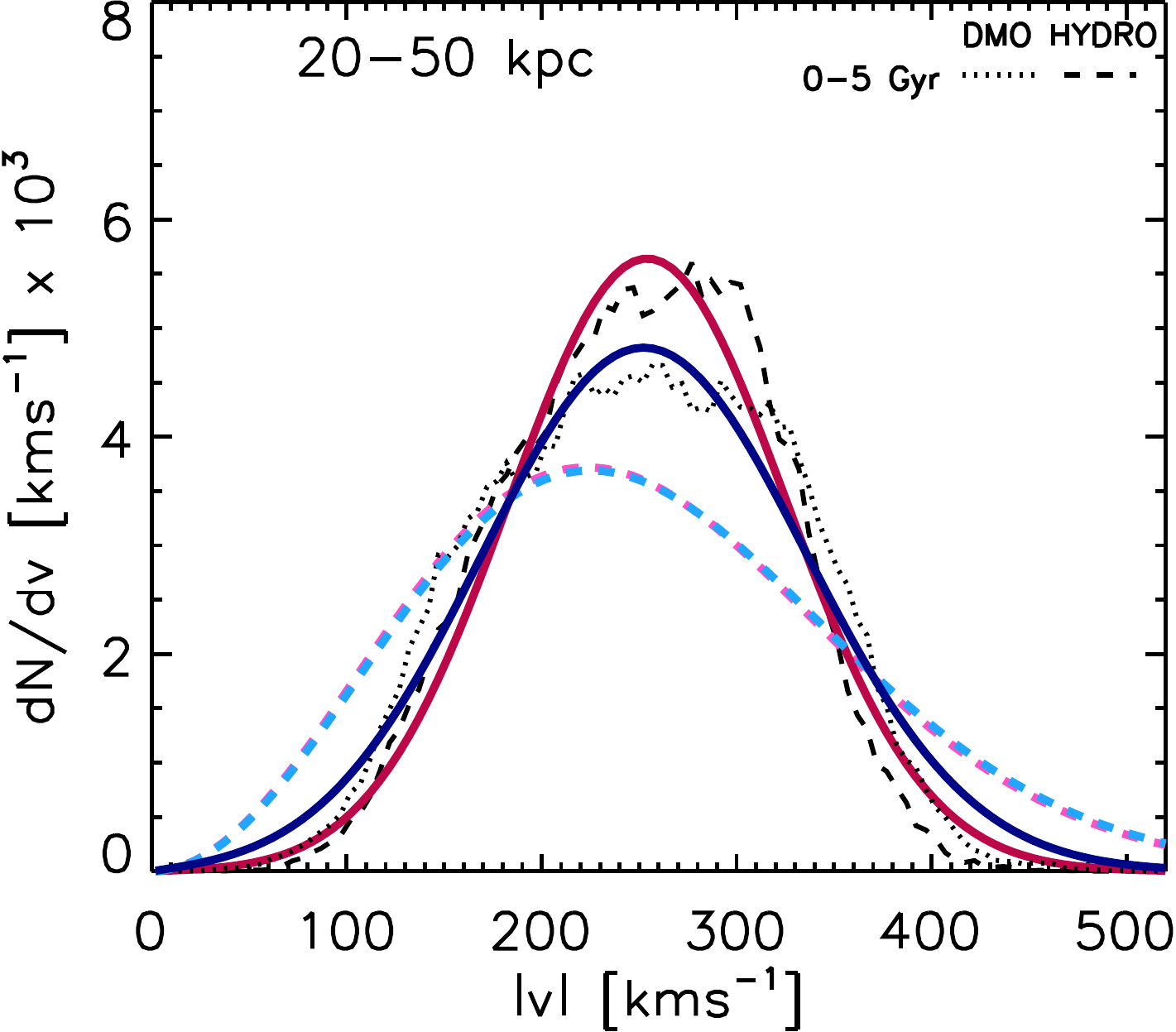} 
    \includegraphics*[trim = 0mm 0mm 0mm 0mm, clip, height = 0.305\textwidth]{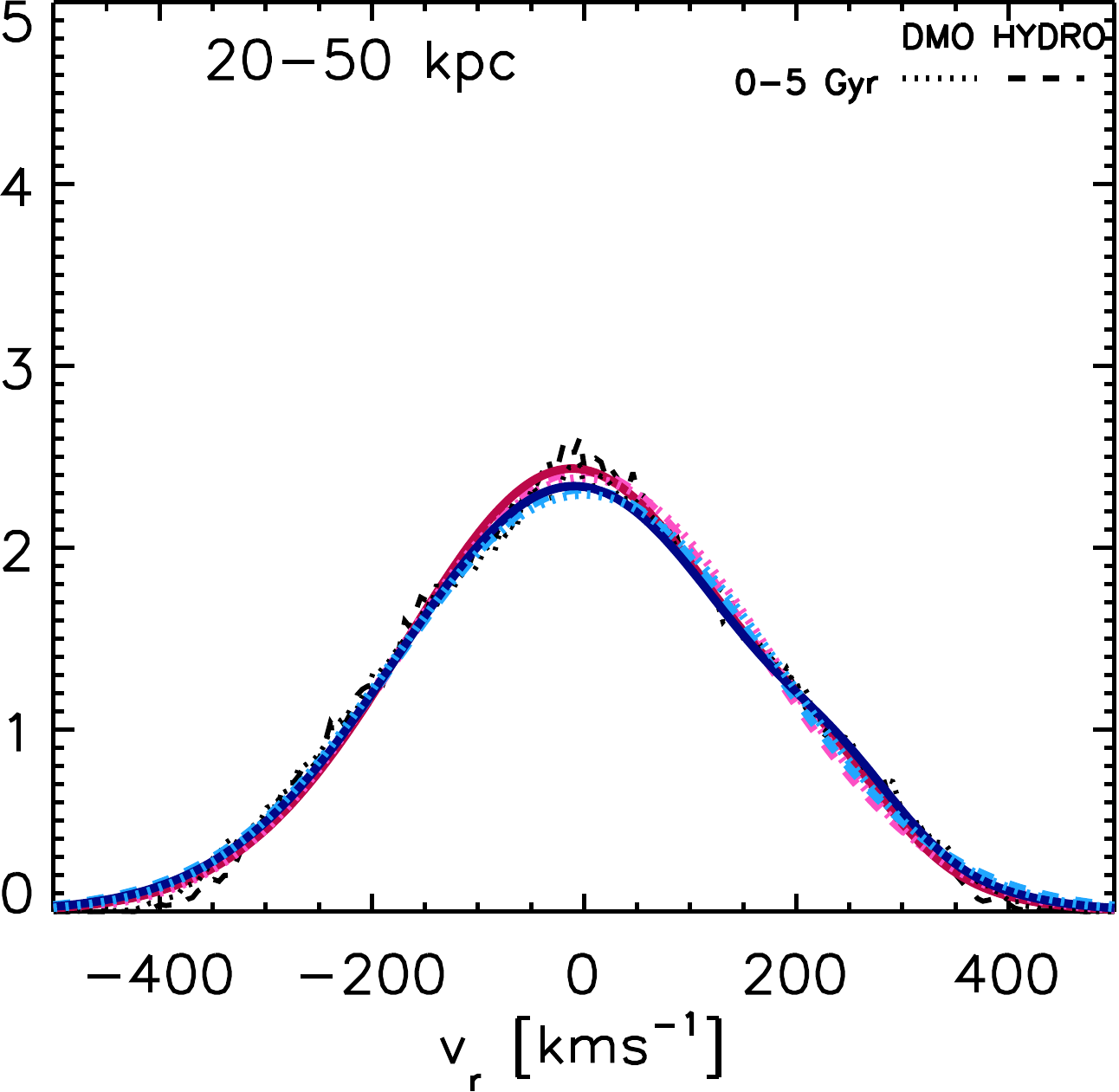} 
    \includegraphics*[trim = 0mm 0mm 0mm 0mm, clip, height = 0.305\textwidth]{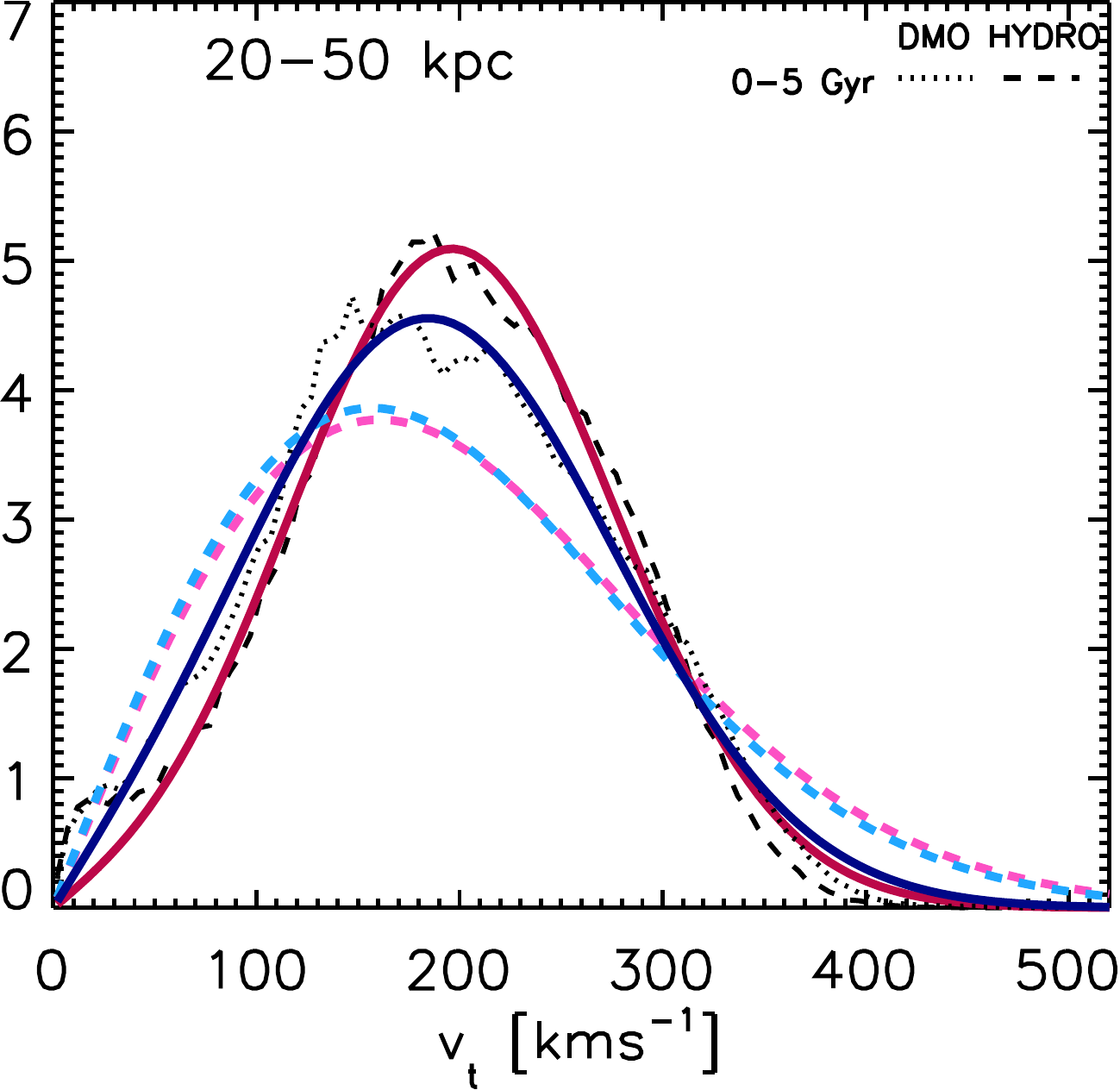} \\ 
    \includegraphics*[trim = 0mm 0mm 0mm 0mm, clip, height = 0.305\textwidth]{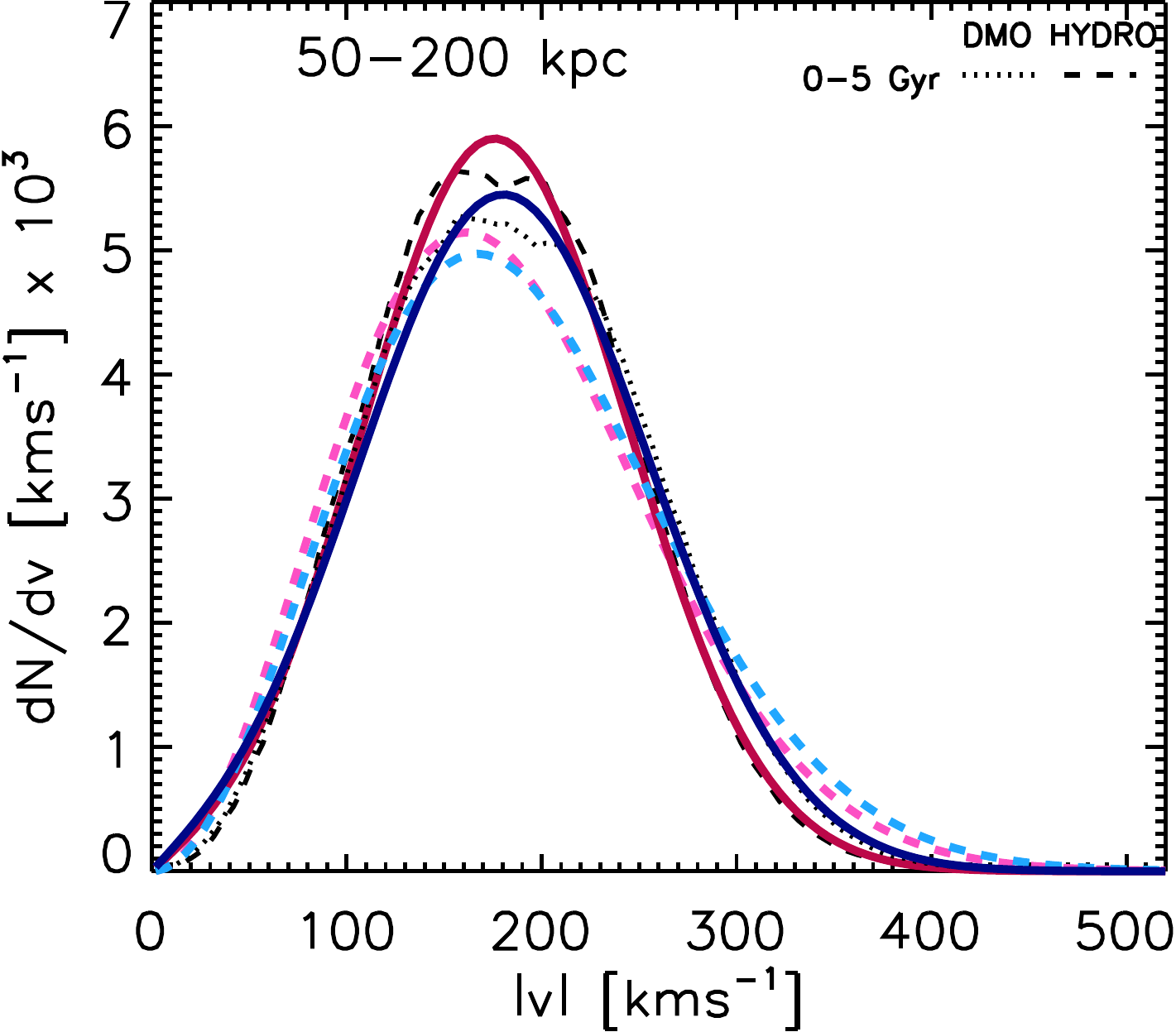} 
    \includegraphics*[trim = 0mm 0mm 0mm 0mm, clip, height = 0.305\textwidth]{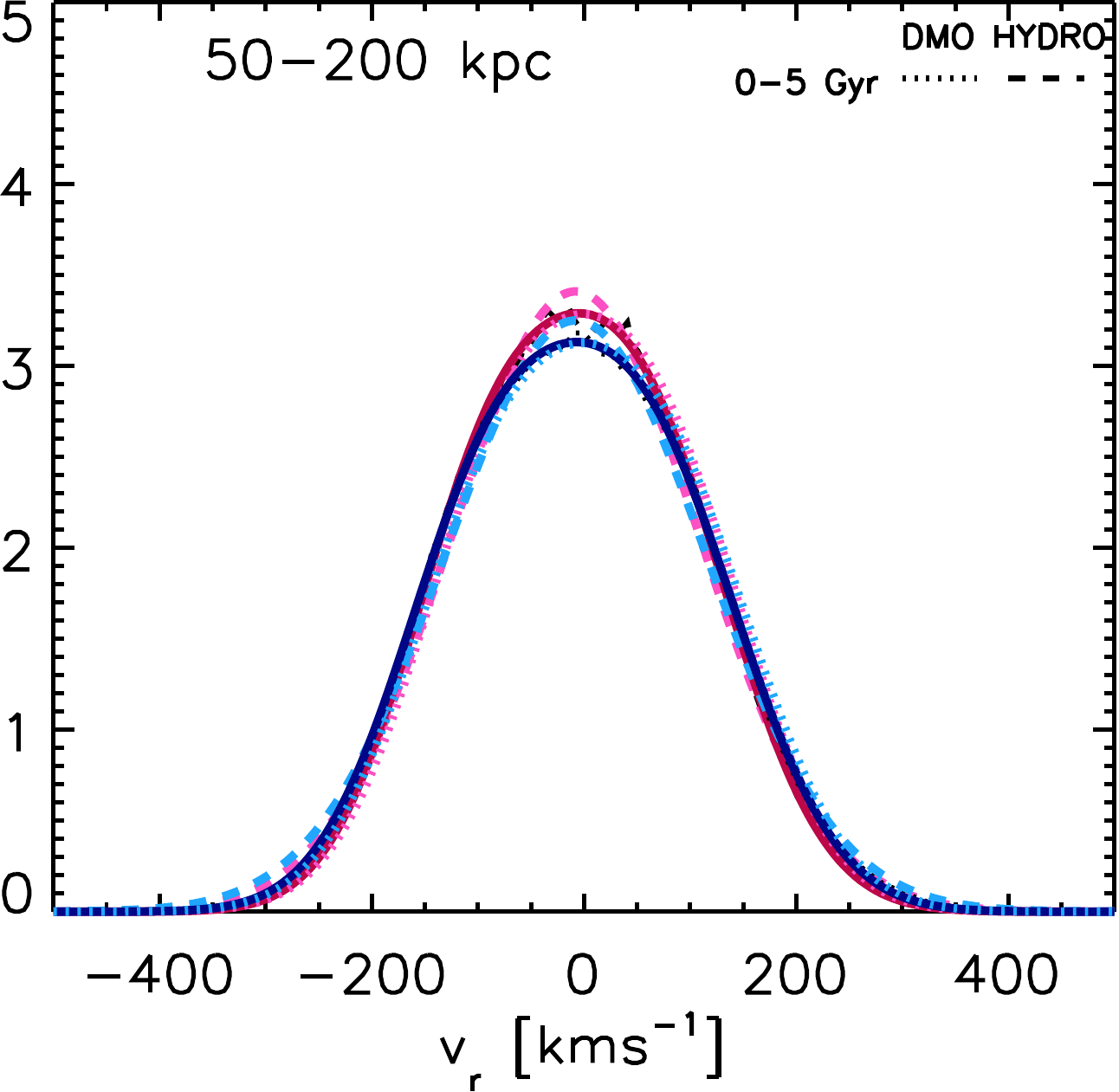} 
    \includegraphics*[trim = 0mm 0mm 0mm 0mm, clip, height = 0.305\textwidth]{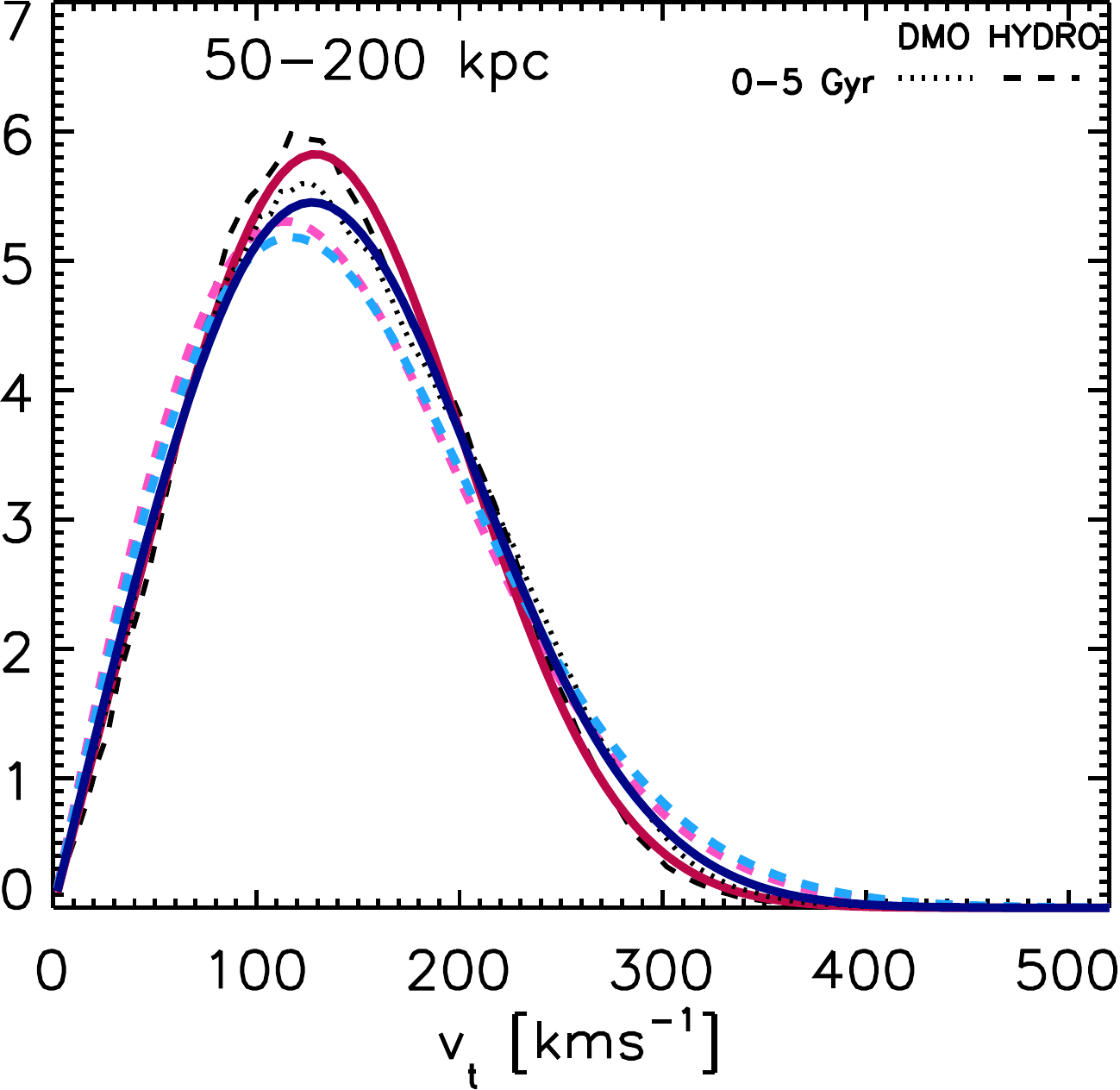} \\ 
  \end{center}
\vspace{-1.5mm}
\caption{Probability density functions (PDFs) for $|v|$ (left column),
  $v_r$ (middle column) and $v_t$ (right column), as shown in
  Figure~\ref{fig:velocities}, averaged over 5 Gyr in lookback
  time. Black dotted and dashed lines show our simulation results in
  the DMO and hydrodynamic simulations, respectively. Dark blue and
  dark red solid lines show the fits to Double-Gaussian or Rician
  distribution functions to the DMO and hydrodynamic data,
  respectively as described in Section~\ref{sec:velocities}. Lighter,
  dashed coloured lines show the corresponding fits to 3D-Maxwellians
  (for $|v|$, left column), a single Gaussian with free parameters
  $\mu$ and $\sigma$ (for $v_r$, middle column), and to 2D-Maxwellians
  (for $v_t$, right column).
      \label{fig:velocities-appendix}}
\label{lastpage}
\end{figure*}

In Figure~\ref{fig:velocities-appendix}, we repeat the time-averaged
probability density functions for the subhalo velocities, as shown in
Figure~\ref{fig:velocities}, and compare the fits used in this work to
Gaussian and Maxwellian fits. It can be seen that, at large radii, the
Rician and Double-Gaussian distributions approach the Gaussian and
Maxwellian approximations, but at small radii, the latter completely
fail to reproduce the data.

\end{document}